\documentclass[sigconf]{acmart}
\usepackage{leftidx}
\usepackage{dsfont}
\usepackage{array}
\usepackage{amsmath}
\usepackage[linesnumbered,ruled,vlined]{algorithm2e}
\usepackage{relsize}
\usepackage{multirow}
\usepackage{booktabs}
\usepackage{graphicx}
\usepackage{diagbox}
\usepackage{xcolor}
\usepackage{colortbl} 
\usepackage{enumitem}
\usepackage{pifont}
\usepackage{tikz}
\usepackage{listings}
\usepackage[dvipsnames]{xcolor}
\usepackage{xspace}
\usepackage{rotating}
\usepackage{adjustbox} 
\usepackage{soul}
\usepackage{lscape}
\usepackage{pgfplots}
\usepackage{pgfplotstable}
\usepackage{bbm}
\usepackage[font=small]{subcaption}
\usepackage[font=small]{caption}
\usepackage{xstring}
\usepackage{subcaption}
\usetikzlibrary{patterns}
\usepackage{lipsum}
\usepackage{pdflscape}
\usepackage[most]{tcolorbox}

\SetKwInOut{Input}{Input}
\SetKwInOut{Output}{Output}

\SetCommentSty{mycommfont}

\newtheorem{example}{Example}[section]
\newtheorem{problem}{Problem}[section]
\usetikzlibrary{shapes.geometric, arrows, positioning, calc}
\SetKwComment{tcc}{\textcolor{darkgray}{\texttt{\slash * }}\textcolor{darkgray}{}}{ \textcolor{darkgray}{*\slash}}

\newcommand{\refsub}[2]{\ref{#1}(\subref{#2})}

\newcommand{\sysName}{SAGE\xspace}
\newcommand{\sysNameP}{$\text{SAGE}^{+}$\xspace}
\newcommand{\mycircle}[1]{{\Large\textcircled{{\small#1}}}}
\DeclareMathOperator*{\argmax}{arg\,max} 

\AtBeginDocument{%
  }

\setcopyright{acmlicensed}
\copyrightyear{2026}
\acmYear{2026}
\acmDOI{XXXXXXX.XXXXXXX}

\acmConference[SIGMOD '26]{International Conference on Management of Data}{May 31--June 5, 2026}{Bengaluru, India}
\acmISBN{978-1-4503-XXXX-X/2018/06}

\begin{document}
	
\title[Data-Semantics-Aware Recommendation of Diverse Pivot
Tables]{Data-Semantics-Aware Recommendation of Diverse
Pivot Tables}
%

\author{Whanhee Cho}
\affiliation{%
  \institution{University of Utah}
  \city{Salt Lake City}
  \state{Utah}
  \country{USA}}
\email{whanhee@cs.utah.edu}

\author{Anna Fariha}
\affiliation{%
  \institution{University of Utah}
  \city{Salt Lake City}
  \state{Utah}
  \country{USA}}
\email{afariha@cs.utah.edu}


\newcommand{\green}{Green}
\newcommand{\blue}{blue}
\newcommand{\red}{red}
\newcommand{\purple}{Magenta}

\renewcommand{\green}{black}
\renewcommand{\blue}{black}
\renewcommand{\red}{black}
\renewcommand{\purple}{black}

\newcommand{\reviseone}[1]{{\color{\green} #1}}
\newcommand{\revisetwo}[1]{{\color{\blue} #1}}
\newcommand{\revisethree}[1]{{\color{\red} #1}}
\newcommand{\revisemeta}[1]{{\color{\purple} #1}}

\begin{abstract}

Data summarization is essential to discover insights from large datasets. In
spreadsheets, \emph{pivot tables} offer a convenient way to summarize tabular
data by computing aggregates over some attributes, grouped by others. However,
identifying attribute combinations that will result in \emph{useful} pivot
tables remains a challenge, especially for high-dimensional datasets. We
formalize the problem of automatically recommending \emph{insightful} and
\emph{interpretable} pivot tables, eliminating the tedious manual process. A
crucial aspect of recommending a \emph{set} of pivot tables is to
\emph{diversify} them. Traditional work inadequately address the
table-diversification problem, which leads us to the problem of \emph{pivot
table diversification}.

We present \sysName, a data-\underline{s}emantics-\underline{a}ware system for
recom\-mendin\underline{g} k-budgeted div\underline{e}rse pivot tables,
overcoming the shortcomings of prior work for top-k recommendations that cause
redundancy. \sysName ensures that each pivot table is \emph{insightful},
\emph{interpretable}, and \emph{adaptive} to the user's actions and preferences,
while also guaranteeing that the set of pivot tables are different from each
other, offering a \emph{diverse} recommendation. We make two key technical
contributions:
(1)~a \emph{data-semantics-aware model} to measure the utility of a single pivot
table and the diversity of a set of pivot tables, and
(2)~a \emph{scalable greedy algorithm} that can efficiently select a set of
diverse pivot tables of high utility, by leveraging data semantics to
significantly reduce the combinatorial search space.
Our extensive experiments on four real-world datasets show that \sysName
outperforms alternative approaches, and efficiently scales to accommodate
high-dimensional datasets. Additionally, through multiple case studies, we
demonstrate \sysName's qualitative superiority over existing tools, and through
a user study, we validate its practical usefulness and alignment with user
preferences.
%

\end{abstract}


\maketitle

\section{Introduction} 
\label{sec:introduction}
Data is at the heart of data-driven decision making. We rely on trends observed
in the data to obtain \emph{insights}~\cite{QuickInsightsDing19} that help us
make informed decisions. However, due to limitations in human
comprehensibility, data must be
\emph{summarized}~\cite{DBLP:journals/tkde/JoglekarGP19,
DBLP:journals/vldb/Sarawagi01, DBLP:journals/pvldb/GebalyAGKS14} to enable
humans discover insights from the summaries, either directly or via
visualizations~\cite{SeedbVartak15} over the summaries. One of the most common
techniques to summarize data is \emph{aggregation}. Simple aggregations involve
functions (e.g, {\small \texttt{SUM}}) to aggregate all rows. More nuanced
aggregations involve multiple groupings of the entities (e.g., {\small
\texttt{GROUP BY GENDER, EDUCATION}}) and then aggregating each group
separately.


While SQL provides functionalities for any custom aggregation query, it is not
suitable for novices due to interface-related limitations. Thanks to the
ubiquity of spreadsheet software---such as Microsoft
Excel~\cite{microsoft_excel}, Google Sheets~\cite{google_sheets}, Apple
Numbers~\cite{apple_numbers}, etc.---a substantial portion of businesses
(around 60\%~\cite{shadow_it_spreadsheets}) and about 2 billion
people~\cite{mooc_excel_business} use spreadsheets for data management and
analysis. These users rely on \emph{pivot tables}, a summary of tabular data
that computes aggregates over a few data attributes, grouped by other data
attributes. Most commercial spreadsheets include a built-in and user-friendly
mechanism to construct pivot tables. Spreadsheet pivot tables are particularly
suitable for novices, where they can rearrange, group, and aggregate data using
intuitive interfaces such as drag-and-drop. Unlike SQL aggregates, spreadsheet
pivot tables offer dynamic user interactions, allowing interactive exploration
such as drilling down, filtering, sorting, etc.

\looseness-1 In an exploratory setting with the goal to discover interesting
data trends, a key challenge in constructing insightful pivot tables lies in
selecting the right \emph{parameters}, i.e., determining which attributes to
use for groupings and aggregations. This task becomes even harder for users
\revisetwo{who lack knowledge of how the data is represented (especially when
the data contains missing or cryptic attribute names)}, lack domain knowledge,
or work with high-dimensional data. In such cases, users must manually explore
a vast space of parameter combinations through a tedious trial-and-error
process. This involves experimenting with various combinations of (1)~grouping
attributes, (2)~aggregation attributes, and (3)~aggregation functions, then
manually assessing the insightfulness of the resulting pivot tables. We
illustrate this challenge with Example~\ref{ex:manual_search}.


\begin{figure}[t]
        \centering
    \resizebox{0.45\textwidth}{!}{%
        \begin{tabular}{llrrrllr}
            \toprule
            \multicolumn{1}{c}{\textbf{ID}} &
            \multicolumn{1}{c}{\textbf{Gender}} &
            \multicolumn{1}{c}{\textbf{Age}} &
            \multicolumn{1}{c}{\textbf{Experience}} &
            \multicolumn{1}{c}{\revisetwo{\dots}} &
            \multicolumn{1}{c}{\textbf{Degree}} &
            \multicolumn{1}{c}{\textbf{Department}} &
            \multicolumn{1}{c}{\textbf{Salary}} \\ \midrule 
			1 & Male 	& 48 & 3  & \revisetwo{\dots} & PhD 	& IT 	& \$50,000 \\
            2 & Female 	& 32 & 1  & \revisetwo{\dots} & MS	 	& Sales & \$20,000 \\
            3 & Male 	& 45 & 12 & \revisetwo{\dots} & PhD     & HR 	& \$100,000\\
            \bottomrule
        \end{tabular}%
    }
     \vspace{-3mm}
	 \caption{A sample table from an employee compensation dataset.}
     \vspace{2mm}	
    \label{tab:salary_database}	
	
	\centering
	\includegraphics[width=0.85\columnwidth]{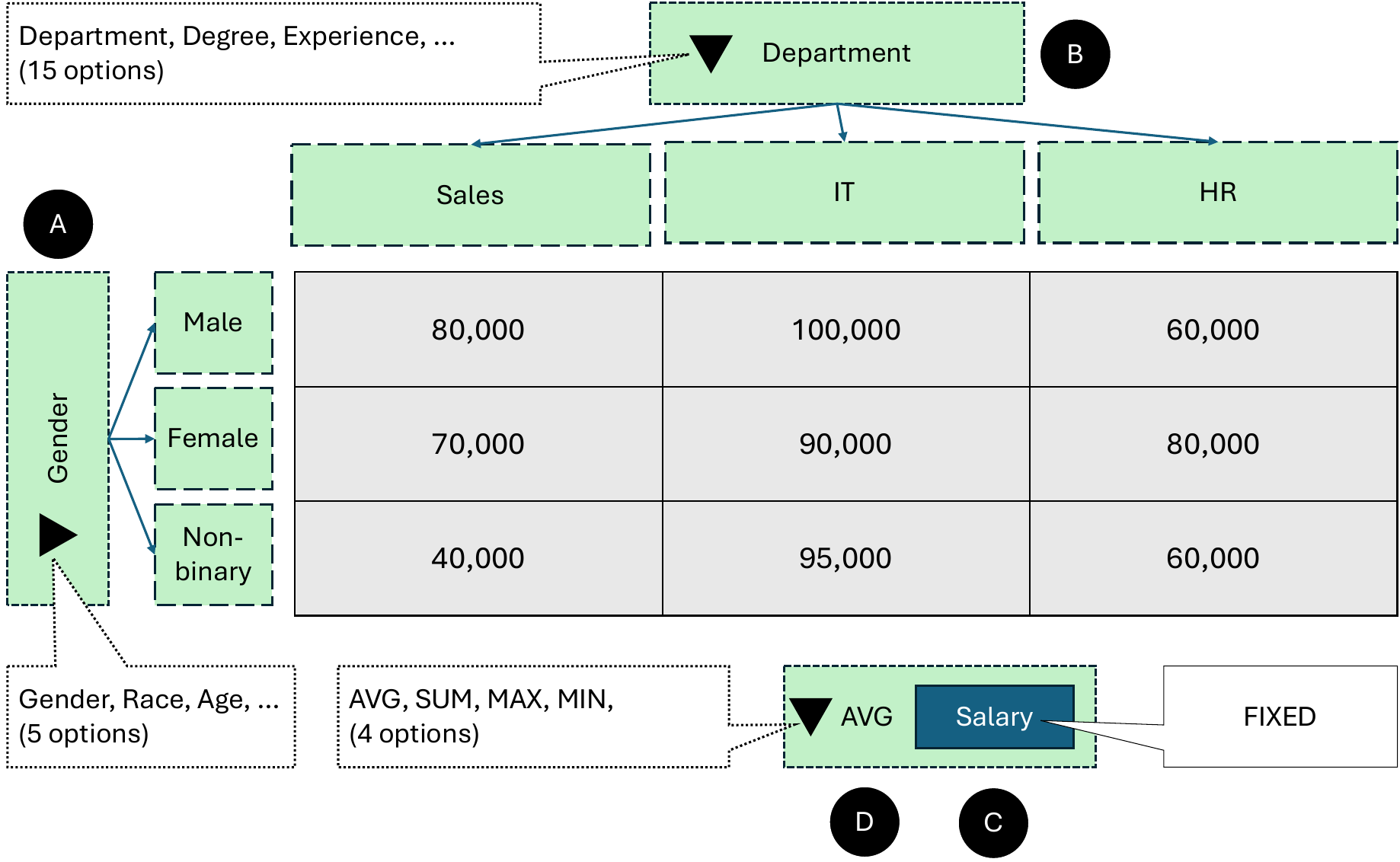}
	 \vspace{-4mm}
	 \caption{\small A pivot table requires 4 parameters: 
	 \mycircle{A}~row-groups,
	 \mycircle{B}~column-groups, 
	 \mycircle{C}~aggregate attributes, and 
	 \mycircle{D}~aggregate functions for each aggregate attribute. 
	 Users can choose multiple values for each parameter. In
	 Example~\ref{ex:manual_search}, Sasha has fixed \mycircle{C}, but still needs
	 to explore \mycircle{A} (5 options), \mycircle{B} (15 options),
	 and \mycircle{D} (4 options).}
	 \vspace{-4mm} \label{fig:ex_pivot_table} 

\end{figure}
 

\begin{example} \label{ex:manual_search} 
\looseness-1 Sasha is investigating potential factors affecting salary across
various groups in an employee compensation dataset over $21$ attributes
including \texttt{ID}, \texttt{Gender}, \texttt{Age}, \texttt{Experience},
\texttt{Degree}, \texttt{Depart\-ment}, \texttt{Salary}, etc.\
(Figure~\ref{tab:salary_database}). With an aim to discover salary
discrepancies across various group combinations, she starts with the pivot
table shown in Figure~\ref{fig:ex_pivot_table}: she puts \texttt{Gender} in the
{row-groups} \mycircle{A} and \texttt{Department} in the column-groups
\mycircle{B}; and chooses \texttt{Salary} as an {aggregate attribute}
\mycircle{C} and \texttt{Average} as the {aggregate function} \mycircle{D}.
 
Sasha is interested in \texttt{Salary} discrepancies, so her choice for
\mycircle{C} is fixed. However, she still needs to explore various combinations
for \mycircle{A}, \mycircle{B}, and \mycircle{D}. Sasha wishes to put
demographic attributes (e.g., \texttt{Gender}, \texttt{Race}, \texttt{Age},
\texttt{Marital Status}, etc.) in the row-groups, as any discrepancy across
different rows will indicate discrimination, and all other attributes in the
column-groups. 

Sasha decides to explore $5$ demographic and $15$ non-demographic attributes,
as well as $4$ aggregation functions: \texttt{MAX}, \texttt{MIN},
\texttt{AVERAGE}, and \texttt{SUM}. This leaves her $\;5 \times 15 \times 4 =
300$ possible combinations,\footnote{Sasha chose only one option for each
parameter. Multiple options (e.g., \texttt{Gender} and \texttt{Race} for
row-groups) will further increase the search space of possible pivot tables.}
and she must carefully inspect each pivot table to identify salary
discrepancies, by manually contrasting the pivot table cells. Assuming each
pivot table has $10$ cells on average and it takes about $2$ minutes to examine
each pivot table, Sasha needs $300 \times 2 = 600$ minutes ($10$ hours)!

\end{example}

\subsubsection*{\textbf{Recommending Pivot Tables.}}
Example~\ref{ex:manual_search} highlights the need for a \emph{recommendation}
system that can automatically suggest the ``best'' pivot tables. While existing
spreadsheet software, such as Microsoft Excel and Google Sheets, are equipped
with features for automatic pivot table recommendation, they have several
shortcomings, which we show next in Example~\ref{ex:recommendation}.


\begin{example}\label{ex:recommendation} Frustrated by manual exploration,
Sasha tries the pivot table recommendations in Google Sheets
(Figure~\ref{fig:software_problems}), obtaining three
recommendations.\footnote{Test conducted in February 2025.} However, the
recommended tables often include too many aggregated attributes beyond her
desired \texttt{Salary}, resulting in convoluted and large tables. Sasha also
observes that most recommendations are redundant---they default to the
groupings by \texttt{Gender} or \texttt{Department}---and lack diversity,
causing her to miss out on insights involving other data attributes. While
Microsoft Excel provides nine recommendations,\footnote{Tested on Microsoft
Excel (Windows) version 2501, February 2025.} it utilizes only 10 out of 21
possible attributes. Additionally, it suggests meaningless aggregations like
{\small \texttt{Sum(Employed Year)}} and {\small \texttt{Sum(Age)}}, revealing
its shortcoming in grasping the semantics. For both MS Excel and Google Sheets,
Sasha failed to specify \texttt{Salary} as her intended aggregate attribute,
restricting her ability to steer the recommendations towards her needs. Lastly,
Sasha asks ChatGPT for three ``insightful'' and ``diverse'' pivot tables,
focusing on average \texttt{Salary}. Apparently reasonable at first, she soon
realizes that the values of the pivot tables are hallucinated,\footnote{Tested
on ChatGPT (GPT-4o), February 2025.} exposing ChatGPT's lack of access to the
actual data and absence of result validation. Sasha concludes that LLMs are
ill-suited for this task, as they do not explicitly enumerate and evaluate all
possible options.

\end{example}

\begin{figure}[t]

    \resizebox{0.46\textwidth}{!}{%
    \centering
    \begin{tabular}{ll@{ }p{9cm}} 
		\toprule 
		\textbf{Tool} & \multicolumn{2}{l}{\textbf{Recommended Pivot Tables}} \\ 
		\midrule
		
		\multirow{7}{*}{Google Sheets~\cite{google_sheets}} & 
		(1) & Average of Age, Years of Experience, Annual Bonus, Overtime
		Hours, Sick Days, Training Hours, Satisfaction Score, \#Projects,
		\#Promotions by Gender \\ &
		(2) & Average of Performance Rating for each Gender by Department\\ &
		(3) & Average of Age, Years of Experience, Performance Rating, Salary,
		Annual Bonus, Overtime Hours, Sick Days, Training Hours, Satisfaction Score,
		\#Projects, \#Promotions by Gender \\
		
		\midrule
		
        \multirow{9}{*}{Microsoft Excel~\cite{microsoft_excel}} & 
		(1) & Count of ID by Degree\\ &
		(2) & Count of ID by Department\\ &
		(3) & Sum of Employed Year by Department\\ &
		(4) & Sum of \#Promotions by Employed Year and Degree\\ &
		(5) & Sum of Age, Children, Performance Rating by Degree\\ &
		(6) & Sum of Children, Performance Rating, Salary by Degree\\ &
		(7) & Sum of Children, Performance Rating, Salary by Department\\ &
		(8) & Sum of Salary by Employed Year and Gender\\ &
		(9) & Sum of Employed Year by Gender and Degree\\
		
		\midrule 

        \multirow{3}{*}{ChatGPT~\cite{OpenAIChat}} & 
		(1) & Average Salary by Years of Experience and Training Hours \\ &
        (2) & Average Salary by Department and Children\\ &
        (3) & Average Salary by Age and Satisfaction Score\\ 
		
		\bottomrule
		
    \end{tabular}}
	 \vspace{-3mm}
	 \caption{\small Google Sheets recommendations are redundant and convoluted;
	 Microsoft Excel includes meaningless recommendations such as
	 \texttt{SUM(Age)}; while ChatGPT recommendations look reasonable, they are
	 data-content-unaware as the pivot table values are hallucinated.}
	 \vspace{-4mm}	 
    \label{fig:software_problems}
\end{figure}


%

Example~\ref{ex:recommendation} highlights several key limitations of existing
tools for automatic recommendation of pivot tables.
First, they do not cater to the user needs for a \emph{focused} and
\emph{adaptive} recommendation of pivot tables.
Second, they focus on top-k recommendations~\cite{deshpande2004item,DAISYVLDB24Junjie,AutoSuggestSIGMODE2020Cong} and do not
consider \emph{diversification}~\cite{DrosouSIGMOD10Search}, which may cause the
users to miss certain data insights.
Finally, existing approaches do not fully leverage the data and its semantics to
ensure that the suggested pivot tables are \emph{useful}, i.e.,
\emph{insightful} and \emph{interpretable}.
We propose \sysName, a data-\underline{s}emantics-\underline{a}ware system for
recom\-mendin\underline{g} k-budgeted div\underline{e}rse pivot tables, which
overcomes the shortcomings of the existing approaches. We summarize the
limitations of currently available tools and research work in
Figure~\ref{fig:related_work} to contrast them against \sysName, and defer a
detailed discussion to \S\ref{sec:related_works}.



\subsubsection*{\textbf{Problem.}} The problem we study in this paper is
recommending a \emph{diverse set} of pivot tables, under a \emph{size}
constraint, while ensuring that each recommended pivot table is \emph{useful},
meaning it is \emph{insightful} and \emph{interpretable}. Furthermore, we want
to achieve two usability goals during recommendation: (1)~\emph{adaptivity},
which takes into consideration already explored pivot tables by the users, and
(2)~\emph{customizability}, which enables the users to guide the recommendation
process by specifying certain data attributes to prioritize.

\subsubsection*{\textbf{Challenges.}} \label{sec:challenge} We now highlight
three key associated challenges that are associated with the problem:

\smallskip \noindent\emph{Challenge 1: semantic modeling of pivot table
utility.} A useful pivot table must be \emph{insightful}, to inform users
\revisetwo{of meaningful and non-obvious patterns}, and \emph{interpretable},
\revisetwo{so that users can easily and quickly extract insights from it}.
\revisemeta{Prior work~\cite{SeedbVartak15, HarrisWWW23SpotLight,
ForesightDemiralp17, QuickInsightsDing19} ignore the \emph{semantic aspect};
they use purely statistical measures to model insightfulness, without
considering interpretability and semantics of the insight. For instance, the
aggregate $\texttt{SUM(Birth\_Year)}$ is semantically meaningless, even if it
indicates strong statistical insight. \label{challenge1} } Furthermore,
semantically modeling insightfulness and interpretability of a pivot table in a
\emph{multi-group setting} (e.g., group by \texttt{Gender},
\texttt{Department}, \texttt{Degree}) is non-trivial and is not addressed in
prior work. In summary, how to model insightfulness and interpretability of a
pivot table while remaining aware of the data semantics is a key challenge.


\smallskip \noindent\emph{Challenge 2: modeling table diversity.} Beyond
recommending insightful and interpretable pivot tables, our goal is to also
\emph{diversify} the set of pivot tables. To the best of our knowledge, the
notion of diversity in the context of pivot tables is not defined in prior work.
Existing diversification approaches~\cite{DrosouSIGMOD10Search,
DrosouBigData17Diversity, DrosouVLDB12DisC, BorodinPODS12MaxSum}
do not trivially extend for ``table diversification'', where the items under
consideration are entire tables rather than individual tuples. Prior work for
recommending insightful data summaries~\cite{DAISYVLDB24Junjie} or
visualizations~\cite{SeedbVartak15} do not consider diversity. For pivot table
diversification, the key challenge is to develop an appropriate distance metric
to model both the syntactic (e.g., attribute coverage) and semantic (e.g.,
provided insights) distances between a pair of pivot tables.

\smallskip \looseness-1 \noindent\emph{Challenge 3: developing an efficient
system.} Our goal is to recommend highly insightful and interpretable pivot
tables, while ensuring diversity among them. Unlike insightfulness and
interpretability, which can be measured for each pivot table in isolation,
diversity requires considering a \emph{set} of pivot tables. \revisetwo{Note
that the number of candidate pivot tables grows exponentially with the number
of data attributes. Furthermore, finding the best fixed-sized set of pivot
tables from these candidates is identical to the minimum set-cover problem, due
to the search space growing exponentially with the number of candidates.
Consequently, the problem is NP-hard in the number of candidates, which grows
exponentially with the number of data attributes. \label{challenge3}}
Furthermore, evaluating insightfulness of a pivot table requires its
materialization, which adds to the computational complexity. While greedy
approaches with approximation guarantees \cite{DrosouSIGMOD10Search,
CarbonellSIGIR98MMR, RadlinskiSIGIR06PersonalWebSearch, BorodinPODS12MaxSum,
DBLP:conf/icdt/Moumoulidou0M21} can alleviate the problem of combinatorial
search, the requirement of materializing candidate pivot tables remain. Even
with approximation algorithms~\cite{CormodeAlgorithm05CountMin,
AgarwalVLDB96MutliAgg, Hoeffding1994ProbabilityInequalities, WeiPODS11Beyond}
for efficient materialization, without aggressive pruning before
materialization, far too many candidates become the bottleneck. Therefore, a
key challenge here is to develop mechanisms that can leverage semantic
understanding of the data to prune unpromising pivot tables and avoid
unnecessary materialization. Another challenge is to discover effective
techniques that can ``push down''~\cite{DBLP:journals/pacmmod/YanLH23}
components of the diversity requirements to the search process to further
prevent unnecessary pivot table materialization.


\subsubsection*{\textbf{Contributions.}} \label{sec:contrib} Our main
contribution is development of a novel system \sysName, for recommendation of a
diverse set of useful pivot tables under a budget (size) constraint. Below, we
provide the key contributions we make in this paper:


\begin{itemize}[leftmargin=*]
    
	 \item We motivate and \emph{formalize the problem} of budgeted
	 recommendation of a diverse set of useful pivot tables, model it as a
	 \emph{constrained optimization problem}, and establish its \emph{desiderata}
	 (\S\ref{sec:two}).


	 \item We provide a formal model to measure the \emph{utility} of a pivot
	 table in terms of insightfulness and interpretability. Unlike prior work, our
	 utility model leverages data semantics (\S\ref{sec:three}).
	 

	 
     \item We establish the notion of \emph{pivot-table diversification}, a key
     component of the problem we study in this paper. Our contribution lies in
     the formulation of a suitable \emph{distance metric}---which considers
     both structural and semantic properties of pivot tables---and its
     application to diversifying a set of pivot tables
     (\S\ref{sec:four}).

     \item To ensure \sysName's efficiency and practicality, we must tackle the
     NP-hardness of the problem. To reduce the search space, we introduce
     \emph{aggressive semantic pruning}. To expedite the recommendation
     process, we leverage offline computation and ``push down'' diversity
     requirements to the search process (\S\ref{sec:five}).

     \item Through an empirical analysis over 4 real-world datasets and case
     studies, we show that \sysName outperforms prior approaches while staying
     scalable and efficient (\S\ref{sec:experiments}).
	 
	 \item \revisemeta{We present a user study that validates our utility model
	 and demonstrates \sysName's superiority over competing baselines based on
	 human perception (\S\ref{sec:userstudy}).}

	
\end{itemize}


\newcommand{\always}{\protect \tikz \fill (0,0) circle (1mm);}
\newcommand{\sometimes}{
	\protect
    \begin{tikzpicture}
        \protect\draw (0,0) circle (1mm);
		\protect\fill[black] (-1mm,0) arc[start angle=180, end angle=360, radius=1mm] -- cycle;		
    \end{tikzpicture}
}

\newcommand{\rarely}{
    \protect\begin{tikzpicture}
        \protect\draw (0,0) circle (1mm);
		\protect\fill[black] (0,0) -- (1mm,0) arc[start angle=0, end angle=-90, radius=1mm] -- cycle;
    \end{tikzpicture}
}

\newcommand{\never}{\protect\tikz \protect\draw (0,0) circle (1mm);} 

\newcommand{\unknown}{\protect\begin{tikzpicture}
    \protect\draw (0,0) circle (1mm);  
    \protect\draw (-0.7mm,-0.7mm) -- (0.7mm,0.7mm);  
    \protect\draw (-0.7mm,0.7mm) -- (0.7mm,-0.7mm); 
    \end{tikzpicture}
}   

\begin{figure}[t]
    \centering
    \resizebox{0.85\columnwidth}{!}{ \small
    \setlength{\tabcolsep}{1pt}
    \centering
    \begin{tabular}{|l|l|cccc|cc|ccc|c|}
    \cline{3-12}
	\multicolumn{1}{c}{}&{\cellcolor{white!100}}&
	\multicolumn{4}{c|}{\tiny {Commercial}} & 
	\multicolumn{2}{c|}{\tiny {Research}} & 
	\multicolumn{3}{c|}{\tiny {LLMs}} & 
	\multicolumn{1}{c|}{\tiny {This}} \\[-1.5mm]
	\multicolumn{1}{c}{}&\multicolumn{1}{c|}{}& 
	\multicolumn{4}{c|}{\tiny {software}} & 
	\multicolumn{2}{c|}{\tiny {work}} & 
	\multicolumn{3}{c|}{} & 
	\multicolumn{1}{c|}{\tiny {work}} \\
    \cline{3-12}
	\multicolumn{1}{c}{}&
		\begin{tabular}{ll}
		\multicolumn{2}{p{38mm}}{\footnotesize{\underline{\textbf{Legends}}}}\\			
        \always 	 	& \footnotesize{Always}\\
        \sometimes 		& \footnotesize{Partially}\\
        \never 	 		& \footnotesize{Not supported}\\
        \unknown 	 	& \footnotesize{Unknown }\\
						& \\		 	 		 
    	\end{tabular}
    & \rotatebox{90}{\hspace{-9mm}Microsoft Excel~\cite{microsoft_excel}}
    & \rotatebox{90}{\hspace{-9mm}Google Sheets~\cite{google_sheets}}
    & \rotatebox{90}{\hspace{-9mm}PowerBI~\cite{PowerBI,QuickInsightsDing19}} 
    & \rotatebox{90}{\hspace{-9mm}Tableau~\cite{Tableau}}
    & \rotatebox{90}{\hspace{-9mm}DAISY~\cite{DAISYVLDB24Junjie}} 
    & \rotatebox{90}{\hspace{-9mm}AutoSuggest~\cite{AutoSuggestSIGMODE2020Cong}} 
    & \rotatebox{90}{\hspace{-9mm}ChatGPT~\cite{OpenAIChat}} 
    & \rotatebox{90}{\hspace{-9mm}Llama3-instruct~\cite{Llama3}\phantom{.}} 
    & \rotatebox{90}{\hspace{-9mm}TableGPT~\cite{TableGPT}} 
    & \rotatebox{90}{\hspace{-9mm}\textbf{\sysName}} \\ 
    \cline{3-12}
	\addlinespace[0.8mm]
	\cline{1-12}
	\multirow{13}{*}{\rotatebox{90}{\textbf{Desirable Properties}}}
	&&&&&&&&&&& \\[-3mm]
	& Budgeted recommendations        				& \never     & \never     & \never     & \never     & \never     & \never     & \always    	& \always    & \always    & \always    \\
    & Guarantees syntactic validity					& \always 	 & \always 	  & \always    & \always 	& \always    & \always    & \never 	 	& \never 	 & \never 	  & \always    \\
    & Guarantees semantic validity					& \never 	 & \never 	  & \always    & \always 	& \always    & \always    & \never 	 	& \never 	 & \never 	  & \always    \\
    & Ensures interpretability				   		& \always    & \never     & \always    & \always    & \unknown   & \always    & \always    	& \sometimes & \sometimes & \always    \\
    & Adaptive to user actions        				& \never     & \never     & \never     & \never     & \never     & \always    & \always    	& \always    & \always    & \always    \\
    & Allows user specifications      				& \never     & \never     & \never     & \always    & \never     & \never     & \always    	& \always    & \always    & \always    \\
    & Ensures diversity                 			& \never     & \never     & \never     & \never     & \never     & \never     & \always    	& \always    & \always    & \always    \\
    & Attribute-name semantics aware  				& \always    & \always    & \always    & \never     & \never     & \always    & \always    	& \always    & \always    & \always    \\
    & Attribute-order insensitive     				& \never     & \never     & \always    & \always    & \never     & \never     & \never     	& \never     & \never     & \always    \\
    & Data-semantics aware            				& \sometimes & \never     & \always    & \never     & \always    & \never     & \sometimes 	& \sometimes & \sometimes & \always    \\
    & No additional requirements      				& \always    & \always    & \always    & \never     & \never     & \never     & \never     	& \never     & \never     & \always    \\
    & Low-cost 										& \always    & \always    & \always    & \always    & \always 	 & \always 	  & \never     	& \always 	 & \always 	  & \always    \\
    & Open-source 				      				& \never     & \never     & \never     & \never     & \never     & \never     & \never     	& \sometimes & \sometimes & \always    \\
    \hline
    \end{tabular}
}
\vspace{-2.5mm}
\caption{\small \sysName satisfies all desirable properties. While PowerBI,
Tableau, DAISY, and AutoSuggest do not directly/always recommend pivot tables,
we include them since they recommend summaries. Code for DAISY and AutoSuggest
is unavailable, thus we rely on the papers, and mark some things as unknown.
LLMs are not designed to directly recommend pivot tables but can be prompted to
do so.}
\vspace{-3mm}
\label{fig:related_work}
\end{figure}

\begin{figure*}[t]
    \centering
    \small    
    \begin{subtable}{0.24\textwidth} 
        \centering
        \resizebox{\textwidth}{!}{
            \begin{tabular}{lrrr}
                \toprule
                \textbf{} & \multicolumn{3}{c}{\textbf{Degree}} \\
				\cline{2-4}
                \textbf{Gender}         & \textbf{BS} 	&  \textbf{MS} & \textbf{PhD} \\ 
				\midrule
                \textbf{Male}   		& 200K     		& 300K 		   & 1000K	\\
                \textbf{Female}         & 100K          & 200K		   & 300K 	\\ 
				\bottomrule
            \end{tabular}
        }
        \caption{\footnotesize Avg.\ Salary by Gender and Degree}
        \label{tab:a}
    \end{subtable}%
    \hfill
    \begin{subtable}{0.25\textwidth}
        \centering
        \resizebox{0.77\textwidth}{!}{
            \begin{tabular}{lrr}
                \toprule
                \textbf{} &
                \multicolumn{2}{c}{\textbf{Department}} \\ 
				\cline{2-3}
                \textbf{Gender}         & \textbf{IT}     & \textbf{Sales}\\
				\midrule
                \textbf{Male}   		& 1000K  	      & 500K\\
                \textbf{Female}         & 400K            & 200K\\
                \bottomrule
            \end{tabular}
        }
        \caption{\footnotesize Avg.\ Salary by Gender and Dept.}
        \label{tab:b}
    \end{subtable}%
    \hfill
    \begin{subtable}{0.28\textwidth}
        \centering
        \resizebox{1.05\textwidth}{!}{
            \begin{tabular}{lcccccc}
                \toprule
                \textbf{} & \multicolumn{6}{c}{\textbf{Degree, Department}} \\ 
				\cline{2-7}
                     & \multicolumn{2}{c}{\textbf{BS}} & \multicolumn{2}{c}{\textbf{MS}} & \multicolumn{2}{c}{\textbf{PhD}} \\
					 \cline{2-7}
				\textbf{Gender} & \textbf{IT} & \textbf{Sales} & \textbf{IT} & \textbf{Sales} &\textbf{IT} & \textbf{Sales}\\
				\midrule
                \textbf{Male}	     & 4 & 1 & 8 & 1 & 10 & 1\\
                \textbf{Female}      & 2 & 8 & 2 & 3 &  1 & 2\\ 
				\bottomrule
            \end{tabular}
        }
        \caption{\footnotesize Count ID by Gender, Degree, and Dept.}
        \label{tab:c}
    \end{subtable}%
    \hfill
    \begin{subtable}{0.222\textwidth}
        \centering
        \resizebox{0.69\textwidth}{!}{
            \begin{tabular}{lrr}
                \toprule
                \textbf{} &
                \multicolumn{2}{c}{\textbf{Department}} \\ 
				\cline{2-3}
                \textbf{Degree}         & \textbf{IT}         	& \textbf{Sales} \\
				\midrule 
				\textbf{BS}     		& 200K 					& 100K\\
                \textbf{MS}       		& 300K 					& 200K\\
                \textbf{PhD}   			& 900K   				& 400K\\
                \bottomrule
            \end{tabular}
        }
        \caption{\footnotesize Avg.\ Salary by Degree and Dept.}
        \label{tab:d}
    \end{subtable}
	
	\vspace{-3mm} 
	 
    \caption{Four pivot tables over the dataset of
    Figure~\ref{tab:salary_database}. While~(a) and~(b) indicate gender-based
    salary gap, (c) and~(d) add additional context.}
	\vspace{-3mm}
    \label{fig:div_adap}
\end{figure*}

\section{Recommending a Diverse set of Pivot Tables}
\label{sec:two}

In this section, we motivate the need for diversity and adaptivity during
recommending pivot tables (\S\ref{two:one}). Then we develop the
desiderata for the problem (\S\ref{two:two}) and formalize it
(\S\ref{sec:problem_statement}).

\subsection{The Need for Diversity and Adaptivity} \label{two:one}
\looseness-1 A key limitation of top-k recommendation is that it may provide
redundant information, causing the users to miss out on relatively less useful,
but complementary data insights. Such lack of \emph{diversity} may even mislead
the users to believe in partial insights that are ``half-true''. Another
shortcoming of existing approaches is that they are not \emph{adaptive} to user
actions, i.e., when the user acknowledges a recommendation, it should be
excluded from the subsequent iterations. However, commercial pivot table
recommendation features are not adaptive (Figure~\ref{fig:related_work}). We
proceed to provide an example to highlight the need for \emph{diverse}
recommendation to help the users get a broader picture of the dataset, and
\emph{adaptive}~\cite{SarawagiVLDB00Adaptive, LiWWW17Query} recommendation to
enable the user guide the recommendation process.

\begin{example}
\label{ex:diversity} 
Recall from Example~\ref{ex:manual_search} that Sasha is interested in salary
discrepancies. She initially finds two pivot tables
(Figure~\ref{fig:div_adap}~(a) \&~(b)) suggesting gender-based pay gap.
However, a deeper pattern emerges when she expands her analysis using other
aggregate functions (e.g., \verb|COUNT|) and discovers the pivot tables shown
in Figure~\ref{fig:div_adap}~(c) \&~(d), which provide her an additional
context that the discrepancy stems from the hiring process: employee counts are
uneven across degrees and departments. Sasha also notes that IT employees earn
more than those in Sales, and PhDs earn more than others. This indicates
degree- and department-based discrepancies, which are expected and acceptable.
Sasha concludes that males earn more on average not due to gender bias, but
largely because more male PhDs work at IT.\footnote{This phenomenon is known as
Simpson's paradox~\cite{paradox, paradoxBerkely}. While our focus is not to
expose Simpson's paradox, we show this as a motivating use-case.}

\looseness-1 Furthermore, in an incremental setting where Sasha iteratively
requests for recommendations of a few pivot tables at a time, she expects the
system to adapt to her actions. For instance, after she accepts or rejects the
suggestions of Figure~\ref{fig:div_adap}~(a) and~(b), the system should avoid
recommending redundant pivot tables that reiterate the same concept
(gender-based salary gap) across other aspects (e.g., marital
status).\footnote{\revisetwo{This example does not dispute any existing fact
regarding the relationship between marital status and gender-based salary
gaps.\label{footnote6}}}
\end{example}

\subsection{Desiderata} \label{two:two}
We now list key desiderata for a pivot-table recommender:


\begin{itemize}[leftmargin=6mm]

\item[\textbf{D1.}] Each recommended pivot table must provide
\emph{insightful}~\cite{SeedbVartak15,ForesightDemiralp17} and
\emph{semantically interesting} information. For instance, a significant gap in
average salary across genders provides insight into gender-based pay gap.

\item[\textbf{D2.}] Each pivot table must be \emph{interpretable}, ensuring
ease of comprehension by humans. For instance, a concise table with 10 cells is
more interpretable than one with 1000 cells.

\item[\textbf{D3.}] While insightfulness and interpretability model the
goodness of a single pivot table, a desirable property for a set of pivot
tables is \emph{diversity}. Thus, the recommended set of pivot tables must
minimize redundancy, covering various data aspects.

\item[\textbf{D4.}] The system for pivot table recommendation must allow
(I)~\emph{customizability}---allowing users to specify the desired size of the
recommendation set, degree of diversity, data scope, etc.---and
(II)~\emph{adaptiveness} to user actions.

\item[\textbf{D5.}] Finally, the system must be \emph{efficient} and
\emph{scalable}---to ensure handling large and high-dimensional data
effectively---and \emph{accessible}---in terms of cost and availability.

\end{itemize}

\subsection{Problem Formulation}\label{sec:problem_statement} We now formalize
our problem for a single-relation database instance (dataset) $D$
by defining a pivot table. 

\newcommand{\pivottable}{\ensuremath{T(\mathbf{F(V)}, \mathbf{G})}\xspace}
\newcommand{\pivottabletwo}{\ensuremath{T(F(V), \mathbf{G})}\xspace}

\begin{definition}[Pivot Table]
    Given a dataset $D$ over a set of attributes $\mathbf{A}$ and the domain of
    aggregate functions $\mathcal{F} = \{\texttt{COUNT}, \texttt{SUM},$
    $\texttt{AVG}, \texttt{ MIN}, \texttt{ MAX}\}$, a pivot table
    \pivottable takes the following form:
	{\small
	$$\texttt{SELECT } \mathbf{F}(\mathbf{V}) \texttt{ FROM } D \texttt{ GROUP
	BY } \mathbf{G}$$}

	\noindent where, $\textbf{G} {=} \{G_1, G_2, \dots\} {\subseteq} \mathbf{A}$ is a
	 subset of attributes for grouping; 
	 $\textbf{V} {=} \{V_1, V_2, \dots\} {\subseteq} \mathbf{A}$ is a subset of
	 attributes for computing aggregates over;
	 $\textbf{G} \cap \textbf{V} = \emptyset$ ensures that no attribute is used
	 for both grouping and aggregation;
	 $\textbf{F} {=} \{F_1, F_2, \dots\}$ is a set of aggregate functions where
	 $F_i {\in} \mathcal{F}$ and $|\mathbf{F}| {=} |\mathbf{V}|$; 
	 and with slight abuse of notation, $\mathbf{F}(\mathbf{V})$ denotes
	 $F_1(V_1), F_2(V_2), F_3(V_3) \dots$

\end{definition}
\subsubsection*{Tabular representation of a pivot table} For
\pivottable, with $|\mathbf{G}| \ge 2$, we fix as row-groups and column-groups
(Figure~\ref{fig:ex_pivot_table}) two non-empty sets $\mathbf{R}, \mathbf{C}
\subset \mathbf{G}$, respectively, where $\mathbf{R} \cup \mathbf{C} =
\mathbf{G}$, $\mathbf{R} \cap \mathbf{C} = \emptyset$.

\begin{example}
	Figure~\ref{fig:div_adap}(c) represents a possible tabular 
	representation for the pivot table:
{
\small
\texttt{SELECT COUNT(ID) FROM D GROUP BY Gender, Degree, Department}.
}
Here, $\mathbf{R} {=} \{\mathtt{Gender}\}$ and $\mathbf{C} {=} \{\mathtt{Degree,\; Dept.}\}$
\end{example}

\subsubsection*{Pivot table canonicalization} The above mechanism allows
structurally different tabular representations of semantically equivalent
tables. However, we only care about the semantics of a table, not its tabular
orientation. Thus, we canonicalize a pivot table \pivottable by
lexicographically sorting $\mathbf{F}(\mathbf{V})$ and $\mathbf{G}$ to obtain
$\mathbf{F}(\mathbf{V})_{\le}$ and $\mathbf{G}_{\le}$, respectively, and derive
the canonical pivot table $T(\mathbf{F}(\mathbf{V})_{\le}, \mathbf{G}_{\le})$.
Furthermore, we obtain a \emph{canonical tabular representation} of a pivot
table \pivottable by assigning to the row-groups ($\mathbf{R}_{\le}$) the first
$\lceil \frac{|\textbf{G}|}{2}\rceil$ elements of $\mathbf{G}_{\le}$ and the
remaining elements to the column-groups ($\mathbf{C}_{\le}$). This process
ensures that pivot tables remain \emph{organization-invariant} (e.g.,
transpose-invariant), i.e., \pivottable and all its variants derived from
different permutations of $\mathbf{F}(\mathbf{V})$ and $\mathbf{G}$ result in
an identical canonical tabular representation.

\begin{example}
	The canonical pivot table for Figure~\ref{fig:div_adap}(c) is:
	{
	\small
	\texttt{SELECT COUNT(ID) FROM D GROUP BY Degree, Department, Gender}.
	}
	 The canonical tabular representation is obtained
	 by setting $\mathbf{R}_{\le} {=} \langle\mathtt{Degree,\; Dept.}\rangle$ and
	 $\mathbf{C}_{\le} {=} \langle\mathtt{Gender}\rangle$, which is simply the
	 transpose of Figure~\ref{fig:div_adap}(c).
\end{example}

\smallskip

Based on the desiderata of \S\ref{two:two}, we set our goal to find a
bounded sized (D4-I) set of pivot tables such that the overall utility (D1 \&
D2) of the pivot tables are maximized while the set of pivot tables meet the
minimum diversity requirement (D3).

\newcommand{\Universe}{\ensuremath{\mathcal{T}_{\mathbf{A}}}\xspace}
\newcommand{\UniverseUser}{\ensuremath{\mathcal{T}_{\mathbf{A}_u}}\xspace}
\newcommand{\Solution}{\ensuremath{\mathbf{T}^{*}}\xspace}
\newcommand{\Candidate}{\ensuremath{\mathbf{T}}\xspace}
\newcommand{\Explored}{\ensuremath{\mathbf{T}_{u}}\xspace}
\newcommand{\UserSpecifiedAttributes}{\ensuremath{\mathbf{A}_u}\xspace}

\begin{problem}[Recommending a set of Pivot Tables]
\label{def:vanilla_recommendation} 
Given 
(i)~a set of possible pivot tables \Universe over a dataset $D$ with attributes $\mathbf{A}$,
(ii)~a function $Utility: \Universe \mapsto[0, 1]$ that returns the utility of a pivot table $T \in \Universe$,
(iii)~a function $Diversity: 2^{\Universe} \mapsto [0, 1]$ that returns the diversity of a set of pivot tables $\mathbf{T} \subseteq \Universe$,
(iv)~a budget $k \in \mathbb{N}^+$, and
(v)~a threshold $\theta \in [0, 1]$,
find a set of pivot tables $\Solution \subseteq \Universe$ s.t:

\begin{tabular}{ll}
	 (\textbf{objective}) 			& $\Solution =\underset{\Candidate \subseteq \Universe}{\argmax} \sum_{T \in \Candidate} Utility(T)$,\\
     (\textbf{size constraint}) 	& $|\Solution| \leq k$, and\\
	 (\textbf{diversity constraint})& $Diversity(\Solution) \geq \theta$\\
\end{tabular}
\label{eq:objective_function}
\end{problem}


Problem~\ref{def:vanilla_recommendation} balances utility and diversity by
maximizing utility while putting a constraint on diversity. Other variants of
this problem are possible such as maximizing a linear combination of the
objective and the diversity constraint. More details are in
\S\ref{sec:five}.


\subsubsection*{Adaptive Recommendation of a set of Pivot Tables.}
\label{adaptivereco} In the adaptive version (D4-II), we discard the already
explored pivot tables by the user \Explored from the set \Universe to obtain
$\Universe {-} \Explored$. When the user highlights a data scope (D4-I) by
specifying a subset of attributes $\UserSpecifiedAttributes {\subseteq}
\mathbf{A}$ they want to focus on, we set the possible pivot tables to
\UniverseUser.

\subsubsection*{Considerations.} Multiple aggregates within a pivot table is
essentially equivalent to concatenating the corresponding single-aggregate
pivot tables, i.e., $\pivottable \equiv \textstyle\bigcup_{F, V \in
\mathbf{F, V}} \pivottabletwo$.
Therefore, for simplicity and to promote interpretability, we limit each pivot
table to have exactly one aggregate. We use $F$ and $V$ to denote the
aggregation function and attribute, respectively. We summarize the notations
used in the rest of this paper in Figure~\ref{fig:notation_pivot}.

\subsubsection*{A Note on Generalizability.} While our work focuses on
recommending pivot tables in spreadsheet environments, the techniques can be
generalized to recommend aggregate queries in relational databases. Pivot tables
can be represented by SQL aggregate queries involving \texttt{Group-by},
allowing \sysName's adaptation in RDBMS.

\begin{figure}[t]
    \centering
    \resizebox{0.95\linewidth}{!}{%
    \begin{tabular}{ll}
    \toprule
    \textbf{Symbol} & \textbf{Description} \\
    \midrule
    $D, \mathbf{A}$, $D[\mathbf{A}]$     					& Database, attributes, possible unique value combinations\\
    $\mathbf{G}, \mathbf{R}, \mathbf{C}$ 	& Grouping attributes, row-groups, column-groups; $\mathbf{R} \cup \mathbf{C} = \mathbf{G}$ \\
    $F, V$		                			& Aggregation function and attribute \\
    \pivottabletwo or $T$ 					& A pivot table for the query \texttt{SELECT} $F(V)$ \texttt{FROM} $D$ \texttt{GROUP BY} $\mathbf{G}$ \\
	$T_{r_i}$/$T^{c_j}$                     & The row/column of $T$ with row/column header $r_i$/$c_j$\\
	$T_{r_i}^{c_i}$                     	& The pivot table cell with row header $r_i$ and column header $c_j$\\
    $n, m$                                  & The cardinality of $D[\mathbf{R}]$, $D[\mathbf{C}]$\\

    \bottomrule
    \end{tabular}
    }
	\vspace{-4mm}
    \caption{\small Table of notations. We use bold letters to denote sets.}
	\vspace{-3mm}
    \label{fig:notation_pivot}
\end{figure}

\section{Utility of a Pivot Table} \label{sec:three}

Based on the desiderata of \S\ref{two:two}, a pivot table has high
utility if it offers \emph{insights} (D1) while being easily
\emph{interpretable} by humans (D2). Thus, we use insightfulness
(\S\ref{sec:insightfulness}) and interpretability
(\S\ref{sec:interpretability}) as the two building blocks to model utility of a
pivot table.

\subsection{Insightfulness}\label{sec:insightfulness}
Intuitively, an insightful pivot table must involve attributes that are
\emph{significant}, i.e., inherently interesting and relevant
(\S\ref{attsigsec}). Furthermore, it should satisfy at least one of the
following criteria:
(1)~provide high \emph{informativeness} (\S\ref{infsec}),
(2)~highlight meaningful \emph{trends} (\S\ref{trendsec}),
or (3)~reveal \emph{surprising}~\cite{DBLP:journals/vldb/Sarawagi01,
ValuesChen21, SupportingJasim22} findings (\S\ref{sursec}).
We build on prior work~\cite{SeedbVartak15, HarrisWWW23SpotLight,
QuickInsightsDing19, ForesightDemiralp17, ValuesChen21, SupportingJasim22} that
model insightfulness based on only statistical properties, but significantly
extend it by taking a \emph{semantics-aware} approach, enabled by
LLMs~\cite{Llama3}.


\subsubsection{Attribute significance}\label{attsigsec}
Typically, not all data attributes are of interest by humans. E.g., grouping
data by \texttt{Name} is typically much less insightful than by
\texttt{Gender}. However, semantic understanding is required to figure out
attribute significance. To this end, we consult an LLM~\cite{Llama3} to
determine the significance for an attribute $A$. When attribute name is missing
or semantically meaningless (e.g., ``Column 1''), we first query an LLM to
suggest appropriate names for attributes by providing it with a small sample of
the data. LLM's semantic-reasoning capability allows us to achieve this without
any domain-specific pre-configuration. To avoid noise, we employ multiple
paraphrased prompts while querying the LLM. In this work, we condition the LLM
to return a simple binary answer (yes $\rightarrow$ 1/no$\rightarrow$ 0).
However, this component can be replaced by a domain-aware model that can return
the likelihood of an attribute being significant for a specific context. We
compute attribute significance of a pivot table $T$, $S_{\text{sig}}:\Universe
\mapsto [0, 1]$ as follows:
\begin{equation}\label{eq1}
	\small
S_{\text{sig}}(T) = \prod_{A \in \{V\} \cup\, \mathbf{G}} \text{Significance}(A)
\end{equation}
Here, $\text{Significance}: \mathbf{A} \mapsto [0, 1]$ denotes the probability
that an attribute $A \in \mathbf{A}$ is a significant attribute w.r.t human
interest.

\begin{example} \label{sigattr}
	For Figure~\refsub{fig:div_adap}{tab:d}, \texttt{Degree},
\texttt{Department}, and \texttt{Salary}, all are significant attributes. Thus,
$S_{\text{sig}}(T)= 1 \times 1 \times 1 = 1$.
\end{example}



\newcommand{\infscore}{\ensuremath{S_{\text{inf}}}\xspace}
\subsubsection{Informativeness} \label{infsec} A statistical way to measure
informativeness within data is to measure spread of the values. Intuitively, if
values in a pivot table deviate from each other significantly, the ``entropy''
is high, and so is the informativeness. In this work, we use \emph{deviation}
across different groups to model informativeness. Unlike prior
work~\cite{SeedbVartak15, HarrisWWW23SpotLight, ForesightDemiralp17} that only
consider a two-group setting (Male vs Female), we consider a multi-group
setting.

Given a database $D$ and a pivot table $T$ with row-groups $\mathbf{R}$ and
column-groups $\mathbf{C}$, let $D[\mathbf{R}]$ and $D[\mathbf{C}]$ be the set
of row and column headers for $T$, respectively. E.g., for
Figure~\refsub{fig:div_adap}{tab:c}, the row headers are $\{$\texttt{Male},
\texttt{Female}$\}$ and the column headers are
$\{$(\texttt{BS}, \texttt{IT}),
(\texttt{BS}, \texttt{Sales}), 
(\texttt{MS}, \texttt{IT}), 
(\texttt{MS}, \texttt{Sales}),
(\texttt{PhD}, \texttt{IT}), 
(\texttt{PhD}, \texttt{Sales})$\}$. 
We use $T_{r_i}$ ($T^{c_i}$) to denote the row (column) of $T$ with row (column)
header $r_i$ ($c_i$). We use $n$ and $m$ to denote the number of rows
$|D[\mathbf{R}]|$ and columns $|D[\mathbf{C}]|$ in $T$, respectively. We compute
the row-wise and column-wise informativeness scores
$S_{\text{inf}}^{\text{row}}$ and $S_{\text{inf}}^{\text{col}}$ as follows:
\vspace{-1mm}
\[
\small
\begin{aligned}
	\small
\infscore^{\text{row}}(T) &= 
	\frac{1}{\binom{n}{2}} 
	\sum_{\substack{r_i,r_j \in D[\mathbf{R}] \text{ s.t } i < j}} 
	\frac{||T_{r_i}, T_{r_j}||_2}{\gamma \cdot m}
\\[-4pt]
\infscore^{\text{col}}(T) &= 
	\frac{1}{\binom{m}{2}} 
	\sum_{\substack{c_i,c_j \in D[\mathbf{C}] \text{ s.t } i < j}} 
	\frac{||T^{c_i}, T^{c_j}||_2}{\gamma \cdot n}
\end{aligned}
\]

\noindent Here, $\gamma$ is a normalization parameter, set to $\max(T) -
\min(T)$, ensuring that $\infscore^{\text{row}}$ and $\infscore^{\text{col}}$
are bounded between $0$ and $1$. Also note that while we use Euclidean distance
($L_2$ distance), any other distance function such as $L_1$ distance can be used
here. We compute the informativeness score $\infscore: \Universe \mapsto [0, 1]$
by taking the maximum of the row-wise and column-wise informativeness scores:
\begin{equation}\label{eq2}
	\small
\infscore(T) = \max (\infscore^{\text{row}}(T), \infscore^{\text{col}}(T))
\end{equation}


\begin{example}\label{infexample}
%
We first compute the pairwise distances along the rows of Figure~\refsub{fig:div_adap}{tab:d}: 
${||T_{\text{BS}}, T_{\text{MS}}||_2}  {=} 141.4K$, 
${||T_{\text{BS}}, T_{\text{PhD}}||_2} {=} 761.6K$, and
${||T_{\text{MS}}, T_{\text{PhD}}||_2} {=} 632.5K$.
We normalize using $\gamma {=} 900K {-} 100K {=} 800K$ and $m {=} 2$, resulting
in normalized distances of $[0.088, 0.476, 0.395]$. Taking an average gives us
$\infscore^{\text{row}}(T) {=} 0.32$. We similarly compute
$\infscore^{\text{col}}(T) {=} 0.22$ and obtain $\infscore(T) {=} \max(0.32,
0.22) {=} 0.32$.
 
%
\end{example}

\newcommand{\trendcor}{\ensuremath{S_{\text{cor}}}\xspace}
\newcommand{\trendrat}{\ensuremath{S_{\text{ratio}}}\xspace}
\newcommand{\trendscore}{\ensuremath{S_{\text{trend}}}\xspace}

\subsubsection{Trend}\label{trendsec}

Trends observed in a pivot table provide insights. However, the degree of
insightfulness hinges on two key factors: the \emph{magnitude} of the trend
metric and how \emph{atypical} or rare it is. For instance, a positive
correlation between income and years of service is generally
expected---employees with longer tenures typically earn more. In contrast, a
trend showing that new hires earn more on average than long-serving employees
contradicts this expectation and thus is particularly insightful.

We use two metrics to quantify the magnitude of a trend: \emph{correlation} and
\emph{ratio}. Furthermore, to assess the degree of a trend's rarity, we query an
LLM, which is aware of a broader semantic context. Thus, our definition of the
\emph{trend score} for a pivot table combines (1)~purely statistical insights,
reflected in high correlation and consistent ratio across pivot table values and
(2)~semantic insights, captured through the LLM's assessment of the trend's
atypicality.

\smallskip\noindent\emph{Correlation.} \looseness-1 We use $\rho_{i,j}$ to
denote the Pearson correlation coefficient between the rows $T_{r_i}$ and
$T_{r_j}$. Since consulting LLMs is costly, we only consider significant
correlations and require the magnitude to be at least $\tau_{\rho}$, a
customizable threshold parameter, with a default value of 50\%. The indicator
function $[\![ |\rho_{i,j}| {\geq} \tau_{\rho} ]\!]$ denotes if the correlation
between the rows $T_{r_i}$ and $T_{r_j}$ is significant. We compute the row-wise
correlation-trend score $\trendcor^{\text{row}}: \Universe \mapsto [0, 1]$ as
follows:
\vspace{-1mm}
\[
\small
\begin{aligned}
\trendcor^{\text{row}}(T) = 
	\frac{1}{\binom{n}{2}} 
	\sum_{r_i,r_j \in D[\mathbf{R}] \text{ s.t. } i < j} 
	|\rho_{i,j}| 
    \cdot [\![ |\rho_{i,j}| \geq \tau_{\rho} ]\!] 
    \cdot \widetilde{Pr}_{\text{cor}}(r_i, r_j)
\end{aligned}
\]
Here, $\widetilde{Pr}_{\text{cor}}(r_i, r_j)$ is the likelihood of \emph{not}
observing a high correlation between the groups $r_i$ and $r_j$ determined by an
LLM. We prompt an LLM ``In a five-point scale from \emph{very likely} to
\emph{very unlikely}, how likely is it that the <{$F(V)$}> for <{$r_i$}> and
<{$r_j$}> are <{positively/negatively}> correlated across
<{$D[\mathbf{C}]$}>?'', where each part within <> is replaced with actual values
such as ``Average Salary'' for <$F(V)$>. We then map the LLM-provided likelihood
to a numerical value using the mappings: \textit{very likely} $\rightarrow$
20\%, \textit{likely} $\rightarrow$ 40\%, \textit{neutral} $\rightarrow$ 60\%,
\textit{unlikely} $\rightarrow$ 80\%, and \textit{very unlikely} $\rightarrow$
100\%. The intuition behind this \emph{inverse} mapping is that the more
unlikely it is to observe a trend, the more insightful it is. We compute
$\trendcor^{\text{col}}(T)$ similarly and set the correlation-trend score $
\trendcor(T) = \max(\trendcor^{\text{row}}(T), \trendcor^{\text{col}}(T))$.

\begin{example} 
    \label{ex:cor-trend}
	In Figure~\refsub{fig:div_adap}{tab:d}, $T_{\text{BS}}$ and $T_{\text{PhD}}$
exhibit a positive correlation of 98\%, which is \textit{likely} (from LLM
consultation), leading $\widetilde{Pr}_{\text{cor}}(\text{BS}, \text{PhD})$ to
be $40\%$. The correlation between $T_{\text{BS}}$ and $T_{\text{MS}}$ is 100\%
(\textit{very likely}); and $T_{\text{MS}}$ and $T_{\text{PhD}}$ is 100\%
(\textit{likely}). Since all correlations meet the threshold 50\%, we compute
$\trendcor^{\text{row}}(T) = (0.98 \times 0.4 + 1.0\times0.2 + 1.0 \times 0.4)/3
= 0.33$. The column-wise correlation-trend score $\trendcor^{\text{col}}(T) {=}
(0.98 \times 0.4)/1 {=} 0.39$. Thus, $\trendcor(T) {=} \max(0.33, 0.39) {=}
0.39$.
\end{example}
%
%
%

\smallskip\noindent\emph{Ratio.} Since correlation fails to capture the
relative \emph{magnitude}, we use \emph{ratio trends} based on
\emph{persistent} ratios between two groups, e.g., $T_{\text{PhD}}$ earning at
least 5$\times$ more than $T_{\text{BS}}$ across all departments. Like
correlation, LLMs inform us the rarity of ratio trends. We compute the
row-wise ratio-trend score $\trendrat^{\text{row}}: \Universe \mapsto [0, 1]$
as follows:

\phantom{test}
\vspace{-3mm}
\[ 
\small
\begin{aligned}
\trendrat^{\text{row}}(T) = \frac{1}{\binom{n}{2}} \sum_{r_i,r_j \in
D[\mathbf{R}]} 
(1-\frac{1}{\pi_{i,j}}) \cdot 
[\![\pi_{i,j} \geq \tau_{\pi} ]\!] \cdot 
\widetilde{Pr}_{\text{ratio}}(r_i, r_j)
\end{aligned}
\]
Here, $\pi_{i,j}$ denotes the minimum element-wise ratio between $T_{r_i}$ and
$T_{r_j}$, i.e., the smallest factor by which any value in $T_{r_i}$ exceeds its
corresponding value in $T_{r_j}$. To reduce LLM consultation cost, we use a
threshold $\tau_{\pi}$ and require $\pi_{i,j}$ to be at least $\tau_{\pi}$
before consulting an LLM. While we set $\tau_{\pi} {=} 2.0$, it is a
customizable parameter (but must be $\ge$ 1).
$\widetilde{Pr}_{\text{ratio}}(r_i, r_j)$ denotes the LLM-provided likelihood of
\emph{not} observing the ratio trend between $T_{r_i}$ and $T_{r_j}$. Note that
for any $i$ and $j$, at most one of $\pi_{i,j}$ or $\pi_{j, i}$ can contribute
to this score, hence we fix the scaling factor to $\binom{n}{2}$. We normalize
the trend magnitude by subtracting the inverse of $\pi_{i,j}$ from $1$, so that
larger ratios yield higher scores. We compute the column-wise ratio-trend score
$\trendrat^{\text{col}}(T)$ similarly and set the ratio-trend score
$\trendrat(T) = \max(\trendrat^{\text{row}}(T), \trendrat^{\text{col}}(T))$.



\begin{example}
    \label{ex:ratio-trend}
	%
    In Figure~\refsub{fig:div_adap}{tab:d}, the minimum ratio between $T_{MS}$
    and $T_{BS}$ is 1.5; between $T_{PhD}$ and $T_{BS}$ is 4.0; and between
    $T_{PhD}$ and $T_{MS}$ is 2.0. After applying the threshold
    $\tau_{\pi}=2.0$, the ratio trends for ($T_{PhD}$, $T_{MS}$) and ($T_{PhD}$,
    $T_{BS}$) are retained. The LLM returns the likelihoods: $\texttt{[Very
    Unlikely, Unlikely]}\rightarrow[1.0, 0.8]$ for these two trends. This gives
    us the row-wise ratio-trend score $\trendrat^{\text{row}}(T) = (3/4 \times
    1.0 + 1/2 \times 0.8)/3 = 0.37$. For the column-wise ratio-trend score, no
    pair satisfies the threshold requirement and thus the score is $0.0$.
    Therefore, the ratio-trend score $\trendrat(T) = \max(0.37, 0.0) = 0.37$.
\end{example}

%


Finally, we compute the trend score by taking the maximum of the
correlation-trend and ratio-trend scores:
\vspace{-1mm}
{\small
\begin{equation}\label{eq3}
    \trendscore(T) =  
    \max\big( 
        \trendcor(T),\;
        \trendrat(T)
    \big)
\end{equation}
}
\vspace{-6mm}
\begin{example} \label{trendexample}
For the pivot table of Figure~\refsub{fig:div_adap}{tab:d}, we computed the
correlation-trend score as 0.39 in Example~\ref{ex:cor-trend} and the
ratio-trend score as 0.37 in Example~\ref{ex:ratio-trend} Thus, $\trendscore(T)=
\max(0.39, 0.37) = 0.39$.
%
%
\end{example}

\newcommand{\surscore}{\ensuremath{S_{\text{sur}}}\xspace}
\subsubsection{Surprise}\label{sursec}

\looseness-1 Surprising values or \emph{outliers} often indicate insights.
E.g., in Figure~\refsub{fig:div_adap}{tab:d}, $T_{\text{PhD}}^{\text{IT}}$ is
exceptionally high (900K) in its column. While such outliers can be insightful,
not all are. Some, like this one, are expected: IT is high-paying, and PhDs
earn more. In contrast, an unusually high $T_{\text{BS}}^{\text{Sales}}$ would
be surprising, and thus insightful. Beyond simply identifying outliers, we
incorporate the unexpectedness of observing outliers using LLM's semantic
knowledge. We compute the row-wise surprise score $\surscore^{\text{row}}:
\Universe \mapsto [0, 1]$ as follows:
%
{\small
    \begin{align*}
    &\surscore^{\text{row}}(T) = 
    \frac{1}{n}{\textstyle\sum_{r_i \in D[\mathbf{R}]}
	OutlierScore(T_{r_i})} &\\
	\text{where,}&\;
	OutlierScore(T_{r_i}) = 
	\begin{cases}
		1 - \frac{\sum_{c \in \mathcal{O}_{r_i}} \widetilde{Pr}_{\text{outlier}}(r_i, c, T_{r_i}^{c})}{|\mathcal{O}_{r_i}|+1}, & \text{if } |\mathcal{O}_{r_i}| > 0  \\
		0, & \text{otherwise }
	\end{cases}\\
	\phantom{\text{where,}}&\;
	\mathcal{O}_{r_i} = \{c_j \in D[\mathbf{C}] \;s.t.\ \; |T_{r_i}^{c_j} - \mu(T_{r_i})| \geq \tau_{\mathcal{O}} \cdot \sigma(T_{r_i})\}
    \end{align*}
}
\looseness-1 \noindent $\mathcal{O}_{r_i}$ is the set of column headers for each
outlier in $T_{r_i}$. E.g., $\mathcal{O}_{\text{PhD}} = \{\text{IT}\}$, if
$900K$ is an outlier for the row $T_{\text{PhD}}$. 
We ensure that the score increases with the number of outliers by subtracting
the inverse of their count from 1. Since even a single outlier matters, we add
1 to the denominator to ensure that even one outlier results in a score
multiplier of 0.5. The threshold $\tau_{\mathcal{O}}$ is set to $4$, since,
assuming normal distribution, 99.99\% of the population is expected to lie
within 4 standard deviations from the
mean~\cite{ConformanceSIGMOD23Anna}\footnote{Data in pivot tables may not be
normally distributed. We use this heuristic for simplicity and the user can
replace this with other methods.}; and anything outside this range is an
outlier. $\widetilde{Pr}_{\text{outlier}}(r_i, c, T_{r_i}^{c})$ denotes the
LLM-obtained likelihood of $T_{r_i}^{c}$ \emph{not} being an outlier w.r.t
$T_{r_i}$. We compute $\surscore^{\text{col}}(T)$ similarly and compute the
surprise score:
\begin{equation}\label{eq4}
	\small
\surscore(T) = \max (\surscore^{\text{row}}(T), \surscore^{\text{col}}(T))
\end{equation}

\subsubsection*{\textbf{Computing Insightfulness.}}
\label{sec:computing_insightfulness} A pivot table is insightful if it exhibits
any of the characteristics: informativeness, trend, or surprise. Thus, we take
the \emph{maximum} of $\infscore$, $\trendscore$, and $\surscore$
(Equations~\ref{eq2}, \ref{eq3}, and~\ref{eq4}) to compute $Insightfulness:
\Universe \mapsto [0, 1]$. To prioritize pivot tables involving significant
attributes, we scale this score by $S_{\text{sig}}$.
\begin{equation}\label{insightfulness-score}
    \small
    Insightfulness(T) = S_{\text{sig}}(T) \cdot 
    \max\big( 
        \infscore(T),\;
        \trendscore(T),\;
        \surscore(T)
    \big)
\end{equation}

\begin{example} 
	For the pivot table of Figure~\refsub{fig:div_adap}{tab:d}, the attribute
significance score $S_{\text{sig}}(T) {=} 1$ (Example~\ref{sigattr}). The
informativeness score $\infscore(T) {=} 0.32$ (Example~\ref{infexample}), the
trend score $\trendscore(T) {=} 0.32$ (Example~\ref{trendexample}), and since
there is no outlier in the pivot table, the surprise score $\surscore(T) {=}
0.0$. Thus $Insightfulness(T) {=} 1 \times \max (0.32, 0.39, 0.0) {=}
0.39$.
\end{example}

\noindent \emph{Remark 1.} Presence of outliers can inflate the normalization
factor $\gamma$ (\S\ref{infsec}), causing $\infscore$ to shrink
significantly due to the compression of the value range of non-outliers.
However, such cases typically yield a high $\surscore$, complementing the low
$\infscore$. 

\revisethree{\smallskip \noindent \emph{Remark 2.} \label{remark2} The Max
operator used in Equation~\ref{insightfulness-score} may favor the component
with the highest mean, without considering their distributions. This a common
challenge in multi-criteria decision making~\cite{multicriteria}. However,
standardizing all components to the same mean can distort their relative
importance. For example, if all pivot tables have uniformly low informativeness
(with low variance) but uniformly high trend (with low variance), normalization
would artificially inflate informativeness and suppress trend, leading to
incorrect selections. The Max operator retains the strongest signal without
distorting component distributions through forced standardization. An
alternative is using a weighted linear combination of components, with weights
learned from labeled datasets via machine learning, which is orthogonal to our
research. Nevertheless, \sysName is agnostic to the choice of the formula to
compute \emph{Insightfulness} and any linear formula would work.}

\subsection{Interpretability}
\label{sec:interpretability}
A key measure of a pivot table's utility is \emph{interpretability}, which
takes into account the cognitive constraints of humans. Consider the aggregate
\texttt{SUM(AGE)}, row group \texttt{Degree} (3 values), and column groups
\texttt{Employed\_Year} (14 values) and \texttt{Department} (2 values), which
results in an 84-cell pivot table. Its interpretability suffers due to three
reasons: (1)~High sparsity resulting from many value combinations yielding
empty sets---e.g., no MS hire in 2011 for the IT department (\S\ref{densec}),
(2)~Semantically invalid aggregate \texttt{SUM(AGE)} (\S\ref{semsec}), and
(3)~Excessive \#columns ($14 \times 2 = 28$) from fine-grained yearly grouping,
compromising conciseness (\S\ref{consec}). We proceed to describe three
desirable interpretability properties.

\newcommand{\denscore}{\ensuremath{S_{\text{den}}}\xspace}
\subsubsection{Density}\label{densec}
Since each pivot table cell maps to a data subset under a specific value
combination (e.g., MS hires in IT in 2011), empty subsets can occur. When
aggregated, these empty subsets produce \textit{null} values. However,
excessive nulls hinder interpretability~\cite{microsoft_forum_null_pivot}, as
humans struggle to draw insights from sparse tables.
This motivates a key interpretability criterion: high \emph{density}. We
compute the \emph{density score} $\denscore: \Universe \mapsto [0, 1]$ as
follows:
\begin{equation}
    \label{eq:denscore}
    {\small
    \denscore(T) = 
    \frac{\sum_{(r_{i}, c_{j}) \in D[\mathbf{R}] \times D[\mathbf{C}]} [\![T_{r_i}^{c_j}\neq\textit{null}]\!]}{n \cdot m}}
\end{equation}
\begin{example}
The pivot table of Figure~\refsub{fig:div_adap}{tab:d} has 3 rows and 2 columns
(total 6 cells) and no null values. Hence, $\denscore(T) = \frac{6}{2 \times 3}
= 1.0$.
\end{example}

\newcommand{\quescore}{\ensuremath{S_{\text{sem}}}\xspace} 

\subsubsection{Semantic validity}\label{semsec}
Row and column headers in a pivot table represent unique values of the grouping
attributes in $\mathbf{G}$. For interpretability, these headers must be
semantically meaningful~\cite{CharacterizingCGF19Leilani}. E.g., \texttt{Degree}
with values \texttt{\{MS, BS, PhD\}} is interpretable, while a functionally
equivalent \texttt{Degree\_ID} with values \texttt{\{1, 2, 3\}} is not, due to
the lack of direct semantic meaning~\cite{ClassificationQQ18Anna}. Similarly,
the aggregate function $F$ must be semantically valid w.r.t $V$: \texttt{AVG} is
semantically valid for \texttt{AGE}, but \texttt{SUM} is
not~\cite{DataWarehouseWiley13Kimball}. Though intuitive for humans, such
judgments require domain knowledge. Thus, we leverage an LLM to mimic human
reasoning and assess aggregation semantics. We define the \emph{semantic
validity score} of \pivottabletwo based on two criteria: (1)~whether the data
types of attributes in $\mathbf{G}$ are textual, and (2)~the extent to which $F$
is semantically valid w.r.t $V$.
\begin{equation}
    \label{eq:quescore}
    \small
    \quescore(T) = 
		\frac{
		|\{A \in \mathbf{G} \text{ s.t } \mathtt{DataType}(A) \text{ is Text}\}|
		}
		{|\mathbf{G}|}  
	\cdot 
    Pr_{\text{agg}}(F, V)    
\end{equation}

\noindent We compute $Pr_{\text{agg}}(F, V)$ from an LLM-generated ranking of $F
\in \mathcal{F}$ based on its semantic validity w.r.t $V$. We score the best
function $1.0$, the next $0.8$, and so on, which ensures that $\quescore(T) \in
[0, 1]$.

%
%
%
%

\begin{example}\label{ex:38}
The pivot table of Figure~\refsub{fig:div_adap}{tab:d}, contains only textual
headers. LLM responds to our prompt ``Rank the functions {\texttt{COUNT},
\texttt{AVG}, \texttt{SUM}, \texttt{MIN}, and \texttt{MAX}}, based on their
appropriateness for analyzing \texttt{Salary}'' with \{\texttt{AVG}, \dots\}.
Thus, $\quescore(T) = \frac{2}{2} \times 1.0 = 1.0$.
\end{example}

\newcommand{\conscore}{\ensuremath{S_{\text{con}}}\xspace}
\subsubsection{Conciseness}\label{consec}

\looseness-1 While multiple grouping attributes may enhance insightfulness, too
many cells reduce comprehensibility~\cite{KeepingIEEE18Zening}, and, thus,
interpretability. To model this, we define conciseness score $\conscore:
\Universe \mapsto [0, 1]$ using a piecewise function~\cite{DashBotIEEE23Dazhen}:

\begin{minipage}{0.32\textwidth}
\begin{equation}
    \label{eq:conscore}
	\hspace{-5mm}
    {\small
    \conscore(T) = 
    \begin{cases}
        1 - z|T|, & \text{if } |T| \leq \tau_{c} \\
        (1 - z\tau_{c}) e^{-\lambda (|T| - \tau_{c})}, & \text{if } |T| > \tau_{c}
    \end{cases}}
	\hspace{0mm}
    \end{equation}
\end{minipage}%
\hspace{2mm}
\begin{minipage}{0.9\textwidth}
\begin{tikzpicture}
  \begin{axis}[
      xlabel={$|T|$},
      ylabel={$\conscore(T)$},
      xlabel near ticks,
      ylabel near ticks,
      xlabel style={font=\scriptsize, inner sep=-3pt},
      ylabel style={font=\scriptsize, inner sep=0pt},
      tick label style={font=\scriptsize},
      xtick={0, 8, 30},
      xticklabels={$0$, $\tau_c$, $\infty$},
      ytick={0, 1},
      yticklabels={$0$, $1$},
      height=2.5cm,
      width=3cm,
      grid=none,
      domain=0:30,
      samples=200,
      enlarge x limits=false,
      enlarge y limits=false,
      clip=false,
      ymin=0,
      ymax=1,
    ]
    \addplot[
      thick,
      blue,
    ]
    {x <= 8 ? 1 - 0.03*x : (1 - 0.03*8)*exp(-0.5*(x - 8))};
    \addplot[dashed, gray] coordinates {(8,0) (8,1)};
  \end{axis}
\end{tikzpicture}
\vspace{1mm}
\end{minipage}
\noindent \looseness-1 Here, $|T|$ denotes the number of cells in $T$. This
formula captures the intuition that interpretability declines gradually at
first, but drops sharply once $|T|$ exceeds a threshold $\tau_c$, set to $16$.
We apply a 3\% linear decrease ($z = 0.03$) until $|T|$ exceeds $\tau_c$, and an
exponential decay at a rate of 50\% ($\lambda = 0.5$) beyond that. This is
grounded in cognitive load theory~\cite{miller1956magical, barrouillet2007time},
which states that performance declines sharply when cognitive demand exceeds
working-memory capacity.


\begin{example}
The pivot table of Figure~\refsub{fig:div_adap}{tab:d} has $6$ cells. Since $6
{<} 16$, we compute the linear part: $1 {-} 0.03{\times} 6 {=} 0.82$. Thus,
$\conscore(T) {=} 0.82$.
\end{example}



\subsubsection*{\textbf{Computing Interpretability.}} 

Unlike insightfulness, where a strong signal from \emph{any} single type of
insight is sufficient, interpretability demands that \emph{all} criteria be met
simultaneously. Therefore, we compute $Interpretability: \Universe \mapsto [0,
1]$ as the average of the three scores: \denscore, \quescore, and \conscore
(Equations~\ref{eq:denscore}, \ref{eq:quescore}, and \ref{eq:conscore}):
\begin{equation}\label{interpretability-score}
    \small
    Interpretability(T) = \frac{\denscore(T) + \quescore(T) + \conscore(T)}{3}
\end{equation}

\begin{example}
For Figure~\refsub{fig:div_adap}{tab:d}, we obtained values for \denscore(T),
\quescore(T), and \conscore(T) to be 1.0, 1.0, and 0.82, respectively, in the
previous examples. Thus, $Inter\-pretability(T)$ ${=}
(1.0{+}1.0{+}0.82)/3{=}0.94$.
\end{example}

\subsection{Computing Utility}
\label{utilitycomputation}

We now define $Utility: \Universe \mapsto [0, 1]$ of a pivot table $T$ by
combining \emph{Insightfulness} (Eq.~\ref{insightfulness-score}) and
\emph{Interpretability} (Eq.~\ref{interpretability-score}). To balance their
contributions, we introduce a tunable parameter $\alpha$, set to 0.5 by default
to give equal weight to both components. However, $\alpha$ can be adjusted to
reflect application-specific preferences.
\begin{equation*}
\small
Utility(T) = \alpha \cdot Insightfulness(T) + (1-\alpha)  \cdot  Interpretability(T)
\end{equation*} 

\begin{example}
For Figure~\refsub{fig:div_adap}{tab:d}, the $Insightfulness$ and
$Inter\-pretability$ scores are 0.39 and 0.94, respectively. Thus,
$\text{Utility}(T) = 0.5\times 0.39 + 0.5\times 0.94 = 0.67$.
\end{example}

\noindent\revisemeta{{\emph{Need for a new Utility Model.}} \label{needUtility}
Our \emph{Utility} model significantly extends prior work that address related
problems~\cite{SeedbVartak15, HarrisWWW23SpotLight, ForesightDemiralp17,
QuickInsightsDing19, VisGuide, InsightLens}.
\textbf{First}, prior work ignore the \emph{semantic validity} of the summary
\emph{structure}, which directly affects interpretability (e.g.,
{\small\texttt{SUM(Birth\_Year)}} or {\small\texttt{AVG(Zip\_Code)}} are
invalid; {\small\texttt{\{Val\_1, Val\_2\}}} is less interpretable than
{\small\texttt{\{Large, Small\}}}; and {\small\texttt{Gender}} is a more
interesting grouping attribute than {\small\texttt{Employee ID}}).
\textbf{Second}, prior work rely solely on statistical signals. For example,
SeeDB~\cite{SeedbVartak15} uses an EMD-based deviation metric, which is just
one of the seven components of our \emph{Utility} model, while
others~\cite{QuickInsightsDing19, InsightLens} use value-proportion and
statistical-significance tests. Moreover, they do not support multi-group
reasoning, required in our setting involving pivot tables. But most
importantly, they lack \emph{semantic awareness} to validate an apparent data
insight. For instance, ``a significant deviation in average height between
toddlers and adults'' will incur a high score based on their metrics. However,
such deviation is trivially expected. In contrast, even a moderate ``deviation
in average height across socio-economic groups'' indicates an interesting
insight.
\textbf{Third}, these methods ignore quality factors, such as \texttt{NULL}
values in some groups in the data summary (which yields empty bars for
visualization). In general, most prior work do not incorporate various aspects
of interpretability in their utility models.
The above shortcomings of the prior work motivates a \emph{semantics-aware} and
\emph{interpretability-focused} Utility model, like ours, that identifies
insights that are meaningful in real-world contexts, not merely statistically
distinct.}


\section{Diversity in a Set of Pivot Tables} \label{sec:four} 
While utility quantifies the goodness of a single pivot table in isolation,
\emph{diversity} captures how well a \emph{set} of pivot tables,
\emph{collectively}, provide complementary and unique perspectives on the data
(D3). High diversity in a set of pivot tables is achieved when the pivot tables
are distant from each other with respect to data coverage and the insights they
provide.

\subsubsection*{Diversity.} Following Max-Min
diversification~\cite{DBLP:conf/icdt/Addanki0MM22}, we define diversity of a set
of pivot tables $\mathbf{T} = \{T_1, T_2, \dots\}$ by the smallest pairwise
distance between $\mathbf{T}$'s elements. More formally, given a symmetric
distance function $dist: \Universe \times \Universe \mapsto [0, 1]$, we define
$diversity: 2^{\Universe} \mapsto [0, 1]$ of a set of pivot tables $\mathbf{T}
\subseteq \Universe$ as follows:
$$
{
	\small
    Diversity(\mathbf{T}) = \min_{T_{i}, T_{j} \in \textbf{T} \text{ s.t. } i < j} dist(T_i, T_j)
}
$$
%
%
\subsubsection*{Distance between pivot tables.} A simple heuristic to model
$dist$ is the degree of disjointness between the attributes that define the
pivot-table queries---if two pivot tables operate on the same set of attributes,
the distance is $0$; for completely disjoint set of attributes, the distance is
$1$. However, this heuristic fails to account for the \emph{structural
semantics} of the pivot-table queries and the \emph{content semantics} of the
data in the pivot tables. To this end, we employ a semantics-preserving
embedding function $E: \Universe \mapsto [-1, 1]^p$, which maps a pivot table $T
\in \Universe$ to a $p$-dimensional vector. Then, we compute distances between
pivot tables in this embedding space:
$$
{\small
	dist(T_1, T_2) = \frac{1 - cosine\_similarity(E(T_1), E(T_2))}{2}
}
$$
Here, $cosine\_similarity: \mathbb{R}^p \times \mathbb{R}^p \rightarrow [-1, 1]$
is a widely used measure for comparing embeddings~\cite{ReimersEMNLP19SBERT,
SimpeICLR17Sanjeev}. We divide by $2$ to achieve normalization s.t $dist(T_1,
T_2) \in [0, 1]$.

\subsubsection*{Pivot-table embedding} Pivot-table embedding should capture both
the syntactic and semantic characteristics of the pivot table. To this end, we
combine the query embedding $E_Q$ and the content embedding $E_C$ through
concatenation, i.e., $E(T) = [E_Q(T) ; E_C(T)]$. Concatenating embeddings is a
widely used technique in machine learning, natural language processing, and
multi-modal learning~\cite{DBLP:conf/naacl/ZhangRW16}.
  
\subsubsection*{Query embedding} \looseness-1 Queries define the structural
intent of a pivot table. While they reference data attributes and are aware of
the schema, they are agnostic to the pivot table's content. Thus, the same
pivot-table query over different database instances with an identical schema
should yield the same query embedding. Effective query embeddings must also
capture the semantics of the attribute names and reflect the query semantics.
E.g., \texttt{GROUP BY Income} and \texttt{GROUP BY Salary} are semantically
similar and should therefore be close in the query-embedding space. To this end,
we use T5~\cite{T5JMLR20Colin}, a natural-language encoder fine-tuned over a
Text-to-SQL dataset~\cite{SpiderEMLNLP18Tao}, to obtain the query embedding
$E_Q: \Universe \mapsto [-1, 1]^{1024}$.



\subsubsection*{Content embedding} The content embedding must capture both the
statistical and distributional properties of the pivot-table data and structural
relationships among its attributes and tuples. In this work, we leverage
TAPEX~\cite{TAPEXICLR22Qian}, a pre-trained encoder trained on sentence-table
pairs, which is designed to understand both the structure and content of tabular
data. This results in a content embedding $E_C: \Universe \mapsto [-1,
1]^{1024}$.

\section{The \sysName Algorithm} \label{sec:five}


We now present a solution to Problem~\ref{eq:objective_function}, which is to
recommend a k-budgeted set of pivot tables with maximum utility under a
diversity constraint. \revisetwo{This is an instance of the NP-hard Maximum
Weight Independent Set (MWIS) problem~\cite{sakai2003note} with complexity
$O(|\Universe|^{k})$ in the size of the search space (\Universe in our case).
Furthermore, \Universe grows combinatorially with $|\mathbf{A}|$ (the number of
data attributes), leading to an exponential growth. \label{nphard5}}
Approximation techniques exist for special cases of MWIS and related
problems~\cite{sakai2003note, joo2015distributed, nayeem2007genetic,
DBLP:journals/pvldb/BrucatoBAM16} and alternative formulations are possible
such as incorporating diversity into the objective or solving the dual version
that maximizes diversity~\cite{DBLP:conf/icdt/Addanki0MM22} under a utility
constraint. However, in \sysName, we adopt a simple greedy approach
(\S\ref{sec:greedyalgo}) for two key reasons: (1)~Interactive response time is
desirable for our problem settings and a greedy approach achieves linear time
complexity of $O(|\Universe|)$. (2)~As we demonstrate in
\S\ref{sec:experiments}, our greedy approach works remarkably well in practice
over real-world datasets, almost always matching the exact solution.

\subsection{Optimizations} \label{sec:offlineopt}

\looseness-1 The linear-time complexity of the greedy approach is still
prohibitive for practical use for two reasons:
(1)~Computing the \emph{Utility} of candidate pivot tables requires their
materialization, which is computationally expensive---especially since
\Universe grows exponentially with $|\mathbf{A}|$. (2)~Some components of
\emph{Utility} relies on the LLM, whose inference latency is a
bottleneck~\cite{PruneandTuneICLR23Aaquib,wan2024efficient}. To address these
efficiency challenges (D5), we introduce two offline optimizations:
(1)~\emph{Candidate pruning}: based on the query structure of the pivot tables,
we eliminate potentially low-utility ones before materialization, thereby
significantly reducing the search space and avoiding many materializations
(\S\ref{sec5:prune}). (2)~A lightweight \emph{proxy model}: tailored to the
dataset, it approximates LLM inferences efficiently (\S\ref{sec5:cache}).

\subsubsection{Pruning} \label{sec5:prune}
To prune pivot tables likely to yield low utility without materialization, we
leverage three components of \emph{Utility} (\S\ref{utilitycomputation}):
C-I:~\emph{attribute significance} (\S\ref{attsigsec}) used in \emph{Insightfulness}
(Eq~\ref{insightfulness-score}),
C-II:~\emph{semantic validity} (\S\ref{semsec}), and
C-III:~\emph{conciseness} (\S\ref{consec}) used in \emph{Interpretability}
(Eq~\ref{interpretability-score}).
These choices are motivated by computational efficiency, as these components
require knowledge of only the pivot-table query, which defines the structure of
the pivot table---such as which attributes are used, number of cells, etc.---and
do not require a full materialization over the dataset.

\looseness-1 While \emph{Insightfulness} selects only the maximum among multiple
components, \emph{Interpretability} averages out three components---two of which are
C-II and C-III above. This enables effective pruning of pivot tables that
already show low scores for C-II and/or C-III. Furthermore, since C-I interacts
in a multiplicative way with other components of \emph{Insightfulness}
(Eq~\ref{insightfulness-score}), a low value will inevitably result in low
\emph{Insightfulness}, making it an ideal choice. Our pruning algorithm works as
follows: we evaluate scores for the above three components and over-approximate
\emph{Utility} (\S\ref{utilitycomputation}) by assigning maximum possible values for
all other components whose values are unknown. Through this conservative
estimate, we discard candidates with a score below a threshold (a system
parameter set to 0.5).

\setlength{\textfloatsep}{0pt}
\begin{algorithm}[t]
\LinesNumbered
\small{
\Input{
    Database $D$,\\
	Unmaterialized candidate pivot table set \Universe,\\
    Diversity threshold $\theta$, \\
    The desired number of pivot tables $k$
}

\Output{
    A high-utility set $\mathbf{T} {\subseteq} \Universe$ s.t.\  $|\mathbf{T}| {\le} k$ and $Diversity(\mathbf{T}) {\ge} \theta$
}

\tcc{\textbf{Offline phase}: executed once for each dataset. Re-executed if the schema changes or major change happens in the data distribution.}
\tcc{Prune candidate set based on pivot-table structure (\S\ref{sec5:prune})}
$\Universe^{Pr} \leftarrow Prune(\Universe)$\\

\tcc{Building LLM-Proxy (\S\ref{sec5:cache})}
$Q \leftarrow$ list of questions about $D$ 	\tcc*[f]{Generate prompts for the LLM}\\

$R \leftarrow \text{LLM-Response}(Q)$		\tcc*[f]{Get responses from the LLM}\\

$\textit{LLM-Proxy} \leftarrow Train(Q, R)$ \tcc*[f]{Train a proxy prediction model}\\

\vspace{1mm}
\tcc{\textbf{Online phase}: executed whenever the data changes.}
\tcc{Materialize and compute utility based on pivot-table contents}
\ForEach{$T \in \Universe^{Pr}$}{
    Materialize $T$ over $D$ \\
	Compute $Utility(T)$ \tcc*[f]{\S\ref{utilitycomputation}}
}
$\Universe^{Pr} = sorted(\Universe^{Pr})$ \tcc*[f]{Sort by the descending order of Utility}\\

\tcc{Greedy selection (\S\ref{sec:greedyalgo})}
$\mathbf{T} \gets \emptyset$ \tcc*[f]{Initialize an empty set}

\While{$|\mathbf{T}| \le k$}{
    \ForEach{$T \in \Universe^{Pr}$}{
		\tcc{No pivot table in $\mathbf{T}$ is within $\theta$ distance away from $T$}
        \If{$\not\exists T' \in \mathbf{T} \text{ s.t.\ } dist(T, T') < \theta$}{
            $\mathbf{T} \gets \mathbf{T} \cup \{T\}$			
        }
    }
}
\Return $\mathbf{T}$
}
\caption{\sysName algorithm}
\label{alg:llm_proxy_diverse_selection}
\end{algorithm}


\subsubsection{LLM-proxy} \label{sec5:cache}
\looseness-1 Recall that the computation of pivot-table utility requires LLM
inferences (\S\ref{sec:three}), and querying LLMs is time-consuming in
practice. Even if we cache the LLM responses, any change in the underlying data
or user-specified parameters (D4) would require LLM re-consultation. To
expedite this process, we train a cheap but significantly faster decision-tree
classifier to mimic LLM behavior, serving as an ``LLM-proxy'' during the online
phase of \sysName (Algorithm~\ref{alg:llm_proxy_diverse_selection}). We train
the classifier over 10{,}000 LLM prompt-response pairs, where we generate
potential LLM-queries based on the dataset. The proxy model is a simple
predictive model to answer the fixed-template questions discussed in
\S\ref{sec:three}. No re-training of this proxy model is required as long as
the data distribution and overall trends in the dataset remain unchanged. While
training with additional prompt-response pairs or employing more complex
classifiers can further improve the accuracy of this proxy model, we show
empirically in Section~\ref{sec:exp_setup} that 10{,}000 prompt-response pairs
already provide reasonable accuracy for practical applications. 

\subsection{\sysName: Greedy Algorithm} \label{sec:greedyalgo}
Algorithm~\ref{alg:llm_proxy_diverse_selection} shows the \sysName workflow.
Line~1 denotes the offline pruning step and lines~2--4 show the steps for
LLM-proxy training. The online phase is shown in lines 5--14. The pivot tables
that survive the pruning step are materialized in line~6, we then compute their
\emph{Utility} scores (line~7) and sort the pruned candidate set of pivot
tables by descending order of utility (line~8). The greedy selection phase
(lines~9--14) selects pivot tables greedily while ensuring that they satisfy
the diversity constraint w.r.t the already selected ones in $\mathbf{T}$. If a
candidate pivot table is at least $\theta$ away from all the previously
selected tables, we include it to $\mathbf{T}$ (lines~12--13). The algorithm
terminates when we have selected $k$ pivot tables or no more candidates satisfy
the diversity constraint.

\setlength{\textfloatsep}{10pt}

\subsection{\revisemeta{\sysNameP: A Practical Variant}} \label{sec:sageplus}
\revisemeta{Even after the two optimizations, \sysName may not ensure
interactive speed especially when the dataset has millions of tuples or
hundreds of attributes, hurting it's practical adoption. To this end, we
propose \sysNameP, a variant of \sysName that incorporates sampling and
approximation techniques to ensure practically acceptable runtimes in an
interactive setting, which is essential for spreadsheet environments.
Specifically, to handle high-cardinality and high-dimensionality, it (i)~uses
20\% of the data samples when the dataset is large (e.g., contains 1M+ tuples
or 100+ attributes) for candidate pivot-table generation, (2)~uses 10\%
subsampling for approximating \emph{Trend} (\S\ref{trendsec}) and
\emph{Surprise} (\S\ref{sursec}) scores, and (iii)~applies dimensionality
reduction using Johnson-Lindenstrauss lemma~\cite{johnson1984extensions}, which
preserves pairwise distances up to a distortion factor ratio $\epsilon = 0.2$,
for approximating the \emph{Informativeness} score (\S\ref{infsec}).
Furthermore, \sysNameP uses multiple cores ($12$ in our experiments) for
parallelizing computations. Note that \sysNameP does not improve the
theoretical runtime complexity compared to \sysName; instead, it provides
practical enhancements that ensure better suitability for practical use.}


\newcount\valcount
\newcommand{\parsevals}[1]{%
  \valcount=0
  \StrCount{#1}{,}[\commacount]%
  \def\temp{#1,}%
  \loop
    \StrBefore{\temp}{,}[\currentval]%
    \expandafter\xdef\csname valItem\the\valcount\endcsname{\currentval}%
    \advance\valcount by 1
    \StrBehind{\temp}{,}[\temp]%
  \ifnum\valcount<\numexpr\commacount+1\relax
  \repeat
}

\newcommand{\dataheatmap}[2]{%
  \begingroup
  \parsevals{#2}%
  \centering
  \adjustbox{valign=c}{%
    \resizebox{0.03\textwidth}{!}{%
      \begin{tikzpicture}
        \def\s{1}
        \foreach \i in {0,...,#1} {
          \foreach \j in {0,...,#1} {
            \pgfmathtruncatemacro{\idx}{(#1+1)*\i + \j}
            \pgfmathtruncatemacro{\idxInt}{\idx}
            \expandafter\let\expandafter\valraw\csname valItem\idxInt\endcsname
            \pgfmathsetmacro{\val}{\valraw}
			\pgfmathsetmacro{\valclip}{max(0,min(0.5,\val))}
			\pgfmathsetmacro{\gray}{100 * (1 - 2*\valclip)}
            \fill[black!\gray!white] (\j*\s,-\i*\s) rectangle ++(\s,\s);
            \draw[black] (\j*\s,-\i*\s) rectangle ++(\s,\s);
          }
        }
      \end{tikzpicture}
    }%
  }%
  \endgroup
}

\newcommand{\hmTopKThree}{%
  \dataheatmap{2}{0.0000,0.4875,0.3938,
0.4875,0.0000,0.4556,
0.3938,0.4556,0.0000,}%
}
\newcommand{\hmTopKFive}{%
  \dataheatmap{4}{0.0000,0.4875,0.3938,0.4842,0.3250,
0.4875,0.0000,0.4556,0.1092,0.5099,
0.3938,0.4556,0.0000,0.4542,0.3397,
0.4842,0.1092,0.4542,0.0000,0.4945,
0.3250,0.5099,0.3397,0.4945,0.0000,}%
}
\newcommand{\hmTopKTen}{%
  \dataheatmap{9}{0.0000,0.4875,0.3938,0.4842,0.3250,0.2995,0.3024,0.4850,0.4853,0.1079,
0.4875,0.0000,0.4556,0.1092,0.5099,0.5001,0.5008,0.1102,0.1111,0.4951,
0.3938,0.4556,0.0000,0.4542,0.3397,0.4521,0.4394,0.4550,0.4569,0.3985,
0.4842,0.1092,0.4542,0.0000,0.4945,0.5025,0.5051,0.0199,0.0228,0.4903,
0.3250,0.5099,0.3397,0.4945,0.0000,0.4539,0.4483,0.4956,0.4964,0.3315,
0.2995,0.5001,0.4521,0.5025,0.4539,0.0000,0.1676,0.5045,0.5057,0.2931,
0.3024,0.5008,0.4394,0.5051,0.4483,0.1676,0.0000,0.5069,0.5082,0.2939,
0.4850,0.1102,0.4550,0.0199,0.4956,0.5045,0.5069,0.0000,0.0185,0.4912,
0.4853,0.1111,0.4569,0.0228,0.4964,0.5057,0.5082,0.0185,0.0000,0.4917,
0.1079,0.4951,0.3985,0.4903,0.3315,0.2931,0.2939,0.4912,0.4917,0.0000,}%
}
\newcommand{\hmTopKThreeVideo}{%
  \dataheatmap{2}{0.0000,0.3020,0.0561,
0.3020,0.0000,0.3072,
0.0561,0.3072,0.0000,}%
}
\newcommand{\hmTopKFiveVideo}{%
  \dataheatmap{4}{0.0000,0.3020,0.0561,0.3066,0.2956,
0.3020,0.0000,0.3072,0.0544,0.1544,
0.0561,0.3072,0.0000,0.3029,0.3006,
0.3066,0.0544,0.3029,0.0000,0.1621,
0.2956,0.1544,0.3006,0.1621,0.0000,}%
}
\newcommand{\hmTopKTenVideo}{%
  \dataheatmap{9}{0.0000,0.1006,0.4250,0.0870,0.2583,0.0804,0.2537,0.4903,0.5081,0.4245,
                  0.1006,0.0000,0.4319,0.0932,0.2819,0.0886,0.2488,0.4881,0.5095,0.4307,
                  0.4250,0.4319,0.0000,0.4298,0.3772,0.4279,0.4665,0.3631,0.5424,0.0348,
                  0.0870,0.0932,0.4298,0.0000,0.2742,0.0908,0.2499,0.4953,0.5124,0.4284,
                  0.2583,0.2819,0.3772,0.2742,0.0000,0.2533,0.3431,0.4188,0.5100,0.3753,
                  0.0804,0.0886,0.4279,0.0908,0.2533,0.0000,0.2548,0.4802,0.5045,0.4251,
                  0.2537,0.2488,0.4665,0.2499,0.3431,0.2548,0.0000,0.5201,0.5067,0.4669,
                  0.4903,0.4881,0.3631,0.4953,0.4188,0.4802,0.5201,0.0000,0.4879,0.3618,
                  0.5081,0.5095,0.5424,0.5124,0.5100,0.5045,0.5067,0.4879,0.0000,0.5418,
                  0.4245,0.4307,0.0348,0.4284,0.3753,0.4251,0.4669,0.3618,0.5418,0.0000}%
}
\newcommand{\hmTopKThreeHouse}{%
  \dataheatmap{2}{0.0000,0.0950,0.1022,
0.0950,0.0000,0.0916,
0.1022,0.0916,0.0000,}%
}
\newcommand{\hmTopKFiveHouse}{%
  \dataheatmap{4}{0.0000,0.0950,0.1022,0.1000,0.0540,
0.0950,0.0000,0.0916,0.1061,0.0923,
0.1022,0.0916,0.0000,0.0872,0.1133,
0.1000,0.1061,0.0872,0.0000,0.1101,
0.0540,0.0923,0.1133,0.1101,0.0000}%
}
\newcommand{\hmTopKTenHouse}{%
  \dataheatmap{9}{0.0000,0.0950,0.1022,0.1000,0.0540,0.1046,0.4849,0.4924,0.4905,0.4816,
0.0950,0.0000,0.0916,0.1061,0.0923,0.0994,0.4785,0.4837,0.4824,0.4690,
0.1022,0.0916,0.0000,0.0872,0.1133,0.1167,0.4837,0.4855,0.4826,0.4760,
0.1000,0.1061,0.0872,0.0000,0.1101,0.1245,0.4913,0.4936,0.4908,0.4864,
0.0540,0.0923,0.1133,0.1101,0.0000,0.0989,0.4816,0.4880,0.4867,0.4783,
0.1046,0.0994,0.1167,0.1245,0.0989,0.0000,0.4753,0.4810,0.4794,0.4724,
0.4849,0.4785,0.4837,0.4913,0.4816,0.4753,0.0000,0.1847,0.1870,0.1539,
0.4924,0.4837,0.4855,0.4936,0.4880,0.4810,0.1847,0.0000,0.0723,0.1592,
0.4905,0.4824,0.4826,0.4908,0.4867,0.4794,0.1870,0.0723,0.0000,0.1483,
0.4816,0.4690,0.4760,0.4864,0.4783,0.4724,0.1539,0.1592,0.1483,0.0000,}%
}

\newcommand{\hmGreedyKThree}{%
  \dataheatmap{2}{0.0,0.484474,0.449029,
0.484474,0.0,0.454272,
0.449029,0.454272,0.0}%
}
\newcommand{\hmGreedyKFour}{%
  \dataheatmap{3}{0.0,0.475588,0.450467,0.344413,0.386065,
0.475588,0.0,0.430741,0.510448,0.378244,
0.450467,0.430741,0.0,0.474303,0.313723,
0.344413,0.510448,0.474303,0.0,0.459226,
0.386065,0.378244,0.313723,0.459226,0.0,}%
                  }
\newcommand{\hmGreedyKFive}{%
  \dataheatmap{4}{0.0000,0.5043,0.3621,0.4855,0.3039,
                  0.5043,0.0000,0.4387,0.5076,0.4888,
                  0.3621,0.4387,0.0000,0.4456,0.3613,
                  0.4855,0.5076,0.4456,0.0000,0.4540,
                  0.3039,0.4888,0.3613,0.4540,0.0000}%
}
\newcommand{\hmGreedyKTen}{%
  \dataheatmap{9}{0.0,0.462446,0.201592,0.194563,0.154708,0.113466,0.443396,0.446323,0.435595,0.463109,
0.462446,0.0,0.448307,0.447136,0.474154,0.481748,0.167937,0.12136,0.470106,0.253391,
0.201592,0.448307,0.0,0.122714,0.202468,0.204259,0.431004,0.430741,0.450467,0.447522,
0.194563,0.447136,0.122714,0.0,0.210311,0.212188,0.428601,0.429736,0.449786,0.444841,
0.154708,0.474154,0.202468,0.210311,0.0,0.156311,0.458014,0.458143,0.443482,0.478854,
0.113466,0.481748,0.204259,0.212188,0.156311,0.0,0.46625,0.468905,0.45653,0.481736,
0.443396,0.167937,0.431004,0.428601,0.458014,0.46625,0.0,0.177202,0.440501,0.207366,
0.446323,0.12136,0.430741,0.429736,0.458143,0.468905,0.177202,0.0,0.475588,0.272476,
0.435595,0.470106,0.450467,0.449786,0.443482,0.45653,0.440501,0.475588,0.0,0.449029,
0.463109,0.253391,0.447522,0.444841,0.478854,0.481736,0.207366,0.272476,0.449029,0.0,}%
}
\newcommand{\hmGreedyKThreeVideo}{%
  \dataheatmap{2}{0.0000,0.3020,0.3331,
                0.3020,0.0000,0.4159,
                0.3331,0.4159,0.0000,}%
}
\newcommand{\hmGreedyKFiveVideo}{%
  \dataheatmap{4}{0.0000,0.3020,0.3331,0.5262,0.3948,
0.3020,0.0000,0.4159,0.5243,0.4212,
0.3331,0.4159,0.0000,0.5159,0.3089,
0.5262,0.5243,0.5159,0.0000,0.5181,
0.3948,0.4212,0.3089,0.5181,0.0000,
                  }%
}
\newcommand{\hmGreedyKTenVideo}{%
  \dataheatmap{9}{0.0000,0.3020,0.2956,0.3331,0.3482,0.5262,0.3455,0.3097,0.3948,0.3183,
0.3020,0.0000,0.1544,0.4159,0.4178,0.5243,0.3788,0.1054,0.4212,0.4291,
0.2956,0.1544,0.0000,0.4348,0.4403,0.5154,0.4087,0.1887,0.4530,0.4307,
0.3331,0.4159,0.4348,0.0000,0.1175,0.5159,0.2861,0.4219,0.3089,0.2008,
0.3482,0.4178,0.4403,0.1175,0.0000,0.5125,0.2591,0.4209,0.2677,0.2421,
0.5262,0.5243,0.5154,0.5159,0.5125,0.0000,0.5128,0.5392,0.5181,0.5115,
0.3455,0.3788,0.4087,0.2861,0.2591,0.5128,0.0000,0.3607,0.1577,0.3319,
0.3097,0.1054,0.1887,0.4219,0.4209,0.5392,0.3607,0.0000,0.4101,0.4413,
0.3948,0.4212,0.4530,0.3089,0.2677,0.5181,0.1577,0.4101,0.0000,0.3711,
0.3183,0.4291,0.4307,0.2008,0.2421,0.5115,0.3319,0.4413,0.3711,0.0000,
                  }%
}
\newcommand{\hmGreedyKThreeHouse}{%
  \dataheatmap{2}{0.0,0.4753,0.4811,
                  0.4753,0.0,0.212,
                  0.4811,0.212,0.0,
                  }%
}
\newcommand{\hmGreedyKFiveHouse}{%
  \dataheatmap{4}{0.0,0.475275,0.4811,0.201054,0.497741,
                  0.475275,0.0,0.211991,0.506856,0.254021,
                  0.4811,0.211991,0.0,0.495849,0.202623,
                  0.201054,0.506856,0.495849,0.0,0.485648,
                  0.497741,0.254021,0.202623,0.485648,0.0,
                  }%
}
\newcommand{\hmGreedyKTenHouse}{%
  \dataheatmap{9}{0.0,0.104641,0.116693,0.475275,0.48097,0.472405,0.456408,0.475712,0.155819,0.479039,
0.104641,0.0,0.102208,0.484931,0.492375,0.481647,0.480288,0.485114,0.130502,0.493462,
0.116693,0.102208,0.0,0.483703,0.485546,0.476047,0.479191,0.489861,0.110933,0.48931,
0.475275,0.484931,0.483703,0.0,0.18466,0.153905,0.194574,0.184643,0.487198,0.154176,
0.48097,0.492375,0.485546,0.18466,0.0,0.159158,0.167586,0.136634,0.487725,0.174758,
0.472405,0.481647,0.476047,0.153905,0.159158,0.0,0.198877,0.184024,0.481382,0.187778,
0.456408,0.480288,0.479191,0.194574,0.167586,0.198877,0.0,0.129966,0.484086,0.173095,
0.475712,0.485114,0.489861,0.184643,0.136634,0.184024,0.129966,0.0,0.493868,0.157942,
0.155819,0.130502,0.110933,0.487198,0.487725,0.481382,0.484086,0.493868,0.0,0.494938,
0.479039,0.493462,0.48931,0.154176,0.174758,0.187778,0.173095,0.157942,0.494938,0.0,}%
}
\newcommand{\hmBFKThree}{%
  \dataheatmap{2}{0.0000,0.3621,0.5063,
                  0.3621,0.0000,0.4442,
                  0.5063,0.4442,0.0000}%
}
\newcommand{\hmBFKFive}{%
  \dataheatmap{4}{0.0000,0.3570,0.3155,0.5152,0.3004,
                  0.3570,0.0000,0.3317,0.4442,0.3992,
                  0.3155,0.3317,0.0000,0.4923,0.3223,
                  0.5152,0.4442,0.4923,0.0000,0.4995,
                  0.3004,0.3992,0.3223,0.4995,0.0000}%
}
\newcommand{\hmBFKThreeVideo}{%
  \dataheatmap{2}{0,0.425016,0.490274,
                  0.425016,0,0.363122,
                  0.490274,0.363122,0}%
}
\newcommand{\hmBFKFiveVideo}{%
  \dataheatmap{4}{0,0.343145,0.506746,0.466495,0.520107,
                  0.343145,0,0.509967,0.377171,0.418811,
                  0.506746,0.509967,0,0.542396,0.487906,
                  0.466495,0.377171,0.542396,0,0.363122,
                  0.520107,0.418811,0.487906,0.363122,0}%
}
\newcommand{\hmBFKThreePitches}{%
  \dataheatmap{2}{0,0.473129,0.47686,
                  0.473129,0,0.262506,
                  0.47686,0.262506,0}%
}
\newcommand{\hmBFKFivePitches}{%
  \dataheatmap{4}{0,0.473129,0.462939,0.47686,0.477184,
                  0.473129,0,0.281457,0.262506,0.213476,
                  0.462939,0.281457,0,0.20523,0.346807,
                  0.47686,0.262506,0.20523,0,0.331792,
                  0.477184,0.213476,0.346807,0.331792,0}%
}
\newcommand{\hmGsheetsMarketing}{%
  \dataheatmap{1}{0.0000,0.5635,
                  0.5635,0.0000,}%
}
\newcommand{\hmGsheetsVideo}{%
  \dataheatmap{2}{0.0000,0.3176,0.3661,
                0.3176,0.0000,0.3292,
                0.3661,0.3292,0.0000,}%
}
\newcommand{\hmGsheetsPitches}{%
  \dataheatmap{2}{0.00000,0.03565,0.00000,
                  0.03565,0.00000,0.03565,
                  0.00000,0.03565,0.00000}%
}
\newcommand{\hmLLMKThreeMarketing}{%
  \dataheatmap{2}{0.0000,0.5286,0.5338,
                  0.5286,0.0000,0.1556,
                  0.5338,0.1556,0.0000,}%
}
\newcommand{\hmLLMKFiveMarketing}{%
  \dataheatmap{4}{0.0000,0.4054,0.4549,0.3528,0.5641,
                  0.4054,0.0000,0.3066,0.3355,0.4280,
                  0.4549,0.3066,0.0000,0.3494,0.3202,
                  0.3528,0.3355,0.3494,0.0000,0.5178,
                  0.5641,0.4280,0.3202,0.5178,0.0000,}%
}
\newcommand{\hmLLMKTenMarketing}{%
  \dataheatmap{9}{0.0000,0.4351,0.3019,0.5254,0.3029,0.3532,0.5217,0.6058,0.6360,0.5858,
0.4351,0.0000,0.3109,0.5101,0.3438,0.3005,0.2352,0.3757,0.4303,0.5784,
0.3019,0.3109,0.0000,0.5014,0.3697,0.3407,0.4107,0.5529,0.5535,0.5867,
0.5254,0.5101,0.5014,0.0000,0.5047,0.5004,0.5252,0.5692,0.5600,0.4082,
0.3029,0.3438,0.3697,0.5047,0.0000,0.2007,0.4668,0.5096,0.5910,0.5768,
0.3532,0.3005,0.3407,0.5004,0.2007,0.0000,0.4203,0.4484,0.5461,0.5649,
0.5217,0.2352,0.4107,0.5252,0.4668,0.4203,0.0000,0.2809,0.2900,0.5694,
0.6058,0.3757,0.5529,0.5692,0.5096,0.4484,0.2809,0.0000,0.2532,0.5482,
0.6360,0.4303,0.5535,0.5600,0.5910,0.5461,0.2900,0.2532,0.0000,0.5489,
0.5858,0.5784,0.5867,0.4082,0.5768,0.5649,0.5694,0.5482,0.5489,0.0000,}%
}
\newcommand{\hmLLMKThreeVideo}{%
  \dataheatmap{2}{0.0000,0.3778,0.3666,
                0.3778,0.0000,0.3170,
                0.3666,0.3170,0.0000,}%
}
\newcommand{\hmLLMKFiveVideo}{%
  \dataheatmap{4}{00.0000,0.0260,0.3778,0.4008,0.4333,
                  0.0260,0.0000,0.3770,0.4014,0.4327,
                  0.3778,0.3770,0.0000,0.1884,0.5388,
                  0.4008,0.4014,0.1884,0.0000,0.5661,
                  0.4333,0.4327,0.5388,0.5661,0.0000,}%
}
\newcommand{\hmLLMKTenVideo}{%
  \dataheatmap{9}{0.0000,0.1085,0.3377,0.2206,0.1188,0.2258,0.2418,0.2293,0.2170,0.2281,
0.1085,0.0000,0.3635,0.1734,0.1772,0.2478,0.1933,0.1751,0.1657,0.2497,
0.3377,0.3635,0.0000,0.3978,0.3281,0.3625,0.3781,0.3891,0.3854,0.3628,
0.2206,0.1734,0.3978,0.0000,0.2456,0.3330,0.2253,0.1962,0.1974,0.3344,
0.1188,0.1772,0.3281,0.2456,0.0000,0.2707,0.2865,0.2768,0.2701,0.2727,
0.2258,0.2478,0.3625,0.3330,0.2707,0.0000,0.3309,0.3175,0.3158,0.0405,
0.2418,0.1933,0.3781,0.2253,0.2865,0.3309,0.0000,0.0760,0.0726,0.3307,
0.2293,0.1751,0.3891,0.1962,0.2768,0.3175,0.0760,0.0000,0.0607,0.3164,
0.2170,0.1657,0.3854,0.1974,0.2701,0.3158,0.0726,0.0607,0.0000,0.3143,
0.2281,0.2497,0.3628,0.3344,0.2727,0.0405,0.3307,0.3164,0.3143,0.0000,}%
}
\newcommand{\hmLLMKThreeHouse}{%
  \dataheatmap{2}{0.0000,0.0575,0.3716,
                0.0575,0.0000,0.3737,
                0.3716,0.3737,0.0000,}%
}
\newcommand{\hmLLMKFiveHouse}{%
  \dataheatmap{4}{0.0000,0.1313,0.3597,0.3053,0.3176,
0.1313,0.0000,0.3381,0.3237,0.3107,
0.3597,0.3381,0.0000,0.2225,0.1284,
0.3053,0.3237,0.2225,0.0000,0.1628,
0.3176,0.3107,0.1284,0.1628,0.0000,}%
}
\newcommand{\hmLLMKTenHouse}{%
  \dataheatmap{9}{0.0000,0.0741,0.3053,0.4356,0.1913,0.1628,0.3237,0.3275,0.1942,0.2225,
                  0.0741,0.0000,0.2981,0.4268,0.2151,0.1898,0.3334,0.3300,0.2181,0.2474,
                  0.3053,0.2981,0.0000,0.3890,0.3339,0.3176,0.1313,0.1439,0.3347,0.3597,
                  0.4356,0.4268,0.3890,0.0000,0.4092,0.4175,0.4087,0.4136,0.4091,0.4184,
                  0.1913,0.2151,0.3339,0.4092,0.0000,0.0934,0.3232,0.3490,0.0183,0.0765,
                  0.1628,0.1898,0.3176,0.4175,0.0934,0.0000,0.3107,0.3373,0.0929,0.1284,
                  0.3237,0.3334,0.1313,0.4087,0.3232,0.3107,0.0000,0.1584,0.3229,0.3381,
                  0.3275,0.3300,0.1439,0.4136,0.3490,0.3373,0.1584,0.0000,0.3494,0.3691,
                  0.1942,0.2181,0.3347,0.4091,0.0183,0.0929,0.3229,0.3494,0.0000,0.0748,
                  0.2225,0.2474,0.3597,0.4184,0.0765,0.1284,0.3381,0.3691,0.0748,0.0000,}%
}

\newcommand{\hmPowrBIKThreeMarketing}{%
  \dataheatmap{2}{0.0000,0.5671,0.4051,
0.5671,0.0000,0.6242,
0.4051,0.6242,0.0000,}%
}
\newcommand{\hmPowerBIKFiveMarketing}{%
  \dataheatmap{4}{0.0000,0.5671,0.4051,0.3871,0.5781,
0.5671,0.0000,0.6242,0.5043,0.2527,
0.4051,0.6242,0.0000,0.4871,0.6166,
0.3871,0.5043,0.4871,0.0000,0.5365,
0.5781,0.2527,0.6166,0.5365,0.0000,}%
}
\newcommand{\hmPowerBIKTenMarketing}{%
  \dataheatmap{8}{0.0000,0.5671,0.4051,0.3871,0.5781,0.5016,0.4431,0.6022,0.4914,
0.5671,0.0000,0.6242,0.5043,0.2527,0.5527,0.4956,0.1935,0.5578,
0.4051,0.6242,0.0000,0.4871,0.6166,0.4806,0.5277,0.6387,0.4738,
0.3871,0.5043,0.4871,0.0000,0.5365,0.5826,0.4244,0.5507,0.5769,
0.5781,0.2527,0.6166,0.5365,0.0000,0.5382,0.5792,0.1808,0.5466,
0.5016,0.5527,0.4806,0.5826,0.5382,0.0000,0.5635,0.5460,0.0894,
0.4431,0.4956,0.5277,0.4244,0.5792,0.5635,0.0000,0.5891,0.5546,
0.6022,0.1935,0.6387,0.5507,0.1808,0.5460,0.5891,0.0000,0.5559,
0.4914,0.5578,0.4738,0.5769,0.5466,0.0894,0.5546,0.5559,0.0000,}%
}
\newcommand{\hmPowerBIKThreeVideo}{%
  \dataheatmap{2}{0.0000,0.3618,0.3938,
                  0.3618,0.0000,0.4779,
                  0.3938,0.4779,0.0000,}%
}
\newcommand{\hmPowerBIKFiveVideo}{%
  \dataheatmap{2}{0.0000,0.3618,0.3938,
                  0.3618,0.0000,0.4779,
                  0.3938,0.4779,0.0000,}%
}
\newcommand{\hmPowerBIKTenVideo}{%
  \dataheatmap{2}{0.0000,0.3618,0.3938,
                  0.3618,0.0000,0.4779,
                  0.3938,0.4779,0.0000,}%
}
\newcommand{\hmPowrBIKThreePitches}{%
  \dataheatmap{2}{0.0000,0.3582,0.4946,
0.3582,0.0000,0.3803,
0.4946,0.3803,0.0000,}%
}
\newcommand{\hmPowerBIKFivePitches}{%
  \dataheatmap{4}{0.0000,0.3582,0.4946,0.5810,0.6152,
0.3582,0.0000,0.3803,0.5321,0.5207,
0.4946,0.3803,0.0000,0.4808,0.4228,
0.5810,0.5321,0.4808,0.0000,0.4407,
0.6152,0.5207,0.4228,0.4407,0.0000,}%
}
\newcommand{\hmPowerBIKTenPitches}{%
  \dataheatmap{9}{0.0000,0.3582,0.4946,0.5810,0.6152,0.6335,0.2952,0.3314,0.6122,0.5738,
0.3582,0.0000,0.3803,0.5321,0.5207,0.5841,0.1864,0.1979,0.5166,0.5452,
0.4946,0.3803,0.0000,0.4808,0.4228,0.4718,0.4317,0.4009,0.4314,0.4772,
0.5810,0.5321,0.4808,0.0000,0.4407,0.3516,0.5389,0.5268,0.4401,0.1815,
0.6152,0.5207,0.4228,0.4407,0.0000,0.4038,0.5642,0.5439,0.1358,0.4439,
0.6335,0.5841,0.4718,0.3516,0.4038,0.0000,0.6006,0.5918,0.3918,0.3132,
0.2952,0.1864,0.4317,0.5389,0.5642,0.6006,0.0000,0.2158,0.5508,0.5450,
0.3314,0.1979,0.4009,0.5268,0.5439,0.5918,0.2158,0.0000,0.5496,0.5352,
0.6122,0.5166,0.4314,0.4401,0.1358,0.3918,0.5508,0.5496,0.0000,0.4374,
0.5738,0.5452,0.4772,0.1815,0.4439,0.3132,0.5450,0.5352,0.4374,0.0000,}%
}
\newcommand{\hmPowrBIKThreeHouse}{%
  \dataheatmap{2}{0.0000,0.5661,0.5681,
                  0.5661,0.0000,0.1381,
                  0.5681,0.1381,0.0000,}%
}
\newcommand{\hmPowerBIKFiveHouse}{%
  \dataheatmap{4}{0.0000,0.5661,0.5681,0.5840,0.5818,
                    0.5661,0.0000,0.1381,0.2691,0.2952,
                    0.5681,0.1381,0.0000,0.2535,0.3112,
                    0.5840,0.2691,0.2535,0.0000,0.1855,
                    0.5818,0.2952,0.3112,0.1855,0.0000,}%
}
\newcommand{\hmPowerBIKTenHouse}{%
  \dataheatmap{9}{0.0000,0.5661,0.5681,0.5840,0.5818,0.5884,0.2111,0.5910,0.5815,0.5891,
                  0.5661,0.0000,0.1381,0.2691,0.2952,0.3314,0.5238,0.3274,0.3373,0.3283,
                  0.5681,0.1381,0.0000,0.2535,0.3112,0.3254,0.5234,0.3304,0.3565,0.3303,
                  0.5840,0.2691,0.2535,0.0000,0.1855,0.1469,0.5420,0.1651,0.2253,0.1649,
                  0.5818,0.2952,0.3112,0.1855,0.0000,0.2158,0.5465,0.2102,0.1354,0.2087,
                  0.5884,0.3314,0.3254,0.1469,0.2158,0.0000,0.5448,0.1101,0.2025,0.1152,
                  0.2111,0.5238,0.5234,0.5420,0.5465,0.5448,0.0000,0.5514,0.5496,0.5506,
                  0.5910,0.3274,0.3304,0.1651,0.2102,0.1101,0.5514,0.0000,0.1775,0.0992,
                  0.5815,0.3373,0.3565,0.2253,0.1354,0.2025,0.5496,0.1775,0.0000,0.1774,
                  0.5891,0.3283,0.3303,0.1649,0.2087,0.1152,0.5506,0.0992,0.1774,0.0000,}%
}

\newcommand{\hmDAISYThreeMarketing}{%
  \dataheatmap{2}{0.0000,0.1599,0.5472,
0.1599,0.0000,0.5547,
0.5472,0.5547,0.0000,}%
}
\newcommand{\hmDAISYFiveMarketing}{%
  \dataheatmap{4}{0.0000,0.1599,0.5472,0.5857,0.4965,
0.1599,0.0000,0.5547,0.5808,0.4893,
0.5472,0.5547,0.0000,0.5216,0.4584,
0.5857,0.5808,0.5216,0.0000,0.4747,
0.4965,0.4893,0.4584,0.4747,0.0000,}%
}
\newcommand{\hmDAISYTenMarketing}{%
  \dataheatmap{9}{0.0000,0.1599,0.5472,0.5857,0.4965,0.1948,0.6114,0.6028,0.1480,0.2111,
0.1599,0.0000,0.5547,0.5808,0.4893,0.1227,0.6063,0.5976,0.1942,0.1204,
0.5472,0.5547,0.0000,0.5216,0.4584,0.5459,0.5353,0.5445,0.5546,0.5573,
0.5857,0.5808,0.5216,0.0000,0.4747,0.5791,0.3758,0.3542,0.5888,0.5813,
0.4965,0.4893,0.4584,0.4747,0.0000,0.4901,0.5043,0.5065,0.5004,0.4852,
0.1948,0.1227,0.5459,0.5791,0.4901,0.0000,0.5999,0.5937,0.2540,0.1290,
0.6114,0.6063,0.5353,0.3758,0.5043,0.5999,0.0000,0.1466,0.6217,0.5995,
0.6028,0.5976,0.5445,0.3542,0.5065,0.5937,0.1466,0.0000,0.6081,0.5932,
0.1480,0.1942,0.5546,0.5888,0.5004,0.2540,0.6217,0.6081,0.0000,0.2660,
0.2111,0.1204,0.5573,0.5813,0.4852,0.1290,0.5995,0.5932,0.2660,0.0000,}%
}
\newcommand{\hmDAISYThreeVideo}{%
  \dataheatmap{2}{0.00000,0.04102,0.56439,
                  0.04102,0.00000,0.56357,
                  0.56439,0.56357,0.00000}%
}
\newcommand{\hmDAISYFiveVideo}{%
  \dataheatmap{4}{0.00000,0.04102,0.56439,0.56852,0.56793,
                  0.04102,0.00000,0.56357,0.56752,0.56747,
                  0.56439,0.56357,0.00000,0.08153,0.10031,
                  0.56852,0.56752,0.08153,0.00000,0.05886,
                  0.56793,0.56747,0.10031,0.05886,0.00000}%
}
\newcommand{\hmDAISYTenVideo}{%
  \dataheatmap{9}{0.00000,0.04102,0.56439,0.56852,0.56793,0.56616,0.04019,0.56364,0.56447,0.56309,
                  0.04102,0.00000,0.56357,0.56752,0.56747,0.56531,0.02499,0.56312,0.56382,0.56264,
                  0.56439,0.56357,0.00000,0.08153,0.10031,0.10125,0.56408,0.07111,0.10372,0.15432,
                  0.56852,0.56752,0.08153,0.00000,0.05886,0.04579,0.56803,0.10604,0.08448,0.13997,
                  0.56793,0.56747,0.10031,0.05886,0.00000,0.03957,0.56769,0.10657,0.07363,0.10744,
                  0.56616,0.56531,0.10125,0.04579,0.03957,0.00000,0.56570,0.11592,0.08379,0.12600,
                  0.04019,0.02499,0.56408,0.56803,0.56769,0.56570,0.00000,0.56287,0.56356,0.56216,
                  0.56364,0.56312,0.07111,0.10604,0.10657,0.11592,0.56287,0.00000,0.07077,0.11001,
                  0.56447,0.56382,0.10372,0.08448,0.07363,0.08379,0.56356,0.07077,0.00000,0.07588,
                  0.56309,0.56264,0.15432,0.13997,0.10744,0.12600,0.56216,0.11001,0.07588,0.00000}%
}
\newcommand{\hmDAISYThreeHouse}{%
  \dataheatmap{2}{0.0000,0.4894,0.4368,
0.4894,0.0000,0.5904,
0.4368,0.5904,0.0000,}%
}
\newcommand{\hmDAISYFiveHouse}{%
  \dataheatmap{4}{0.0000,0.4894,0.4368,0.3824,0.5166,
0.4894,0.0000,0.5904,0.5409,0.6047,
0.4368,0.5904,0.0000,0.3479,0.2932,
0.3824,0.5409,0.3479,0.0000,0.4214,
0.5166,0.6047,0.2932,0.4214,0.0000,}%
}
\newcommand{\hmDAISYTenHouse}{%
  \dataheatmap{9}{0.0000,0.4894,0.4368,0.3824,0.5166,0.5038,0.5089,0.5059,0.1127,0.4406,
0.4894,0.0000,0.5904,0.5409,0.6047,0.3427,0.3313,0.3614,0.4941,0.2234,
0.4368,0.5904,0.0000,0.3479,0.2932,0.5866,0.5922,0.5979,0.4390,0.5592,
0.3824,0.5409,0.3479,0.0000,0.4214,0.5578,0.5631,0.5628,0.3820,0.5143,
0.5166,0.6047,0.2932,0.4214,0.0000,0.6081,0.6068,0.6151,0.5215,0.5850,
0.5038,0.3427,0.5866,0.5578,0.6081,0.0000,0.1508,0.2291,0.5092,0.3496,
0.5089,0.3313,0.5922,0.5631,0.6068,0.1508,0.0000,0.1535,0.5163,0.3491,
0.5059,0.3614,0.5979,0.5628,0.6151,0.2291,0.1535,0.0000,0.5129,0.3617,
0.1127,0.4941,0.4390,0.3820,0.5215,0.5092,0.5163,0.5129,0.0000,0.4467,
0.4406,0.2234,0.5592,0.5143,0.5850,0.3496,0.3491,0.3617,0.4467,0.0000,}%
}
\newcommand{\hmExcelKThreeMarketing}{%
  \dataheatmap{2}{0.0000,0.1357,0.2557,
0.1357,0.0000,0.2562,
0.2557,0.2562,0.0000,}%
}
\newcommand{\hmExcelFiveMarketing}{%
  \dataheatmap{4}{0.0000,0.1357,0.2557,0.2750,0.2646,
0.1357,0.0000,0.2562,0.2738,0.2650,
0.2557,0.2562,0.0000,0.1282,0.0760,
0.2750,0.2738,0.1282,0.0000,0.1285,
0.2646,0.2650,0.0760,0.1285,0.0000,}%
}
\newcommand{\hmExcelKTenMarketing}{%
  \dataheatmap{6}{0.0000,0.1357,0.2557,0.2750,0.2646,0.2604,0.2537,
0.1357,0.0000,0.2562,0.2738,0.2650,0.2875,0.2641,
0.2557,0.2562,0.0000,0.1282,0.0760,0.1358,0.2951,
0.2750,0.2738,0.1282,0.0000,0.1285,0.1749,0.3308,
0.2646,0.2650,0.0760,0.1285,0.0000,0.1322,0.2957,
0.2604,0.2875,0.1358,0.1749,0.1322,0.0000,0.3222,
0.2537,0.2641,0.2951,0.3308,0.2957,0.3222,0.0000,}%
}
\newcommand{\hmExcelKThreeVideo}{%
  \dataheatmap{2}{0.0000,0.2125,0.2721,
0.2125,0.0000,0.2154,
0.2721,0.2154,0.0000,}%
}
\newcommand{\hmExcelKFiveVideo}{%
  \dataheatmap{4}{0.0000,0.2125,0.2721,0.2486,0.2474,
0.2125,0.0000,0.2154,0.2836,0.2795,
0.2721,0.2154,0.0000,0.1544,0.1534,
0.2486,0.2836,0.1544,0.0000,0.0359,
0.2474,0.2795,0.1534,0.0359,0.0000,}%
}
\newcommand{\hmExcelKTenVideo}{%
  \dataheatmap{8}{0.0000,0.2125,0.2721,0.2486,0.2474,0.1909,0.2719,0.2555,0.0871,
0.2125,0.0000,0.2154,0.2836,0.2795,0.0839,0.2152,0.3132,0.2518,
0.2721,0.2154,0.0000,0.1544,0.1534,0.2455,0.0117,0.2101,0.2899,
0.2486,0.2836,0.1544,0.0000,0.0359,0.3059,0.1544,0.0944,0.2467,
0.2474,0.2795,0.1534,0.0359,0.0000,0.3023,0.1534,0.0898,0.2458,
0.1909,0.0839,0.2455,0.3059,0.3023,0.0000,0.2452,0.3321,0.2363,
0.2719,0.2152,0.0117,0.1544,0.1534,0.2452,0.0000,0.2101,0.2898,
0.2555,0.3132,0.2101,0.0944,0.0898,0.3321,0.2101,0.0000,0.2396,
0.0871,0.2518,0.2899,0.2467,0.2458,0.2363,0.2898,0.2396,0.0000,}%
}
\newcommand{\hmExcelKThreeHouse}{%
  \dataheatmap{2}{0.0000,0.4371,0.3122,
0.4371,0.0000,0.4312,
0.3122,0.4312,0.0000,
}%
}
\newcommand{\hmExcelKFiveHouse}{%
  \dataheatmap{4}{0.0000,0.4371,0.3122,0.3342,0.3170,
0.4371,0.0000,0.4312,0.4248,0.4409,
0.3122,0.4312,0.0000,0.0863,0.0967,
0.3342,0.4248,0.0863,0.0000,0.0810,
0.3170,0.4409,0.0967,0.0810,0.0000,}%
}
\newcommand{\hmExcelKTenHouse}{%
  \dataheatmap{6}{0.0000,0.4371,0.3122,0.3342,0.3170,0.3402,0.4700,
                  0.4371,0.0000,0.4312,0.4248,0.4409,0.3845,0.5696,
                  0.3122,0.4312,0.0000,0.0863,0.0967,0.1307,0.3463,
                  0.3342,0.4248,0.0863,0.0000,0.0810,0.1112,0.3307,
                  0.3170,0.4409,0.0967,0.0810,0.0000,0.1346,0.3049,
                  0.3402,0.3845,0.1307,0.1112,0.1346,0.0000,0.3708,
                  0.4700,0.5696,0.3463,0.3307,0.3049,0.3708,0.0000,}%
}

\newcommand{\hmPGreedyKThree}{%
  \dataheatmap{2}{0.0,0.321754,0.504118,
                  0.321754,0.0,0.413599,
                  0.504118,0.413599,0.0}%
}
\newcommand{\hmPGreedyKFive}{%
  \dataheatmap{4}{0.0,0.303352,0.444028,0.311969,0.545652,
                  0.303352,0.0,0.504118,0.418379,0.538923,
                  0.444028,0.504118,0.0,0.332776,0.553978,
                  0.311969,0.418379,0.332776,0.0,0.545385,
                  0.545652,0.538923,0.553978,0.545385,0.0}%
}
\newcommand{\hmPGreedyKTen}{%
  \dataheatmap{9}{0.0,0.155859,0.321349,0.321303,0.359048,0.324873,0.35217,0.336703,0.380359,0.354507,
                  0.155859,0.0,0.303352,0.300781,0.357832,0.32286,0.322213,0.315582,0.385316,0.358744,
                  0.321349,0.303352,0.0,0.101513,0.259718,0.172394,0.156075,0.165792,0.294947,0.237376,
                  0.321303,0.300781,0.101513,0.0,0.236053,0.151645,0.154205,0.155457,0.278078,0.222145,
                  0.359048,0.357832,0.259718,0.236053,0.0,0.170859,0.245487,0.1952,0.139405,0.201744,
                  0.324873,0.32286,0.172394,0.151645,0.170859,0.0,0.171312,0.127208,0.215973,0.147005,
                  0.35217,0.322213,0.156075,0.154205,0.245487,0.171312,0.0,0.100044,0.271451,0.220921,
                  0.336703,0.315582,0.165792,0.155457,0.1952,0.127208,0.100044,0.0,0.227806,0.184928,
                  0.380359,0.385316,0.294947,0.278078,0.139405,0.215973,0.271451,0.227806,0.0,0.165048,
                  0.354507,0.358744,0.237376,0.222145,0.201744,0.147005,0.220921,0.184928,0.165048,0.0}%
}

\newcommand{\hmPGreedyKThreeVideo}{%
  \dataheatmap{2}{0.0, 0.584408, 0.329837,
                  0.584408, 0.0, 0.563182,
                  0.329837, 0.563182, 0.0}%
                  }
\newcommand{\hmPGreedyKFiveVideo}{%
  \dataheatmap{4}{0.0, 0.584408, 0.329837, 0.552376, 0.225247,
                  0.584408, 0.0, 0.563182, 0.370227, 0.5553,
                  0.329837, 0.563182, 0.0, 0.537874, 0.266369,
                  0.552376, 0.370227, 0.537874, 0.0, 0.524765,
                  0.225247, 0.5553, 0.266369, 0.524765, 0.0}%
                  }
\newcommand{\hmPGreedyKTenVideo}{%
    \dataheatmap{9}{0.0, 0.584408, 0.329837, 0.559422, 0.552376, 0.563149, 0.126216, 0.312028, 0.318685, 0.122653,
                    0.584408, 0.0, 0.563182, 0.167363, 0.370227, 0.349088, 0.587512, 0.567019, 0.579948, 0.583287,
                    0.329837, 0.563182, 0.0, 0.529911, 0.537874, 0.553591, 0.30151, 0.127241, 0.103325, 0.296269,
                    0.559422, 0.167363, 0.529911, 0.0, 0.331881, 0.331118, 0.561109, 0.53628, 0.548363, 0.554919,
                    0.552376, 0.370227, 0.537874, 0.331881, 0.0, 0.11159, 0.552713, 0.545097, 0.557645, 0.550464,
                    0.563149, 0.349088, 0.553591, 0.331118, 0.11159, 0.0, 0.56747, 0.558471, 0.573233, 0.563912,
                    0.126216, 0.587512, 0.30151, 0.561109, 0.552713, 0.56747, 0.0, 0.316453, 0.305172, 0.102393,
                    0.312028, 0.567019, 0.127241, 0.53628, 0.545097, 0.558471, 0.316453, 0.0, 0.108686, 0.313472,
                    0.318685, 0.579948, 0.103325, 0.548363, 0.557645, 0.573233, 0.305172, 0.108686, 0.0, 0.293685,
                    0.122653, 0.583287, 0.296269, 0.554919, 0.550464, 0.563912, 0.102393, 0.313472, 0.293685, 0.0}%
}

\newcommand{\hmPGreedyKThreeHouse}{%
  \dataheatmap{2}{0.0,0.272588,0.219838,
                  0.272588,0.0,0.276159,
                  0.219838,0.276159,0.0}%
}
\newcommand{\hmPGreedyKFiveHouse}{%
  \dataheatmap{4}{0.0,0.272588,0.219838,0.220992,0.354671,
                  0.272588,0.0,0.276159,0.314636,0.418519,
                  0.219838,0.276159,0.0,0.234891,0.40403,
                  0.220992,0.314636,0.234891,0.0,0.299488,
                  0.354671,0.418519,0.40403,0.299488,0.0}%
}
\newcommand{\hmPGreedyKTenHouse}{%
  \dataheatmap{9}{0.0,0.272588,0.183231,0.315076,0.219838,0.199891,0.300239,0.220992,0.149618,0.291861,
                0.272588,0.0,0.330898,0.196904,0.276159,0.289848,0.173295,0.314636,0.30859,0.158063,
                0.183231,0.330898,0.0,0.315519,0.213185,0.19099,0.296032,0.173317,0.134286,0.31262,
                0.315076,0.196904,0.315519,0.0,0.253065,0.276489,0.172935,0.323828,0.29134,0.198988,
                0.219838,0.276159,0.213185,0.253065,0.0,0.140802,0.222026,0.234891,0.18306,0.233,
                0.199891,0.289848,0.19099,0.276489,0.140802,0.0,0.247762,0.206797,0.186854,0.255696,
                0.300239,0.173295,0.296032,0.172935,0.222026,0.247762,0.0,0.269468,0.278996,0.10397,
                0.220992,0.314636,0.173317,0.323828,0.234891,0.206797,0.269468,0.0,0.200685,0.285691,
                0.149618,0.30859,0.134286,0.29134,0.18306,0.186854,0.278996,0.200685,0.0,0.291404,
                0.291861,0.158063,0.31262,0.198988,0.233,0.255696,0.10397,0.285691,0.291404,0.0}%
}

\newcommand{\hmGreedyKThreeCover}{%
  \dataheatmap{2}{0.0, 0.3027, 0.347285,
0.3027, 0.0, 0.316906,
0.347285, 0.316906, 0.0,}%
}
\newcommand{\hmGreedyKFiveCover}{%
  \dataheatmap{4}{0.0, 0.3027, 0.347285, 0.201916, 0.209436,
                  0.3027, 0.0, 0.316906, 0.227074, 0.245471 ,
                  0.347285, 0.316906, 0.0, 0.297212, 0.371538,
                  0.201916, 0.227074, 0.297212, 0.0, 0.206508,
                  0.209436, 0.245471, 0.371538, 0.206508, 0.0}%
}
\newcommand{\hmGreedyKTenCover}{%
  \dataheatmap{9}{0.0, 0.3027, 0.347285, 0.201916, 0.209436, 0.384657, 0.220793, 0.301536, 0.224474, 0.329466,
0.3027, 0.0, 0.316906, 0.227074, 0.245471, 0.252395, 0.317331, 0.340484, 0.290927, 0.279363,
0.347285, 0.316906, 0.0, 0.297212, 0.371538, 0.219813, 0.309125, 0.220863, 0.356001, 0.297187,
0.201916, 0.227074, 0.297212, 0.0, 0.206508, 0.321732, 0.243495, 0.26775, 0.215828, 0.267588,
0.209436, 0.245471, 0.371538, 0.206508, 0.0, 0.338007, 0.286724, 0.348936, 0.242111, 0.361243,
0.384657, 0.252395, 0.219813, 0.321732, 0.338007, 0.0, 0.356652, 0.319266, 0.367296, 0.332635,
0.220793, 0.317331, 0.309125, 0.243495, 0.286724, 0.356652, 0.0, 0.24522, 0.256121, 0.321101,
0.301536, 0.340484, 0.220863, 0.26775, 0.348936, 0.319266, 0.24522, 0.0, 0.345876, 0.305579,
0.224474, 0.290927, 0.356001, 0.215828, 0.242111, 0.367296, 0.256121, 0.345876, 0.0, 0.321819,
0.329466, 0.279363, 0.297187, 0.267588, 0.361243, 0.332635, 0.321101, 0.305579, 0.321819, 0.0
}%
}

\newcommand{\hmPGreedyKThreeCover}{%
  \dataheatmap{2}{0.0, 0.328329, 0.341093,
0.328329, 0.0, 0.316764,
0.341093, 0.316764, 0.0}%
}
\newcommand{\hmPGreedyKFiveCover}{%
  \dataheatmap{4}{0.0, 0.211607, 0.328329, 0.216648, 0.40785,
                  0.211607, 0.0, 0.279854, 0.270717, 0.352415,
                  0.328329, 0.279854, 0.0, 0.246833, 0.278879,
                  0.216648, 0.270717, 0.246833, 0.0, 0.341656,
                  0.40785, 0.352415, 0.278879, 0.341656, 0.0}%
}
\newcommand{\hmPGreedyKTenCover}{%
  \dataheatmap{9}{0.0, 0.211607, 0.328329, 0.216648, 0.40785, 0.228921, 0.341093, 0.316539, 0.355852, 0.288988,
0.211607, 0.0, 0.279854, 0.270717, 0.352415, 0.304136, 0.353535, 0.23112, 0.295758, 0.330845,
0.328329, 0.279854, 0.0, 0.246833, 0.278879, 0.31578, 0.316764, 0.267156, 0.265442, 0.303305,
0.216648, 0.270717, 0.246833, 0.0, 0.341656, 0.213411, 0.273368, 0.326914, 0.287128, 0.202459,
0.40785, 0.352415, 0.278879, 0.341656, 0.0, 0.32611, 0.217487, 0.290593, 0.260904, 0.326644,
0.228921, 0.304136, 0.31578, 0.213411, 0.32611, 0.0, 0.236278, 0.353727, 0.30101, 0.220143,
0.341093, 0.353535, 0.316764, 0.273368, 0.217487, 0.236278, 0.0, 0.372195, 0.288218, 0.256461,
0.316539, 0.23112, 0.267156, 0.326914, 0.290593, 0.353727, 0.372195, 0.0, 0.296384, 0.362581,
0.355852, 0.295758, 0.265442, 0.287128, 0.260904, 0.30101, 0.288218, 0.296384, 0.0, 0.203303,
0.288988, 0.330845, 0.303305, 0.202459, 0.326644, 0.220143, 0.256461, 0.362581, 0.203303, 0.0
}%
}

\section{Experimental Results}
\label{sec:experiments}
We now present experimental results to demonstrate the efficacy of \sysName in
practical settings to address the following questions:

\begin{itemize}[leftmargin=*]

     \item (Q1) How do \sysName's runtime and recommendation quality compare
     quantitatively with those of existing methods? (\S\ref{sec:quality})
	
     \item (Q2) What is the effect of the optimization techniques---pruning and
     LLM-proxy---on \sysName's runtime performance? (\S\ref{sec:opt})
	 
	 \item (Q3) How well does \sysName scale with data growth?
	 (\S\ref{sec:scalability})

 	 \item (Q4) How do key parameters (budget $k$ and diversity threshold
 	 $\theta$) influence the quality of \sysName recommendations?
 	 (\S\ref{sec:param})
	 
	 \item (Q5) How do \sysName recommendations qualitatively compare against
	 commercial software and LLMs over real datasets and \reviseone{how does
	 \sysName adapt based on user feedback?} (\S\ref{sec:case_study})
	
\end{itemize}

\subsection{Experimental Setup} \label{sec:exp_setup}

\subsubsection{Setup.} All experiments were run on machines with 256 GB RAM
running Ubuntu 22.04 LTS with CPU 12 cores and GPU NVIDIA H100 96GB with CUDA
12.8. We implemented our solutions (available publicly~\cite{sagerepo}) with
Python 3.10.3. For embeddings, we utilized TAPEX-large~\cite{tapex-large-model}
and T5 trained by Spider~\cite{t5-text2sql-spider-model}. We employed
Llama-3-7B~\cite{llama-3-1-8b-instruct-model} as the LLM for semantic
consultation.

\subsubsection{\reviseone{Offline Phase}} \label{offlinephase} \reviseone{As
discussed in \S\ref{sec:offlineopt}, \sysName relies on two offline
optimizations: pruning (\S\ref{sec5:prune}) and developing an LLM-Proxy model
(\S\ref{sec5:cache}). The LLM-Proxy model is composed of a decision tree whose
parameters are learned from a training dataset of prompts and their
corresponding answers retrieved from
Llama-3-7B~\cite{llama-3-1-8b-instruct-model}. We configured the decision tree
to have a maximum depth of $15$ and used the Gini function as the splitting
criterion. We generated $10{,}000$ prompts---for each of Correlation
(\S\ref{trendsec}), Ratio (\S\ref{trendsec}), and Surprise
(\S\ref{sursec})---splitting the resulting datasets 80/20 into training and test
sets. On average, across the 4 datasets, the LLM-Proxy model achieved an
accuracy of 90\%, 89\%, and 64\% for Correlation, Ratio, and Surprise,
respectively. On average, the offline phase took about 92 minutes per dataset,
which includes both pruning and LLM-Proxy training.}

\subsubsection{Datasets} 
\label{sec:datasets}
We used \reviseone{four} real-world datasets with varying domains and sizes to
assess \sysName's generalizability.

\smallskip
\noindent$-$ \textbf{Marketing}~\cite{marketing-campaigns-dataset}
contains 2{,}240 tuples and 28 attributes (19 categorical and 9 numerical),
capturing demographic and behavioral information about customers like marital
status \& purchase history.

\noindent$-$\textbf{Video}~\cite{videogamesales-dataset} comprises video
game sales from various countries, platforms, and release years, containing
16{,}600 tuples across 11 attributes (7 categorical and 4 numerical).


\begin{table*}[t]
\renewcommand{\arraystretch}{1.5}
\centering
\resizebox{\textwidth}{!}{
\begin{tabular}{|c|l|rrrrrrc|rrrrrrc|rrrrrrc|}
\hline
\multirow{3}{*}{} & \multirow{3}{*}{} 
& \multicolumn{7}{c|}{\textbf{Marketing}} 
& \multicolumn{7}{c|}{\textbf{Video}} 
& \multicolumn{7}{c|}{\textbf{House}} \\
\cline{3-23}
& & \#PT & T(s) & Ins & Int & Util & \revisetwo{$\underset{\text{m-dist}}{\text{Div}}$} & \revisetwo{$\underset{\text{heatmap}}{\text{Div}}$}
&   \#PT & T(s) & Ins & Int & Util & \revisetwo{$\underset{\text{m-dist}}{\text{Div}}$} & \revisetwo{$\underset{\text{heatmap}}{\text{Div}}$}
&   \#PT & T(s) & Ins & Int & Util & \revisetwo{$\underset{\text{m-dist}}{\text{Div}}$} & \revisetwo{$\underset{\text{heatmap}}{\text{Div}}$} \\
\hline
\multirow{8}{*}{{\rotatebox{90}{$k{=}3$, $\theta {=} 0.30$ or $0.20$}}}
& Top-k             
& 3 & 147  & 2.86 & 2.23 & \textbf{2.55}     & 0.39               & \hmTopKThree         
& 3 & 64   & 1.99 & 1.27 & \textbf{1.63}     & 0.06               & \hmTopKThreeVideo   
& 3 & 2    & 2.26 & 2.13 & \textbf{2.20}     & 0.09               & \hmTopKThreeHouse \\
& DAISY             
& 3 & 23   & 0.54 & 1.48 & 1.01              & 0.16               & \hmDAISYThreeMarketing 
& 3 & 21   & 1.63 & 1.00 & 1.32				 & 0.04               & \hmDAISYThreeVideo     
& 3 & 455  & 0.23 & 1.17 & 0.70              & \textbf{0.44}      & \hmDAISYThreeHouse \\
& LLMs              
& 3 & 15   & 0.16 & 1.75 & 0.96              & 0.16               & \hmLLMKThreeMarketing 
& 3 & 25   & 0.01 & 0.58 & 0.29              & \underline{0.32}   & \hmLLMKThreeVideo     
& 3 & 30   & 0.07 & 0.55 & 0.31              & 0.06               & \hmLLMKThreeHouse \\
& PowerBI           
& 3 & 9    & 0.91 & 0.94 & 0.93              & 0.41               & \hmPowrBIKThreeMarketing 
& 3 & 6    & 0.91 & 0.94 & 0.93              & \textbf{0.36}      & \hmPowerBIKThreeVideo   
& 3 & 15   & 0.50 & 1.76 & 1.13              & \underline{0.36}   & \hmPowrBIKThreeHouse \\
& GSheets           
& 2 & 1    & 0.00 & 1.67 & 0.83              & \textbf{0.56}      & \hmGsheetsMarketing      
& 3 & 1    & 0.04 & 0.59 & 0.31              & \underline{0.32}   & \hmGsheetsVideo          
&-- & --   & --   & --   &   --              & --                 & -- \\
& Excel             
& 3 & 1    & 0.29 & 2.42 & 1.35              & 0.14               & \hmExcelKThreeMarketing 
& 3 & 1    & 0.14 & 2.16 & 1.15              & 0.21               & \hmExcelKThreeVideo     
& 3 & 1    & 0.00 & 1.87 & 0.93              & 0.31               & \hmExcelKThreeHouse \\
& \textbf{\sysName} 
& 3 & 161  & 2.86 & 2.23 & \textbf{2.55}     &\underline{0.45}    & \hmGreedyKThree      
& 3 & 69   & 1.99 & 1.27 & \textbf{1.63}     & 0.30               & \hmGreedyKThreeVideo   
& 3 & 2    & 2.21 & 2.09 & \underline{2.15}  & 0.21               & \hmGreedyKThreeHouse \\
& \revisemeta{\textbf{\sysNameP}} 
& \revisemeta{3} & \revisemeta{23}   & \revisemeta{2.76} & \revisemeta{2.23} & \revisemeta{\underline{2.50}}  & \revisemeta{0.32}    & \hmPGreedyKThree      
& \revisemeta{3} & \revisemeta{17}   & \revisemeta{1.47} & \revisemeta{1.63} & \revisemeta{\underline{1.55}}  & \revisemeta{0.30}    & \hmPGreedyKThreeHouse   
& \revisemeta{3} & \revisemeta{2}    & \revisemeta{2.09} & \revisemeta{2.06} & \revisemeta{2.07}              & \revisemeta{0.22}    & \hmPGreedyKThreeHouse \\
\hline

\multirow{7}{*}{{\rotatebox{90}{$k{=}5$, $\theta {=} 0.30$ or $0.20$}}}
& Top-k            
& 5 & 148  & 4.76 & 3.72 & \textbf{4.24}     & 0.11               & \hmTopKFive          
& 5 & 64   & 3.31 & 2.03 & \textbf{2.67}     & 0.05 			  & \hmTopKFiveVideo     
& 5 & 2    & 3.77 & 3.56 & \textbf{3.66}     & 0.05               & \hmTopKFiveHouse \\
& DAISY             
& 5 & 23   & 0.54 & 2.51 & 1.52              & 0.16               & \hmDAISYFiveMarketing
& 5 & 21   & 2.57 & 1.67 & 2.21              & 0.04               & \hmDAISYFiveVideo     
& 5 & 455  & 0.23 & 1.84 & 1.04              & \underline{0.29}   & \hmDAISYFiveHouse \\
& LLMs              
& 5 & 22   & 0.93 & 2.94 & 1.94              & \underline{0.31}   & \hmLLMKFiveMarketing 
& 5 & 20   & 0.18 & 0.86 & 0.52              & 0.03               & \hmLLMKFiveVideo      
& 5 & 44   & 0.07 & 0.75 & 0.41              & 0.13               & \hmLLMKFiveHouse \\
& PowerBI           
& 5 & 9    & 0.20 & 4.23 & 2.21              & 0.25               & \hmPowerBIKFiveMarketing 
& - & -    & -    & -    & -                 & -                  & -      
& 5 & 15   & 0.50 & 2.99 & 1.74              & \textbf{0.36}      & \hmPowerBIKFiveHouse \\
& Excel               
& 5 & 1    & 0.29 & 3.98 & 2.13              & 0.08               & \hmExcelFiveMarketing 
& 5 & 1    & 0.31 & 3.28 & 1.80              & 0.04               & \hmExcelKFiveVideo    
& 5 & 1    & 0.00 & 3.05 & 1.52              & 0.08               & \hmExcelKFiveHouse \\
& \textbf{\sysName}  
& 5 & 161  & 4.76 & 3.70 & \underline{4.23}  & \textbf{0.33}      & \hmGreedyKFive
& 5 & 69   & 3.31 & 2.03 & \textbf{2.67}     & \textbf{0.30}      & \hmGreedyKFiveVideo  
& 5 & 2    & 3.42 & 3.52 & \underline{3.47}  & 0.20               & \hmGreedyKFiveHouse \\
& \revisemeta{\textbf{\sysNameP}}  
& \revisemeta{5} & \revisemeta{23}   & \revisemeta{4.67} & \revisemeta{3.67} & \revisemeta{4.17            }  & \revisemeta{0.30            }   & \hmPGreedyKFive
& \revisemeta{5} & \revisemeta{17}   & \revisemeta{1.80} & \revisemeta{2.92} & \revisemeta{\underline{2.36}}  & \revisemeta{\underline{0.23}}	& \hmPGreedyKFiveVideo
& \revisemeta{5} & \revisemeta{2 }   & \revisemeta{2.65} & \revisemeta{3.46} & \revisemeta{3.06            }  & \revisemeta{0.22            }   & \hmPGreedyKFiveHouse\\
\hline

\multirow{7}{*}{{\rotatebox{90}{$k{=}10$, $\theta{=}0.10$}}}
& Top-k             
& 10 & 147  & 9.53 & 7.44 & \textbf{8.49}     & 0.02               & \hmTopKTen          
& 10 & 64   & 4.06 & 5.50 & \textbf{4.78}     & 0.05			   & \hmTopKTenVideo     
& 10 & 2    & 7.57 & 7.02 & \textbf{7.30}     & 0.05               & \hmTopKTenHouse \\
& DAISY             
& 10 & 23   & 0.54 & 4.22 & 2.38              & \underline{0.12}   & \hmDAISYTenMarketing
& 10 & 21   & 4.09 & 3.33 & \underline{3.71}  & 0.02               & \hmDAISYTenVideo    
& 10 & 455  & 0.23 & 3.28 & 1.75              & \underline{0.11}   & \hmDAISYTenHouse \\
& LLMs              
& 10 & 32   & 1.70 & 5.67 & 3.69              & \textbf{0.20}      & \hmLLMKTenMarketing 
& 10 & 28   & 0.74 & 2.24 & 1.49              & 0.04               & \hmLLMKTenVideo     
& 10 & 43   & 0.26 & 1.72 & 0.99              & 0.02               & \hmLLMKTenHouse \\
& PowerBI                
& 9  & 9    & 0.20 & 4.88 & 2.54              & 0.09               & \hmPowerBIKTenMarketing  
& -  & -    & -    & -    & -                 & -                  & -      
& 10 & 15   & 0.50 & 5.99 & 3.24              & \textbf{0.14}      & \hmPowerBIKTenHouse \\
& Excel             
& 7 & 1    & 0.29 & 5.30 & 2.79               & 0.08               & \hmExcelKTenMarketing
& 9 & 1    & 0.31 & 4.71 & 2.51               & 0.01               & \hmExcelKTenVideo    
& 7 & 1    & 0.00 & 3.97 & 1.98               & 0.08               & \hmExcelKTenHouse \\
& \textbf{\sysName} 
& 10 & 161 & 5.58 & 9.53 & 7.44               & 0.11 			   & \hmGreedyKTen  
& 10 & 69  & 4.06 & 5.50 & \textbf{4.78}      & \textbf{0.11}      & \hmGreedyKTenVideo  
& 10 & 2   & 7.55 & 6.91 & \underline{7.23}   & 0.10               & \hmGreedyKTenHouse \\
& \revisemeta{\textbf{\sysNameP}}
& \revisemeta{10} & \revisemeta{23}   & \revisemeta{9.49} & \revisemeta{7.44} & \revisemeta{\underline{8.47}}  & \revisemeta{0.10}               & \hmPGreedyKTen
& \revisemeta{10} & \revisemeta{17}   & \revisemeta{0.54} & \revisemeta{5.56} & \revisemeta{3.05   }           & \revisemeta{\underline{0.10}}   & \hmPGreedyKTenVideo  
& \revisemeta{10} & \revisemeta{2 }   & \revisemeta{6.13} & \revisemeta{6.92} & \revisemeta{6.52   }           & \revisemeta{0.10}               & \hmPGreedyKTenHouse \\
\hline
\end{tabular}
}

\caption{Comparison with baselines across three datasets and three values of
$k$. For $k=3$ and $k=5$, we used $\theta=0.30$ on Marketing and Video, and
$\theta=0.20$ on House. For $k=10$, $\theta=0.10$ was used across all datasets.
Columns show the number of pivot tables recommended (\#PT), elapsed time (T, in
seconds), \emph{Insightfulness} (Ins), \emph{Interpretability} (Int),
\emph{Utility} (Util), and \revisetwo{\emph{Diversity} (Div) in terms of
minimum pairwise distance (m-dist) and a distance matrix visualization
(heatmap)}. In the distance matrix, the diagonal is black, denoting 0 self
distance from a pivot table to itself. Lighter colors denote less similar
pairs, offering diversity. For Excel, Power BI, and Google Sheets, $k$ cannot
not be controlled. Therefore, we consider the first-$k$ items when more than
$k$ are returned. The best values in Util and m-dist are marked as bold and the
second best underlined.}
\vspace{-7mm} 
\label{tab:sage-comparison} 
\end{table*}

\noindent$-$\textbf{House}~\cite{house-sale-dataset} contains
information about property sales, comprising 1{,}460 tuples and 81
attributes (58 categorical and 23 numerical).

\noindent$-$ \revisemeta{\textbf{Cover Type}~\cite{covertype-dataset} is a
large dataset with $581{,}000$ tuples and $110$ attributes, containing tree
observations across a National Forest.}

\subsubsection{Baselines}
\label{sec:exp_baselines} 
Below, we list the baselines. \reviseone{To the best of our knowledge, no
existing open-source tool or academic work directly addresses our problem
setting of recommending a set of diversified pivot tables. Therefore, we
include commercially available software as baselines, despite their internal
logic not being open-sourced.}:

\smallskip \noindent{\textbf{Brute-Force}} considers all possible k-sized pivot table sets
and follows an exhaustive approach to solve
Problem~\ref{eq:objective_function}. In the absence of ground truth, its
results can be treated as the optimal solution.

\smallskip\noindent{\textbf{Top-k}} ranks candidate pivot tables in descending
order of their utility scores and selects the top-$k$, without accounting for
diversity. For fair comparison, we applied our optimizations here.

\smallskip\noindent{\textbf{LLM}} refers to Llama-3-8B-Instruct by
Meta~\cite{llama-3-1-8b-instruct-model}, a transformer-based large language
model. We prompted it with a data sample and asked for interesting and diverse
pivot tables in natural language.

\smallskip\noindent{\textbf{DAISY}}~\cite{DAISYVLDB24Junjie} is a query
recommendation system trained on crowdsourced ``interesting'' queries. Due to
the original model and data not being available, we did our own implementation.
We generated insightful pivot tables from
Auto-Suggest~\cite{AutoSuggestSIGMODE2020Cong}, created negative training data
by replacing attributes in the positive tables with random ones, and trained a
binary classifier to distinguish them.

\smallskip\noindent{\textbf{Microsoft Excel}}~\cite{microsoft_excel} is a
widely used commercial spreadsheet software that offers built-in pivot table
recommendations. We used the Windows version 2501.

\smallskip\noindent{\textbf{PowerBI}}~\cite{PowerBI,QuickInsightsDing19}, also
by Microsoft, is a business intelligence software that offers ``quick
insights'' in various forms. For comparison, we only considered ones that align
with our format of pivot table.

\smallskip\noindent{\textbf{Google Sheets}}~\cite{google_sheets} is an online
spreadsheet software known for easy collaboration, which recommends pivot
tables. We used the browser version during February, 2025.

\smallskip\noindent For Excel, PowerBI, and Google Sheets, $k$ cannot be
controlled. Thus, we consider the first-$k$ items. Google Sheets fails to
produce more than 3 recommendations, thus, we exclude it when $k > 3$.

\subsection{Contrasting against Baselines}
\label{sec:quality}

Table~\ref{tab:sage-comparison} contrasts \sysName against the baselines across
three datasets. Our primary metrics for comparison are \emph{Utility} (Util)
and \emph{Diversity} (m-dist). \revisetwo{However, since \emph{Insightfulness}
and \emph{Interpretability} constitute \emph{Utility} (\S\ref{sec:three}), we
report them for additional context}. The results show that \sysName effectively
balances utility and diversity, outperforming approaches like Top-k, which
inherently lack diversity. For Marketing, \sysName achieves a strong overall
performance, ranking either best or second-best in both metrics. Notably, when
\sysName marginally loses in terms of diversity, that is usually complemented
by significantly higher utility. For example, in Marketing, $k = 10$, LLM's
diversity (0.20) is more than \sysName's diversity (0.11). However, LLM's
utility (3.69) is about half of \sysName's utility (7.44). A similar situation
is seen for $k=3$ where GSheets yields only 30\% of \sysName's utility. Recall
that \sysName's goal is to just satisfy the diversity constraint, not maximize
it.

For Video, \sysName outperforms all baselines across all cases in utility and
two cases in diversity. \sysName and Top-k's similar performance can be
attributed to the dataset attributes, such as North-America Sales and
Europe-Sales, which possess similar value ranges, leading to comparable
utility. For House, Top-k marginally outperforms \sysName in utility, but at
the cost of very low diversity. \sysName diversity is relatively poor here,
because the pruning phase retained only 11 of 81 attributes, where others used
all attributes.

\looseness-1 While Top-k achieves high utility by design, it fails to
diversify. LLM performs well on Marketing but struggles on Video and House,
because Marketing has many categorical values, which LLMs are adept at
interpreting and utilizing for diverse recommendations. DAISY shows good
diversity because it predicts most insightful tables, benefitting diversity.
While PowerBI demonstrates good diversity across the board, it typically yields
poor utility. Google Sheets demonstrates good diversity on Marketing and Video,
however, its recommendations are limited to a small set of tables, leading to
low utility. It also fails for House due to high dimensionality (81
attributes), indicating its limitation in handling complex, high-dimensional
datasets. Excel consistently shows reasonable utility, but with low diversity,
due to allowing significant column overlaps.

\revisemeta{In most cases, the practical variant \sysNameP shows slight
reduction in the utility score. However, the runtime gain is significant
compared to \sysName, e.g., from 161s to 23s for Marketing. Since \sysNameP
uses sampling, its scores can be either an over- or under-approximation.
However, empirically we found its scores to be comparable to \sysName's scores
(typically within 10\%). \label{sagepluscomparison}}


\begin{table}[t]
\centering
\resizebox{\columnwidth}{!}{
\begin{tabular}{@{}cl@{}rrrrr@{}}
\toprule

  & \textbf{Approach}
  & \textbf{BruteForce} & \textbf{No PR/PX} & \textbf{No PR} & \textbf{No PX} & \textbf{\sysName} \\
\midrule
 & \#Pivot Tables (PTs)			     & 130   & 130    & 130    & 20     & 20   \\
\midrule
\multirow{5}{*}{\textbf{\rotatebox{90}{K = 2}}}
& \#PT combinations	    		       	& 8,385       & N/A    & N/A    & N/A    & N/A   \\
& Runtime (s)                       & 5,817       & 4,971  & 76     & 492    & 23   \\
& Utility (\%)                      & 100         & 100    & 97     & 100    & 97 \\
& -- Insightfulness (\%)            & 100         & 100    & 94     & 100    & 94 \\
& -- Interpretability (\%)          & 100         & 100    & 100    & 100    & 100 \\
\midrule
\multirow{5}{*}{\textbf{\rotatebox{90}{K = 5}}}
& \#PT combinations				          & 286M		    & N/A    & N/A    & N/A    & N/A   \\
& Runtime (s)                       & 19,604      & 4,971  & 76     & 492    & 24   \\
& Utility (\%)                      & 100         & 100    & 97     & 100    & 97   \\
& -- Insightfulness (\%)            & 100         & 100    & 92     & 100    & 92   \\
& -- Interpretability (\%)          & 100         & 100    & 100    & 100    & 100   \\\\[-3pt]

\hline
\multicolumn{3}{|l}{\small{PR = Pruning optimization}} 	& \multicolumn{4}{l|}{\small{No PR/PX = Plain greedy, no optimization}}\\
\multicolumn{3}{|l}{\small{PX = LLM-Proxy optimization}} & \multicolumn{4}{l|}{\small{\sysName = Greedy + PR + PX}}\\
\hline
\end{tabular}} 
\vspace{1mm}
\caption{Comparison among variants for $\theta=0.1$ on the Marketing dataset
over 5 attributes. We report the runtimes for the online phase for \sysName and
other variants when optimizations are applied.}
\label{tab:optimization_results}
\vspace{-4mm}
\end{table}

\subsection{Effect of Optimizations}
\label{sec:opt}
Table~\ref{tab:optimization_results} demonstrates how the
optimizations---pruning (PR) and LLM-proxy (PX)---significantly improve
\sysName's runtime performance while costing minimal utility loss. For this
experiment, we used a vertical slice of the Marketing dataset over 5 attributes
to allow Brute Force to finish within a reasonable time. We performed an
ablation study where we turned off all the optimizations (No PR/PX = plain
greedy), no pruning (No PR), no LLM-proxy (No PX), and \sysName with both
optimizations. Notably, we find that greedy achieves exact results in par with
Brute Force, validating our choice.

\revisetwo{BruteForce shows an infeasible runtime of over $19K$ seconds (over
$5$ hours) even for a moderate $k=5$, as the number of candidate pivot table
combinations was $\binom{130}{5} \approx 286$M, for only $5$ data attributes.}
LLM-proxy causes slight decrease (3\%) in utility due to less accurate proxy
classifier, however, boosts performance significantly. Pruning does not hurt
utility, due to its conservative filtering of unpromising candidates.
Figure~\ref{fig:opt_param_plot} (left) shows runtime comparison over three
datasets (limited to 5 attributes), where both optimizations offer significant
performance boost across the board. Pruning is significantly beneficial for
high-dimensional datasets, due to the exponential growth of the candidate space
w.r.t attributes.


\begin{figure}
\hspace{-5mm}
\begin{subfigure}[t]{0.52\columnwidth}
\centering
\resizebox{\columnwidth}{!}{

\begin{tikzpicture}
\begin{axis}[
    xlabel style={font=\huge}, 
    ylabel style={font=\huge},
    ybar,
    axis on top,
    bar width=0.2cm,
    width=7cm,
    height=5.5cm,
    ymin=0,
    ymax=100000,
    ymode=log,
    ytick={10,100,1000,10000},
    tick style={major tick length=2pt},
    minor tick style={draw=none},    
    enlarge x limits=false, 
    xmin=0.5,               
    xmax=3.5,               
    xtick={1,2,3},
    xtick style={draw=none},
    xticklabels={Marketing, Video, House},
    x tick label style={rotate=0, anchor=east, yshift=-10pt, xshift=8mm, font=\large}, 
    y tick label style={font=\large},
	ylabel={Runtime (s)},
    ylabel style={font=\large, yshift=-3mm},
    legend style={
        at={(-0.06,1.02)},       
        anchor=south west,  
        legend columns=3,
        draw=none,
        /tikz/every even column/.append style={column sep=0.3cm},
        font=\large
    },
	legend image code/.code={
	  	\draw[draw=black,fill opacity=1] (0cm,-0.08cm) rectangle (0.3cm,0.1cm);
	},
	every node near coord/.append style={font=\large, black, rotate=90, anchor=west},
]

\addplot+[
    ybar,
    fill=black!0,
    draw=black, 
    point meta=explicit,
    nodes near coords*={
	\pgfmathprintnumber[
	    use comma=false,
	    1000 sep={}
	]{\pgfplotspointmeta}	
	},
] coordinates {
    (1, 8446) [8446]
    (2, 493) [493]
    (3, 2183) [2183]
};
\addlegendentry{BF}

\addplot+[
    ybar,
    fill=black!50,
    draw=black, 
    point meta=explicit,
    nodes near coords*={
	\pgfmathprintnumber[
	    use comma=false,
	    1000 sep={}
	]{\pgfplotspointmeta}	
	},
] coordinates {
(1, 4971) [4971]
(2, 106) [106]
(3, 983) [983]
};
\addlegendentry{-PR\&PX}
\addplot+[
    ybar,
	pattern=north east lines,
    pattern color=black,
    draw=black,    
    point meta=explicit,
    nodes near coords*={
	\pgfmathprintnumber[
	    use comma=false,
	    1000 sep={}
	]{\pgfplotspointmeta}	
	},
] coordinates {
(1, 492) [492]
(2, 12) [12]
(3, 7) [7]
};
\addlegendentry{-PX}
\addplot+[
    ybar,
    pattern=dots,   
    pattern color=black,
    draw=black,    
    point meta=explicit,
    nodes near coords*={
	\pgfmathprintnumber[
	    use comma=false,
	    1000 sep={}
	]{\pgfplotspointmeta}	
	},
] coordinates
{
(1, 76) [76]
(2, 36) [36]
(3, 530) [530]
};
\addlegendentry{-PR}
\addplot+[
    ybar,
    fill=black!100,
    draw=black,    
    point meta=explicit,
    nodes near coords*={
	\pgfmathprintnumber[
	    use comma=false,
	    1000 sep={}
	]{\pgfplotspointmeta}	
	},
] coordinates {
(1, 24) [24]
(2, 1) [1]
(3, 1) [1]
};
\addlegendentry{\sysName}
\end{axis}
\end{tikzpicture}
}
\end{subfigure}%
\hspace{2mm}
\begin{subfigure}[t]{0.4\columnwidth}
\centering
\resizebox{\columnwidth}{!}{%
\begin{tikzpicture}
\begin{axis}[
  width=7cm,
  height=5.5cm,
  xlabel style={font=\huge}, 
  ylabel style={font=\huge, yshift=-5mm},
  xlabel={$\theta$}, ylabel={Utility},
  xtick={0.0,  0.2, 0.4, 0.6, 0.8, 1.0},
  xticklabels={0.0,  0.2, 0.4, 0.6, 0.8, 1.0},
  ytick={0,2,...,10},
  x tick label style={/pgf/number format/fixed, font=\huge}, 
  y tick label style={font=\huge},
  scaled ticks=false,
  grid=major,
  legend style={at={(-0.06,1.02)},       
        anchor=south west,  			
        legend columns=2, font=\huge,	draw=none, column sep=6pt},
]
\addplot+[blue, mark=*, mark size=4pt, opacity=0.6] coordinates { (0.0,3.340) (0.1,2.909) (0.2,2.716)
(0.3,2.629) (0.4,1.550) (0.5,1.076) (0.6,0.675) (0.7,0.675) (0.8,0.675)
(0.9,0.675) (1.0,0.675) };
\addlegendentry{$k{=}5$}
\addplot+[red, mark=square*, mark size=4pt, opacity=0.6] coordinates { (0.0,6.211) (0.1,4.906) (0.2,3.885)
(0.3,2.995) (0.4,1.550) (0.5,1.076) (0.6,0.675) (0.7,0.675) (0.8,0.675)
(0.9,0.675) (1.0,0.675) };
\addlegendentry{$k{=}10$}
\addplot+[green!60!black, mark=triangle*, opacity=0.6, mark size=5pt, mark options={fill=green!60!black, draw=green!60!black}] coordinates { (0.0,8.677) (0.1,6.019)
(0.2,3.885) (0.3,2.995) (0.4,1.550) (0.5,1.076) (0.6,0.675) (0.7,0.675)
(0.8,0.675) (0.9,0.675) (1.0,0.675) };
\addlegendentry{$k{=}15$}
\addplot+[orange, mark=diamond*, opacity=0.6, mark size=4pt] coordinates { (0.0,10.819) (0.1,6.019)
(0.2,3.885) (0.3,2.995) (0.4,1.550) (0.5,1.076) (0.6,0.675) (0.7,0.675)
(0.8,0.675) (0.9,0.675) (1.0,0.675) };
\addlegendentry{$k{=}20$}
\end{axis}
\end{tikzpicture}
}
\end{subfigure}%
\vspace{-4mm}

\caption{(Left) Effect of optimization techniques. We used $k{=}4$ and
$\theta=0.1$. (Right) Effect of $\theta$ on \emph{Utility} on the Video
dataset.} \label{fig:opt_param_plot} 

\vspace{-1mm} 
\end{figure}


\begin{figure}[t]
\centering

\begin{subfigure}[t]{0.48\columnwidth}
\centering
\resizebox{\textwidth}{!}{%
\begin{tikzpicture}
\begin{axis}[
  ybar,
  bar width=20pt,
  width=20cm,
  height=10cm,
  ymin=0,
  ymax=500,
  every axis/.append style={font=\fontsize{32}{36}\selectfont},
    xlabel style={font=\fontsize{32}{36}\selectfont, yshift=-20pt},
    ylabel style={font=\fontsize{32}{36}\selectfont, yshift=40pt},
    x tick label style={font=\fontsize{32}{36}\selectfont},
    y tick label style={font=\fontsize{32}{36}\selectfont},
  xlabel={Tuples (\%)}, ylabel={Runtime (s)},
  xtick={0.2,0.4,0.6,0.8,1.0},
  xticklabels={20,40,60,80,100},
  ytick={0,100,...,500},
  grid=major,
  enlarge x limits={abs=0.7},
  legend style={
      at={(-0.05,1.02)}, 
      anchor=south west,
      legend columns=4,
      font=\Huge,
      fill=none,
      draw=none
  },
  grid=major,
  enlarge x limits=0.1,
  xtick pos=left,
  ytick pos=left,
]

\addplot+[fill=green!60!black, draw=black] coordinates {
  (0.2,153.2561)
  (0.4,164.6201)
  (0.6,163.5402)
  (0.8,168.2446)
  (1.0,162.3688)
};
\addplot+[fill=green!20!white, draw=black, postaction={pattern=north east lines}] coordinates {
  (0.2,18.3847)
  (0.4,19.0369)
  (0.6,18.3386)
  (0.8,18.6775)
  (1.0,18.8035)
};
\addplot+[fill=orange!80!black, draw=black] coordinates {
  (0.2,113.4003)
  (0.4,204.8112)
  (0.6,307.0491)
  (0.8,395.9639)
  (1.0,491.3975)
};
\addplot+[fill=orange!30!white, draw=black, postaction={pattern=north east lines}] coordinates {
  (0.2,17.9051)
  (0.4,20.6787)
  (0.6,29.0860)
  (0.8,29.5663)
  (1.0,39.1373)
};

\end{axis}
\end{tikzpicture}
}
\end{subfigure}%
\hfill
\begin{subfigure}[t]{0.48\columnwidth}
\centering
\resizebox{\textwidth}{!}{%
\begin{tikzpicture}
\begin{axis}[
  ybar,
  bar width=20pt,
  width=20cm,
  height=10cm,
  ymin=0,
  ymax=500,
  every axis/.append style={font=\fontsize{32}{36}\selectfont},
    xlabel style={font=\fontsize{32}{36}\selectfont, yshift=-20pt},
    ylabel style={font=\fontsize{32}{36}\selectfont, yshift=40pt},
    x tick label style={font=\fontsize{32}{36}\selectfont},
    y tick label style={font=\fontsize{32}{36}\selectfont},
  xlabel={Attributes (\%)}, ylabel={Runtime (s)},
  xtick={0.2,0.4,0.6,0.8,1.0},
  xticklabels={20,40,60,80,100},
  ytick={0,100,...,500},
  grid=major,
  enlarge x limits={abs=0.7},
  legend style={
      at={(-0.05,1.02)}, 
      anchor=south west,
      legend columns=4,
      font=\Huge,
      fill=none,
      draw=none
  },
  grid=major,
  enlarge x limits=0.1,
  xtick pos=left,
  ytick pos=left,
]


\addplot+[fill=green!60!black, draw=black] coordinates {
  (0.2,8.2704)
  (0.4,8.1352)
  (0.6,14.6462)
  (0.8,35.4537)
  (1.0,161.0900)
};
\addplot+[fill=green!20!white, draw=black, postaction={pattern=north east lines}] coordinates {
  (0.2,12.3594)
  (0.4,12.5614)
  (0.6,12.3169)
  (0.8,13.3029)
  (1.0,18.7541)
};
\addplot+[fill=orange!80!black, draw=black] coordinates {
  (0.2,30.0249)
  (0.4,247.6962)
  (0.6,483.3563)
  (0.8,460.0636)
  (1.0,488.6474)
};
\addplot+[fill=orange!30!white, draw=black, postaction={pattern=north east lines}] coordinates {
  (0.2,12.1158)
  (0.4,23.5718)
  (0.6,35.0790)
  (0.8,38.7217)
  (1.0,40.0975)
};

\end{axis}
\end{tikzpicture}
}
\end{subfigure}%

\vspace{0mm}
\begin{tikzpicture}
\begin{axis}[
  hide axis,
  xmin=0, xmax=1,
  ymin=0, ymax=1,
  area legend,
  legend columns=4,
  legend style={
      /tikz/every even column/.append style={column sep=0.3cm},
      draw=none,
      at={(0.5,1.2)},
      anchor=south,
      font=\scriptsize
  }
]


\pgfplotsset{
  legend cell align={left},
  legend image post style={xscale=0.8},
  legend image code/.code={
    \draw[#1, yshift=-0.2em] (0cm,0cm) rectangle (0.25cm,0.15cm);
  },
  legend style={
    /tikz/every even column/.append style={column sep=2pt},
    row sep=-2pt,
  },
}

\addlegendimage{ybar,fill=green!60!black,draw=black}
\addlegendentry{\sysName (Marketing)}
\addlegendimage{ybar,fill=green!20!white,draw=black,postaction={pattern=north east lines}}
\addlegendentry{\sysNameP (Marketing)}
\addlegendimage{ybar,fill=orange!80!black,draw=black}
\addlegendentry{\sysName (CoverType)}
\addlegendimage{ybar,fill=orange!30!white,draw=black,postaction={pattern=north east lines}}
\addlegendentry{\sysNameP (CoverType)}
\end{axis}
\end{tikzpicture}

\vspace{-7cm}
\vspace{-2mm}
\caption{\revisemeta{\sysName and \sysNameP runtime (s) w.r.t (Left) \#tuples, (Right)
\#attributes. We used $k=5$ and $\theta=0.10$ for these experiments.}}

\label{fig:scalability_plots}
\end{figure}

\subsection{Scalability} \label{sec:scalability}
\revisemeta{\looseness-1 Figure~\ref{fig:scalability_plots} shows the runtimes
of \sysName and \sysNameP w.r.t \#tuples (left) and \#attributes (right),
averaged over three executions. For both \sysName and \sysNameP, runtime
increases linearly with data cardinality and quadratically with dimensionality,
because increasing dimensions significantly expands the combinatorial search
space. \sysNameP shows only small fluctuations as \#tuples or \#attributes
increases, thanks to its sampling and approximation methods. Remarkably,
\sysNameP processed the CoverType dataset (110 attributes, 581K+ tuples) in
under 50 seconds—demonstrating practical scalability. Despite the substantial
runtime reduction, \sysNameP maintained comparable utility and diversity scores
to \sysName (within a 10\% variation).}

\subsection{Parameter Sensitivity}
\label{sec:param}


Figure~\ref{fig:opt_param_plot} (right) depicts the impact of varying the
diversity threshold ($\theta$) on the utility score over the Video dataset. We
observe that as $\theta$ increases, the utility score drops for all $k$. This is
expected because a higher $\theta$ constrains the selection of candidates more
stringently, leading to fewer eligible items. To satisfy this stricter
criterion, the algorithm may be compelled to select items with lower utility to
maintain diversity, thus decreasing the total utility. The utility drop becomes
particularly significant for high values of $k$, as satisfying a high diversity
threshold for a larger result set is more challenging.

\subsection{Case Studies}
\label{sec:case_study}

We now present findings from case studies in which we manually examined the
pivot tables for qualitative insights regarding recommendations made by \sysName
vs.\ commercial software and LLMs.

\subsubsection{Diverse and meaningful aggregates.} \looseness-1 We found
\sysName to consistently recommend diverse and semantically meaningful
aggregates. For Video, \sysName's includes \texttt{COUNT, MAX, MIN, SUM}, and
\texttt{MEAN}. In contrast, Excel, Google Sheets, and PowerBI typically involve
only \texttt{SUM} or \texttt{COUNT}. While LLMs too suggest a variety of
aggregates, they often involve hallucinated attributes. E.g.,
\texttt{LotLocation} is not an attribute of House, which LLM recommends.
Despite requesting to use the specified aggregates, LLM still selects
out-of-scope aggregates such as median or standard deviation. Excel once
suggested \texttt{SUM(Birth\_Year)}, a completely meaningless aggregate.

\subsubsection{Diverse and meaningful attributes.} We found \sysName to select
diverse and semantically meaningful attributes for aggregation and grouping. In
Video, \sysName utilizes \texttt{Publisher, Sales,} and \texttt{Year} for
various aggregations, providing comprehensive coverage of the attribute space.
This contrasts sharply with Excel and Google Sheets, which make repetitive
\texttt{GROUP BY} choices, while LLM consistently focuses only on
\texttt{Sales} for aggregation. In House, \sysName identifies attributes such
as \texttt{KitchenAbvGr}, \texttt{HalfBath}, and \texttt{FullBath} for
aggregations that other baselines ignore. While only \sysName and LLM
appropriately use \texttt{SalesPrice} for aggregation, LLM limits itself to
aggregate only over this attribute. In contrast, \sysName diversifies selection
of attributes, avoiding meaningless choices made by other tools, such as
PowerBI's poor recommendation to \texttt{GROUP BY} \texttt{ID}.

\subsubsection{Avoiding unsurprising summaries.} Since \sysName can assess the
degree of surprisingness in observed trends, it avoids recommending trivial
queries with low utility. Unlike other tools that consistently recommend common
aggregations such as \texttt{SUM(INCOME)}, regardless of context, \sysName
avoids recommendations that provide obvious insights (e.g., people with low GDP
spend less). For the Video dataset, existing baselines typically focused on
aggregations such as \texttt{SUM(SALES)}. In contrast, \sysName included
diverse types of aggregations such as \texttt{MEAN(NA\_Sales)} and
\texttt{MAX(JP\_Sales)}. Additionally, it discovered more surprising patterns,
such as \texttt{COUNT(YEAR) GROUP BY GENRE}, which were not considered by
others.

\subsubsection{Adaptive recommendation} 
\label{sec:adapt}
\reviseone{We evaluated \sysName's adaptability to user needs (desiderata D4)
through a case study on the Video dataset. Initially, the user indicates
\texttt{Genre} as a grouping attribute they are interested in, and requests
three recommendations ($k=3$). In the first round, \sysName suggests pivot
tables showing \texttt{COUNT(Publishers)} and \texttt{AVERAGE(Global Sales)} by
\texttt{Genre} and \texttt{Region}. After the user indicates disinterest in
\texttt{Publishers}, \sysName adapts by recommending \texttt{COUNT(Years)} and
\texttt{AVERAGE(EU Sales)} by \texttt{Genre}, which the user accepts. For the
third recommendation, \sysName presents \texttt{SUM} and \texttt{MEAN} of
\texttt{EU Sales} and \texttt{Global Sales} by \texttt{Genre}, further
diversifying the aggregates while maintaining focus on the user-specified
grouping on \texttt{Genre}. Unlike existing pivot table recommenders that
disallow customizability and lack mechanism for user feedback, \sysName
dynamically adapts to user preferences.}



\section{\revisemeta{User Study}} \label{sec:userstudy}

\revisemeta{To validate whether our utility (\S\ref{sec:three}) and diversity
(\S\ref{sec:four}) models align with real-world user perception, we conducted
a comprehensive user study with 36 participants, 53\% of whom use spreadsheets
daily or several times per week. The study comprised five sections, each
containing 1--4 objective questions followed by an open-ended prompt for
participants to explain their choices. Below, we summarize each section's setup
and key findings based on the responses.}

\smallskip\noindent \revisemeta{\textbf{\#1 Validating Insightfulness.} We
presented four pairs of pivot tables; within each pair, one had a high score
and the other a low score for one of the \emph{Insightful} components:
(a)~Attribute Significance (\S\ref{attsigsec}), (b)~Informativeness
(\S\ref{infsec}), (c)~Trend (\S\ref{trendsec}), and (d)~Surprise
(\S\ref{sursec}). For each pair, participants had to select the more insightful
table or indicate a tie. The percentages of participants who chose the
high-scoring pivot table (declared a tie) were 88.9\% (8.3\%), 91.7\% (8.3\%),
91.7\% (8.3\%), and 75.0\% (16.7\%), respectively. These results show that, on
average, 86.8\% of participants agreed with our notion of pivot-table
insightfulness, demonstrating strong alignment between our model and human
perception.}

\smallskip\noindent \revisemeta{\textbf{\#2 Validating Interpretability.} Setup
of this section was identical to the previous one, but to validate three
\emph{Interpretability} components: (a)~Density (\S\ref{densec}), (b)~Semantic
validity (\S\ref{semsec}), and (c)~Conciseness (\S\ref{consec}). The
percentages of users who identified the high-scoring pivot table as more
interpretable (declared a tie) were 77.8\% (8.3\%), 91.7\% (5.6\%), and 91.7\%
(8.3\%), respectively. These results show that, on average, 87.1\% of
participants agreed with our notion of pivot-table interpretability,
demonstrating strong alignment between our model and human perception.}

\smallskip\noindent \revisemeta{\textbf{\#3 Validating Utility via Ablation.}
This section included two questions, each presenting a pair of pivot tables for
participants to choose the one that is overall more useful---i.e., shows strong
insights while being easy to interpret---or declare a tie. We found that 86.1\%
(8.3\%) of participants preferred (tied with) the balanced summary over a
highly interpretable but low-insight summary, and 69.4\% (13.9\%) preferred
(tied with) the balanced summary over a highly insightful but less
interpretable one. These results establish that both components of our utility
model are essential.}

\smallskip\noindent \revisemeta{\textbf{\#4 Validating Diversity.} We asked
participants to compare two sets of pivot tables, one more diverse than the
other. A striking 91.7\% preferred the diverse set, confirming alignment
between our model and user perception.}

\newcommand{\toolA}{Microsoft Excel} 
\newcommand{\toolB}{Google Sheets}
\newcommand{\toolC}{an LLM-based method} 
\newcommand{\toolD}{Top-K}

\smallskip\noindent \revisemeta{\textbf{\#5 Contrasting with Other Baselines.}
We asked two questions to compare recommendations by \sysName with four other
baselines: \toolA, \toolB, \toolC, and \toolD, with names of the tools
anonymized to avoid any bias. Among the participants, 52.8\% preferred \sysName
over the other tools and 13.9\% noted that all recommendations are equally
good. This result confirms that \sysName consistently provides more
user-aligned recommendations compared to existing baselines.}

\smallskip

\revisemeta{The free-text justifications helped us understand participants'
reasoning and we found several interesting responses. For instance, one wrote,
``Tool A (\sysName) is [\dots] the best option, because each table provides
useful insights. Tool B has the odd sum of birth year attribute, while Tool C
essentially has the same table twice [\dots].'' We provide additional details
of the user study in the Appendix.}

\section{Related Work}
\label{sec:related_works}

\subsubsection*{Data Summarization} \revisemeta{Below, we discuss three major
areas that share the general problem that we address, which is recommending
interesting aggregates or views to help users understand data.}

\smallskip

\revisemeta{\noindent\textbf{OLAP Cube Exploration.} Prior work in OLAP cube
exploration~\cite{SarawagiVLDB00Adaptive, DBLP:journals/vldb/Sarawagi01,
ExploreOLAP, ExplainingDifferencesSarawagi, DiscoverySarawagi} aim to identify
``surprising'' data regions to help users uncover previously unseen and
interesting patterns. These approaches define surprise based on user
familiarity, and interestingness based on purely statistical methods. However,
they lack awareness of the underlying data \emph{semantics}, often resulting in
recommendations that are practically uninteresting and hard to interpret.
Moreover, they typically suggest a single item or a top-$k$ list, without
considering \emph{bundle} suggestions that offer a \emph{set} of diversified
items.}

\smallskip

\revisethree{\noindent\textbf{Insight Generation and Visualization.} A number
of prior work define informative summaries as tables that exhibit significant
statistical disparities among groups~\cite{SeedbVartak15, HarrisWWW23SpotLight,
ForesightDemiralp17, QuickInsightsDing19, VisGuide, InsightLens}. Some measure
the significance of insights using null-hypothesis
testing~\cite{QuickInsightsDing19, VisGuide} while others exploit
LLMs~\cite{InsightLens}. While insight generation and visualization
recommendation~\cite{InsightLens, EDA4Sum, Voyager, VisGuide} are related to
our work, most prior works in this space do not incorporate the three key
dimensions together: insightfulness, interpretability, and diversity. Notably,
recommending a k-budgeted \emph{set} of summaries is a significantly harder
problem than simple top-k recommendation.}

\smallskip

\revisemeta{\noindent\textbf{View or Exploration-step Recommendation.} View
recommendation systems~\cite{ViewSeeker, QAGView, Voyager} explore a related
direction. Voyager~\cite{Voyager} ranks visualizations based on perceptual
effectiveness to improve interpretability whereas ViewSeeker~\cite{ViewSeeker}
compares multiple visualizations using various similarity measures. However,
these systems are not semantics-aware and do not consider diversity.} Smart
Drill-Down~\cite{DBLP:journals/tkde/JoglekarGP19} enables users to discover and
summarize interesting groups of tuples described by rules. However, their unit
is a rule, while ours is a pivot table.
Auto-Suggest~\cite{AutoSuggestSIGMODE2020Cong} and
DAISY~\cite{DAISYVLDB24Junjie} leverage large-scale user logs, SQL queries, or
crowdsourcing to recommend pivot tables based on historical usage patterns.
However, they cannot ensure pivot-table informativeness, as they do not
validate the content of the generated pivot tables.

\smallskip

\smallskip \revisemeta{\looseness-1 In summary, prior work fall short in one of
the following aspects.
\textbf{First}, they do not consider the semantic validity and interpretability
aspects of the summary structure. E.g., (a)~the aggregates {\small
\texttt{SUM(Birth\_Year)}} and {\small \texttt{AVG(Zip\_Code)}} are
semantically invalid, and (b)~an attribute with values {\small
\{\texttt{Val\_1, Val\_2\}}} is not interpretable whereas {\small
\{\texttt{Large, Small}\}} is.
\textbf{Second}, they lack \emph{semantic awareness} to \emph{validate} an
apparent data insight. For instance, they typically determine the degree of
insightfulness purely based on statistical properties, such as whether
aggregated values between two groups significantly deviate from each other,
without factoring in the degree of expectedness of that deviation based on
common knowledge (e.g., a significant deviation in average height between
toddlers and adults is expected). In contrast, \sysName consults with an
LLM---which can mimic a human domain expert who can provide the likelihood of
observing a statistically interesting phenomenon---and effectively prunes
statistically insightful, but practically mundane insights.
\textbf{Third}, beyond simple top-k, they do not focus on the \emph{bundle
recommendation problem}---which involves suggesting a \emph{set} of $k$
results---with \emph{diversity} requirements, which is our focus.
\textbf{Fourth}, while a few works consider diversity in
summarization~\cite{EDA4Sum, QAGView, QAGViewFull, ATENA}, they ignore the
semantic aspect of diversity. DAISY~\cite{DAISYVLDB24Junjie} models diversity
during query collection, not recommendation, while QAGView~\cite{QAGView}
diversifies at a tuple-level granularity. In contrast, we use a semantics-aware
model for the distance metric---using query and contend embedding---to ensure
\emph{semantic} diversity across the recommended pivot tables.}

\subsubsection*{Diversification Algorithms.} Query result
diversification~\cite{DBLP:journals/pvldb/WangMM18} methods define diversity
across three aspects: (1)~content, (2)~novelty, and
(3)~coverage~\cite{DrosouSIGMOD10Search, DrosouVLDB12DisC}. Max-Sum
diversification~\cite{DrosouSIGMOD10Search,CarbonellSIGIR98MMR,
RadlinskiSIGIR06PersonalWebSearch} maximizes the linear combination of
diversity and utility scores, while Max-Min diversification maximizes the
minimum diversity between selected items. However, existing diversification
methods target tuples or documents, not entire tables. Thus, they do not
trivially extend to table recommendations, as in our case.


\section{Conclusions and Future Work}

We presented \sysName to recommend diverse set of pivot tables while balancing
insightfulness and interpretability. We introduced a utility model for a single
pivot table, a diversity metric for pivot-table sets, and a simple greedy
algorithm built on two optimization techniques for efficient recommendation. We
empirically showed that \sysName outperforms baselines in diversity while
maintaining high utility. Our case studies illustrated \sysName's ability to
avoid generic patterns and provide data-semantics-aware suggestions. To the best
of our knowledge, this is the first work to combine diversity and data-semantics
for data summarization.

\sysName currently does not incorporate contextual information, such as the
user's workflow (e.g., their end goals) or broader ecosystem (e.g., tools in
their software stack). Integrating such context could significantly enhance the
quality of recommendations. Another promising direction is to support
alternative forms of data summaries---such as textual descriptions that
highlight key trends---in addition to structured formats like pivot tables.


\bibliographystyle{ACM-Reference-Format}
\bibliography{paper}

\clearpage

\appendix
\renewcommand{\thefigure}{A\arabic{figure}}
\setcounter{figure}{0}
\renewcommand{\thetable}{A\arabic{table}}
\setcounter{table}{0}

\newpage
\section{Appendix}
\subsection{User Study}
\begin{figure*}[t]
  \centering
  \resizebox{0.60\textwidth}{!}{%
  \begin{tabular}{cccc}
    \includegraphics[width=\linewidth, angle=270]{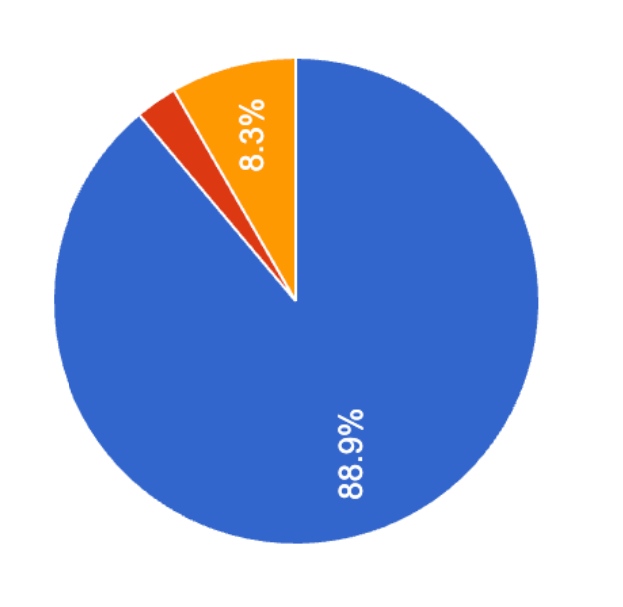} &
    \includegraphics[width=\linewidth, angle=270]{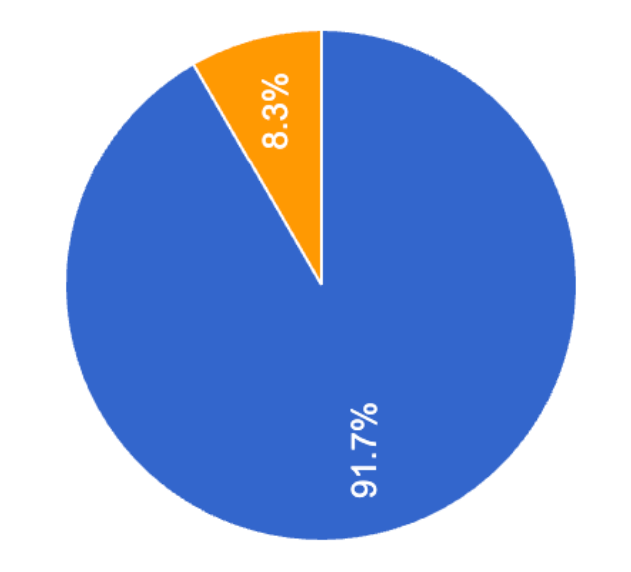} &
    \includegraphics[width=\linewidth, angle=270]{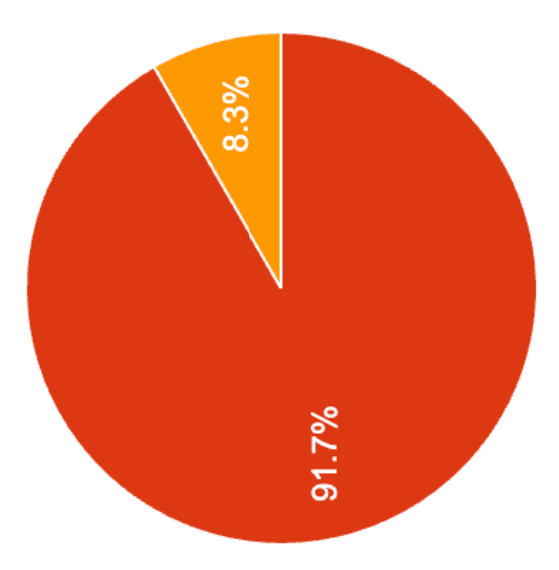} &
    \includegraphics[width=\linewidth, angle=270]{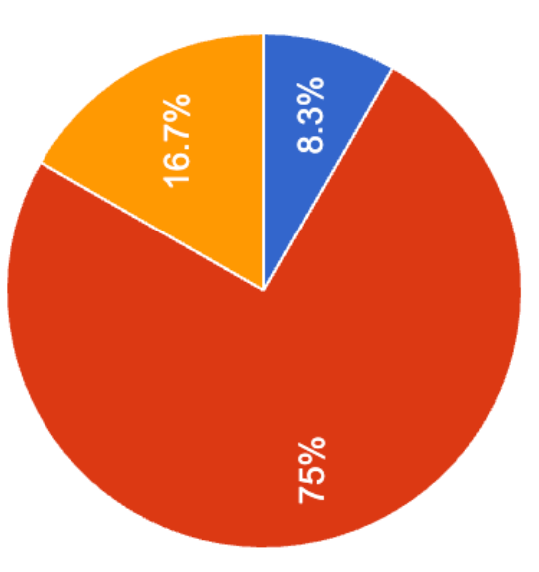} \\[1mm]

    \includegraphics[width=\linewidth, angle=270]{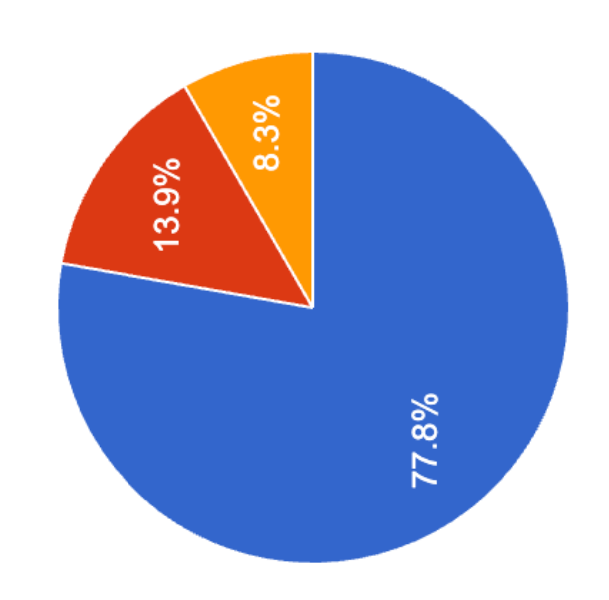} &
    \includegraphics[width=\linewidth, angle=270]{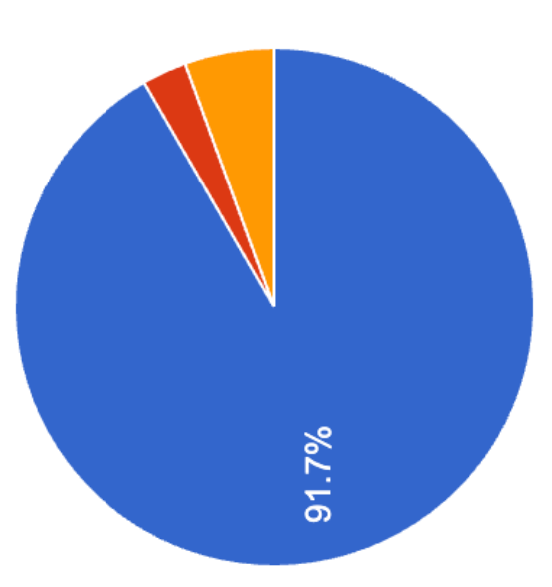} &
    \includegraphics[width=\linewidth, angle=270]{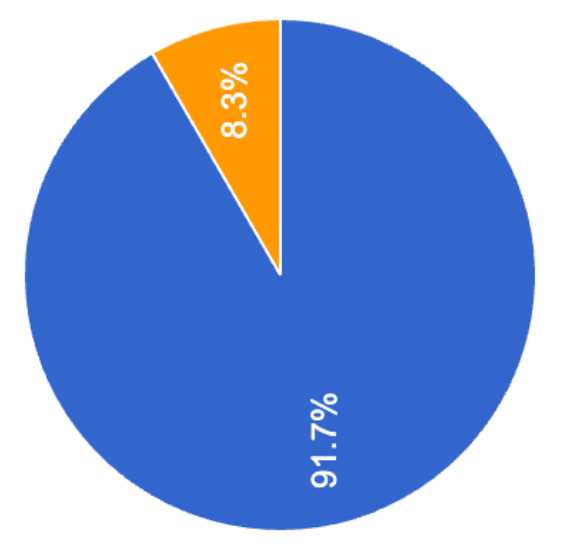} &
    \includegraphics[width=\linewidth, angle=270]{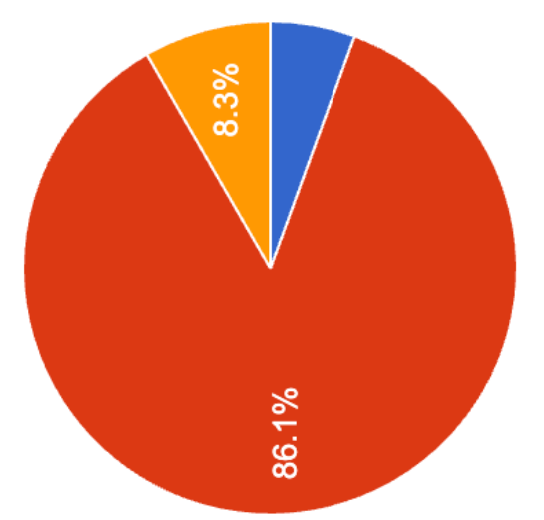} \\[1mm]

    \includegraphics[width=\linewidth, angle=270]{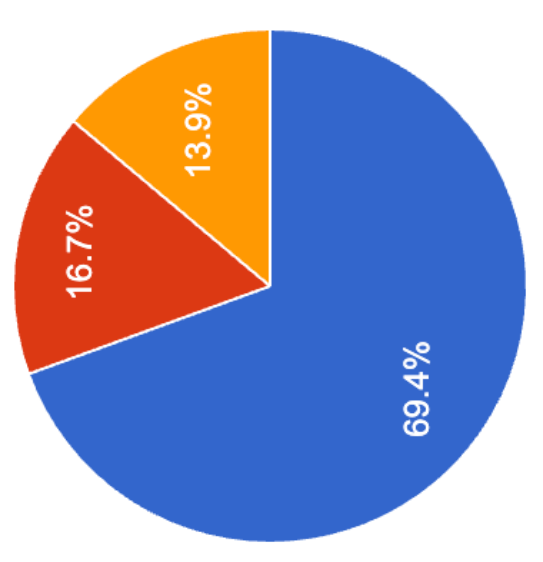} &
    \includegraphics[width=\linewidth, angle=270]{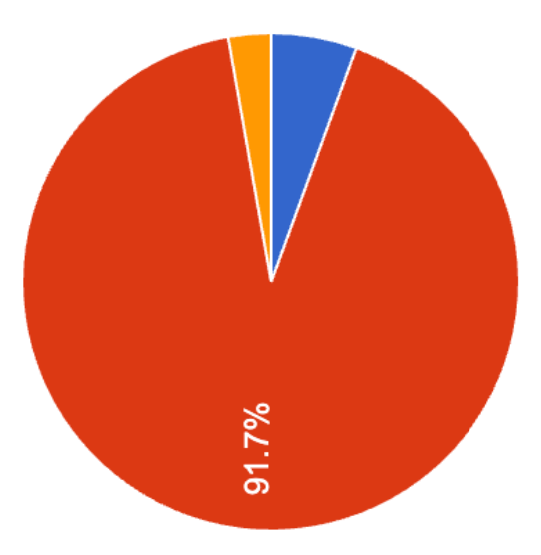} &
    \includegraphics[width=\linewidth, angle=270]{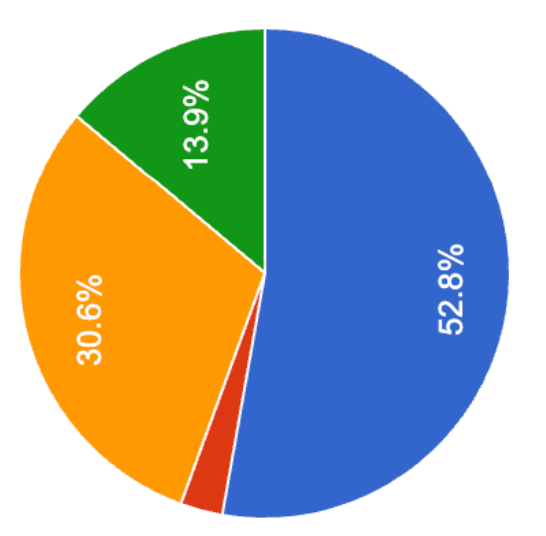} &
    \includegraphics[width=\linewidth, angle=270]{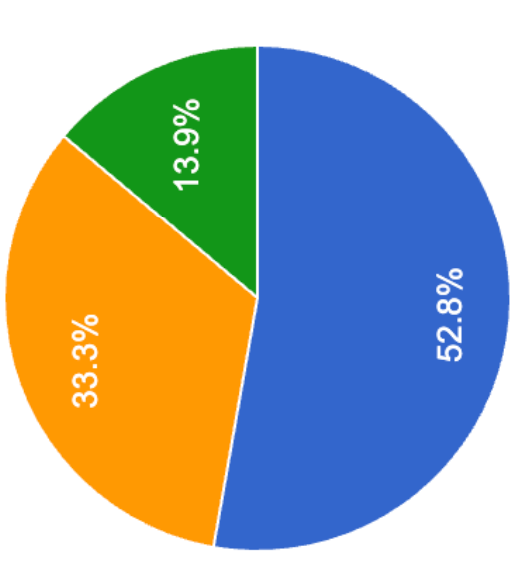}
  \end{tabular}
  }

  \caption{Twelve user study pie charts summarizing participant responses. Each chart shows the proportion of participants selecting our \sysName's result, and values above 50\% indicate that \sysName was preferred.}
  \label{fig:user_study_pies}
\end{figure*}

\begin{figure*}[t]
  \centering
  \includegraphics[width=0.8\linewidth]{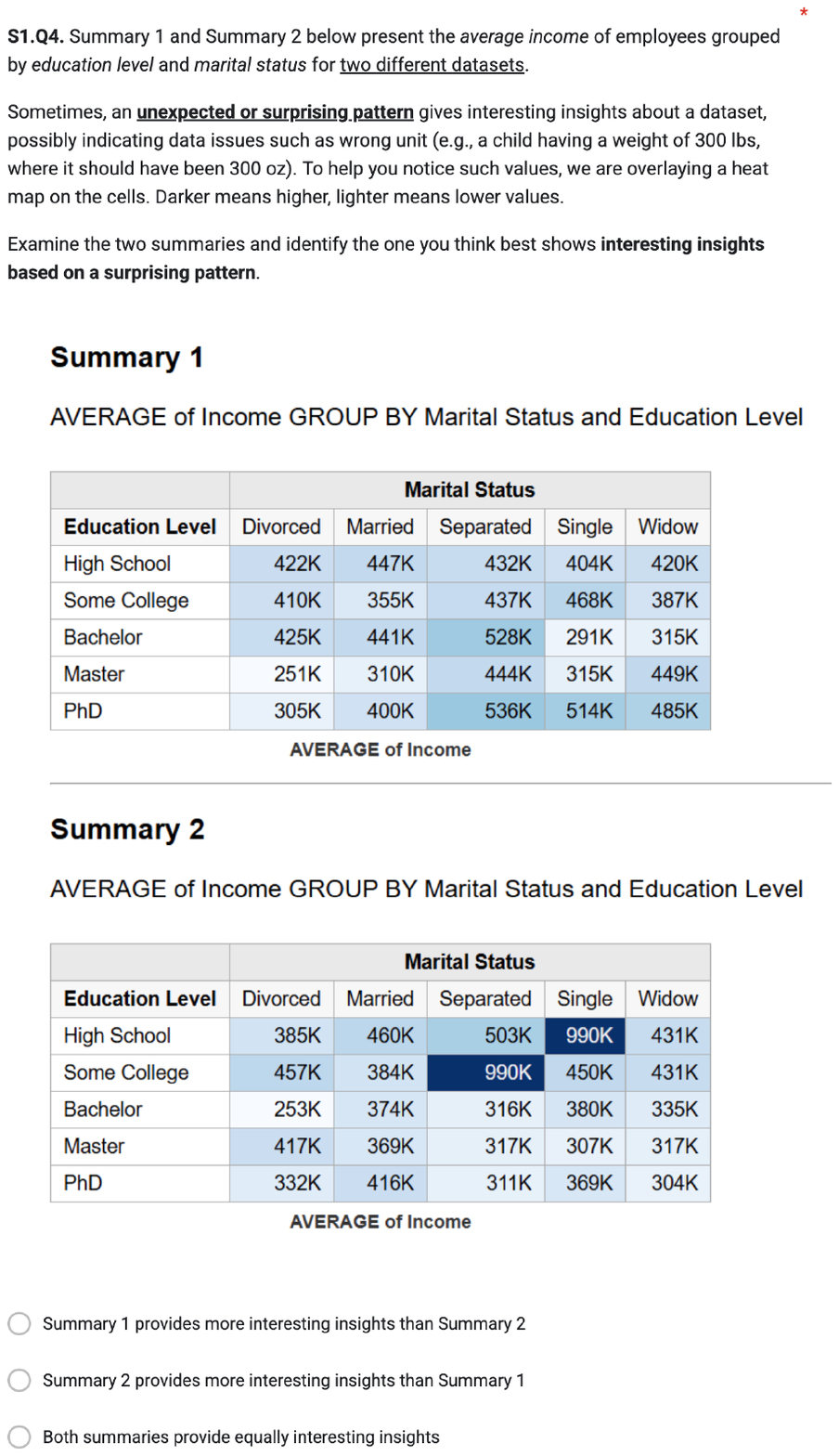}
  \caption{User study question in which participants select the more surprising outlier patterns between two pivot tables.}
  \label{fig:user_study_s1q4}
\end{figure*}

\begin{figure*}[t]
  \centering
  \includegraphics[width=0.8\linewidth]{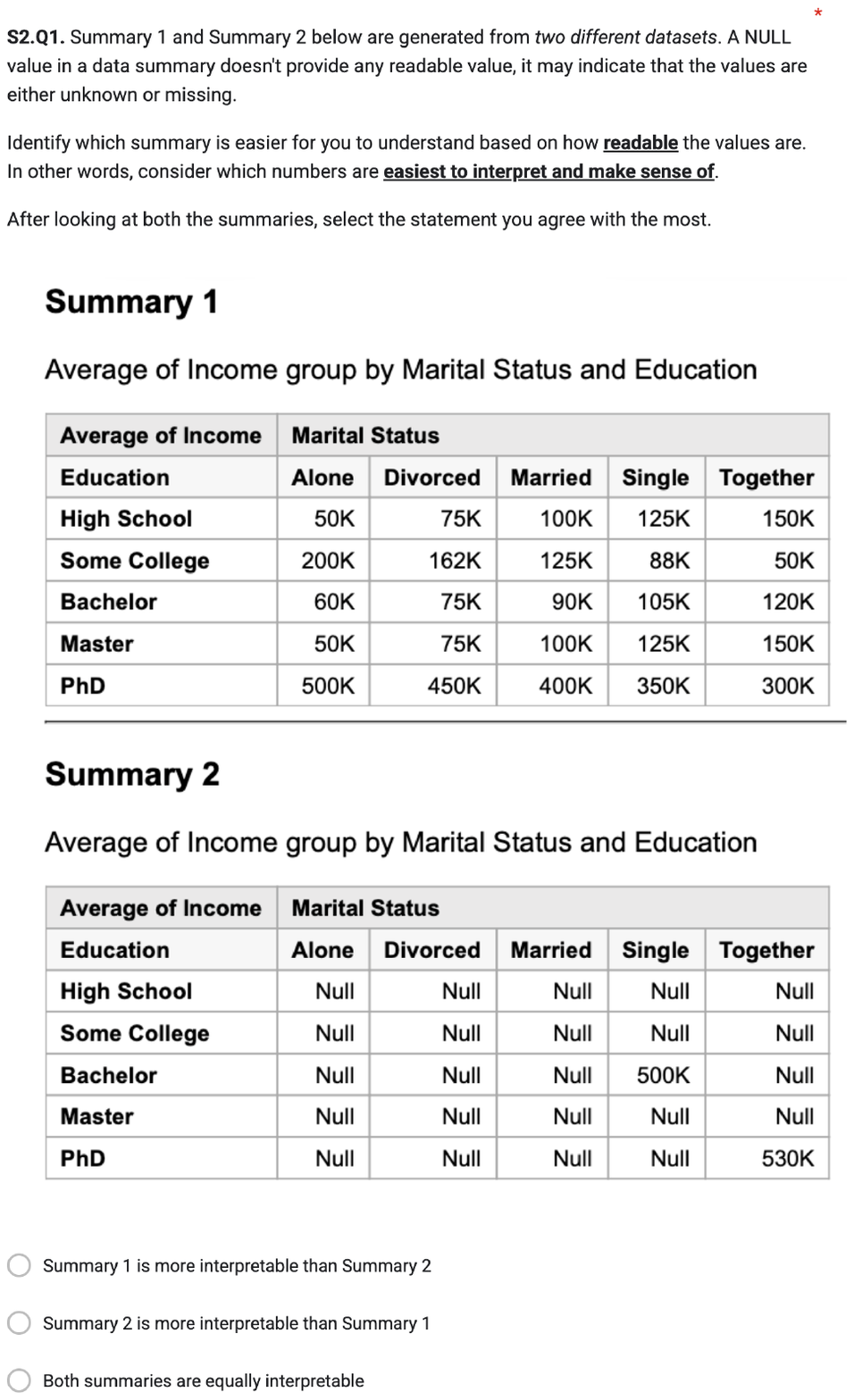}
  \caption{User study question in which participants selected the more readable pivot table.}
  \label{fig:user_study_s2q1}
\end{figure*}

\begin{figure*}[t]
  \centering
  \includegraphics[width=0.8\linewidth]{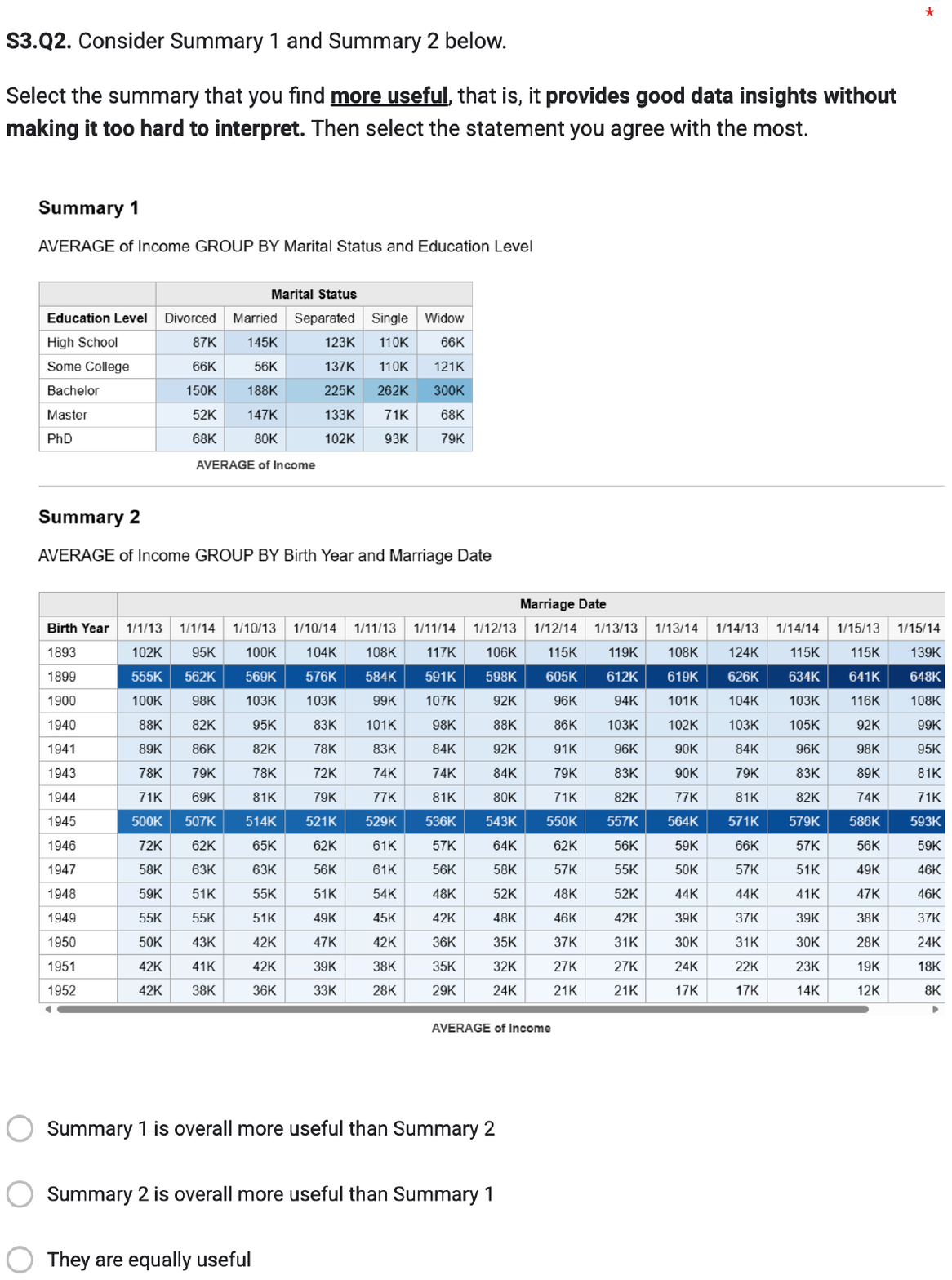}
  \caption{User study question in which participants select the more useful table from two pivot tables.}
  \label{fig:user_study_s3q2}
\end{figure*}

\begin{figure*}[t]
  \centering
   \includegraphics[width=0.6\linewidth, angle=270]{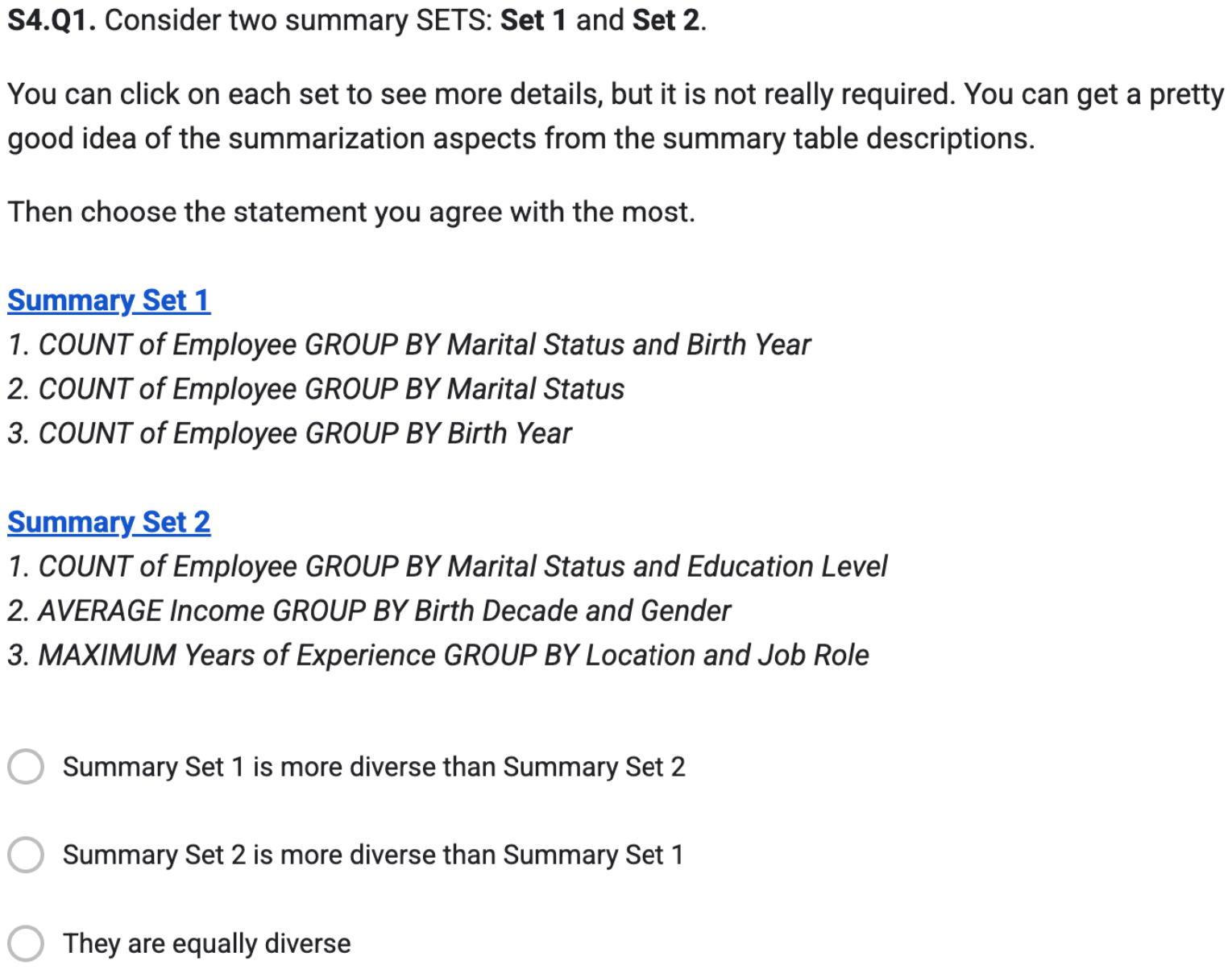}
  \caption{User study question in which participants select the more diverse set from two sets of pivot tables.}
  \label{fig:user_study_s4q1}
\end{figure*}

\begin{figure*}[t]
  \centering
  \includegraphics[width=0.6\linewidth]{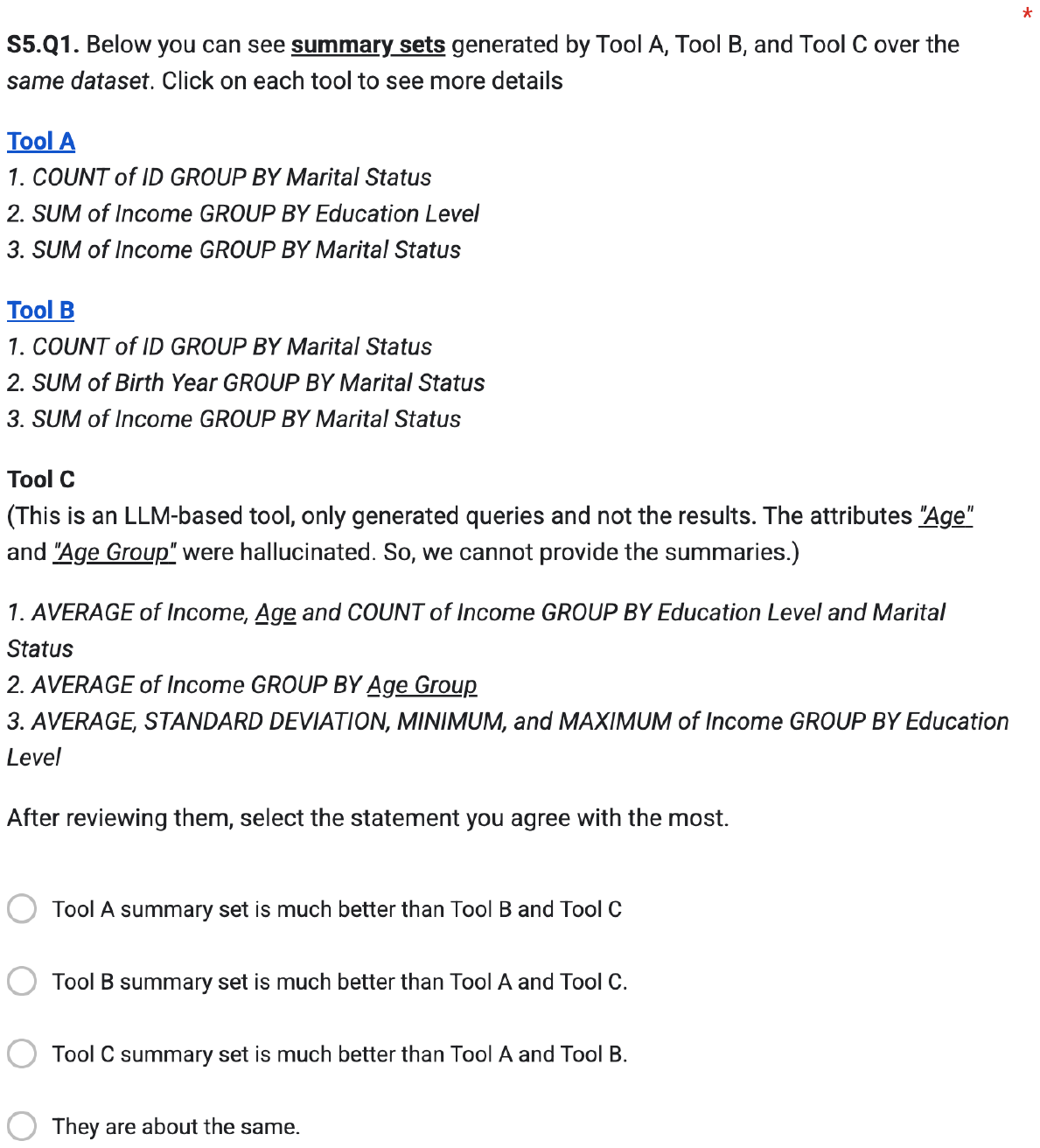}
  \caption{User study question in which participants selected the most diverse and useful set of pivot tables.}
  \label{fig:user_study_s5q1}
\end{figure*}

We provide snapshots of the user study responses and questions used to validate
each component of our model. Figure~\ref{fig:user_study_pies} presents the user
response distributions for all study questions. In each pie chart, a proportion
greater than 50\% indicates that participants selected \sysName's result,
corresponding to the validation score we report.
Figure~\ref{fig:user_study_s1q4} shows a question evaluating insightfulness by
asking participants to select the more surprising outlier.
Figure~\ref{fig:user_study_s2q1} evaluates interpretability by asking users to
choose the more interpretable pivot table. Figure~\ref{fig:user_study_s3q2}
evaluates utility via an ablation setting where Summary 2 has reduced
interpretability. Figure~\ref{fig:user_study_s4q1} evaluates diversity by asking
users to select the more diverse set of pivot tables. Finally,
Figure~\ref{fig:user_study_s5q1} asks participants to choose their preferred
system output across all criteria, allowing comparison against other tools.
Table~\ref{tab:participant_rationales} presents the representative rationales
provided by participants for the snapshot questions.

\subsubsection{Participants' Background and Experience}
\begin{figure*}[t]
  \centering
  \resizebox{1.0\textwidth}{!}{%
  \begin{minipage}{\textwidth}
    \centering

    \begin{minipage}[b]{0.48\textwidth}
      \includegraphics[width=\linewidth]{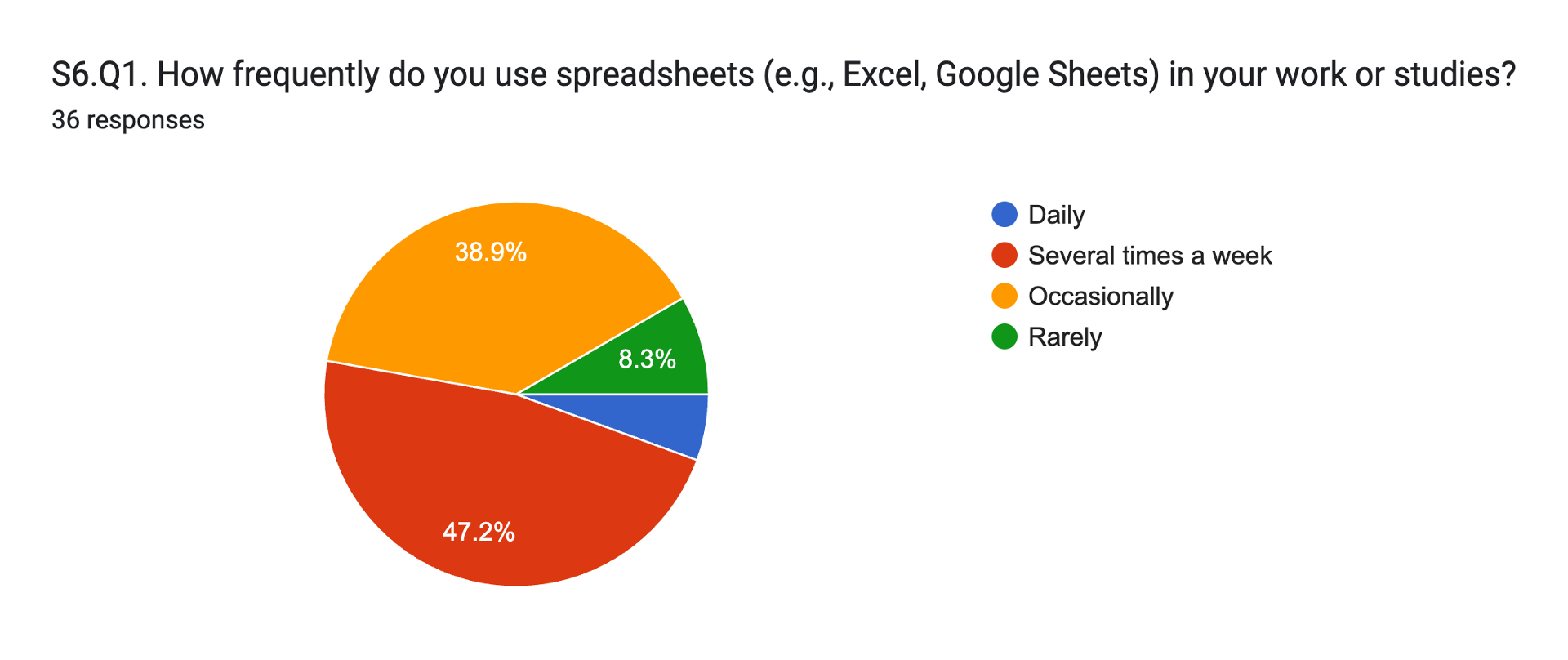}
    \end{minipage}
    \begin{minipage}[b]{0.48\textwidth}
      \includegraphics[width=\linewidth]{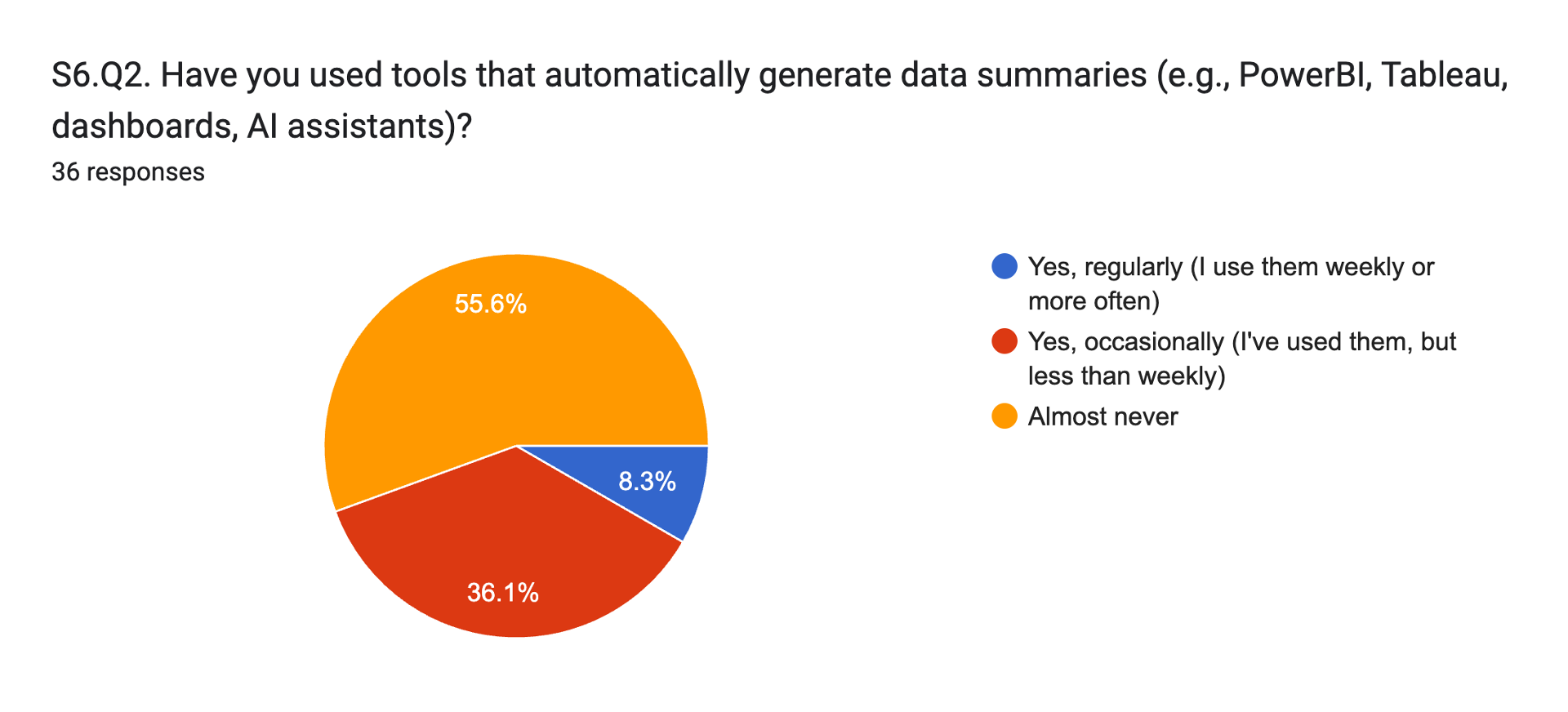}
    \end{minipage}

    \vspace{2mm}

    \begin{minipage}[b]{0.48\textwidth}
      \includegraphics[width=\linewidth]{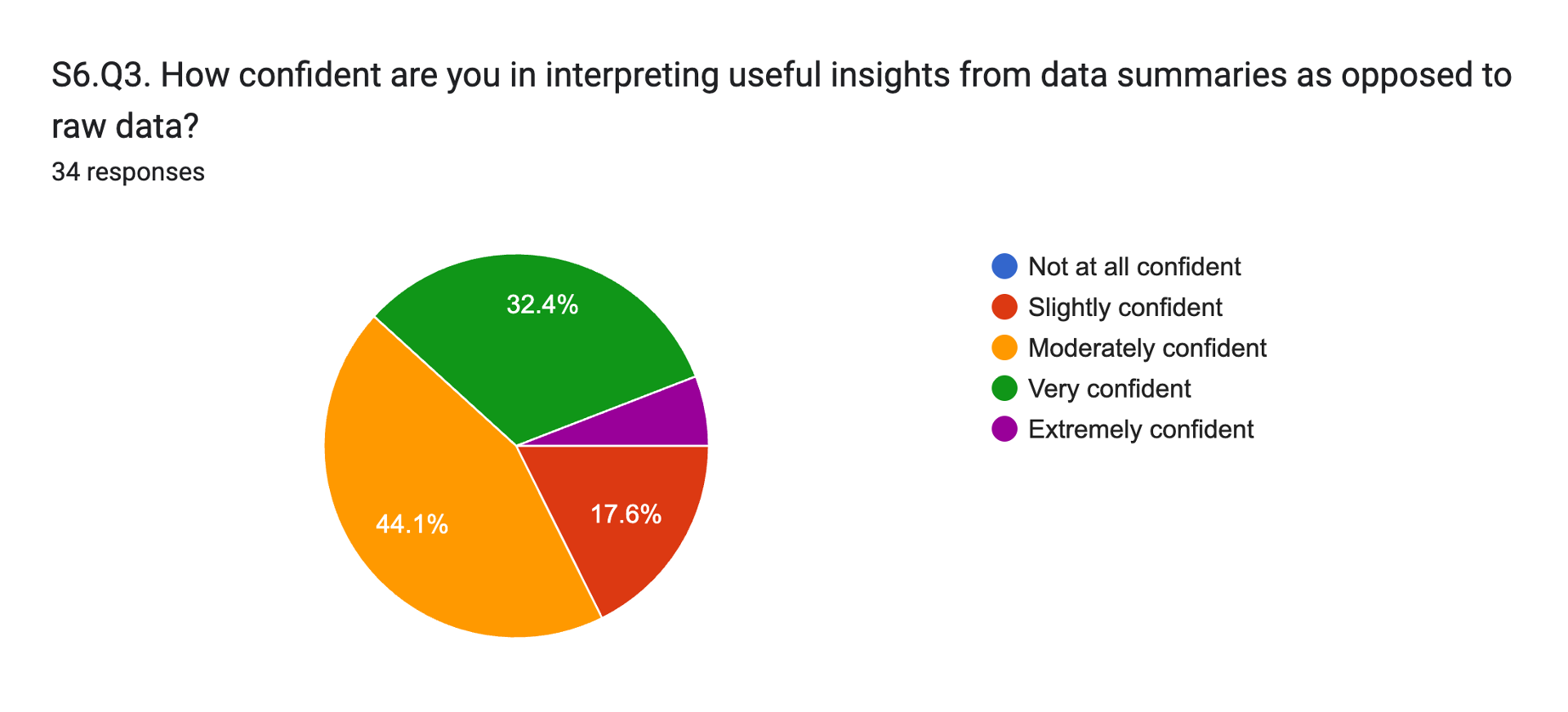}
    \end{minipage}
    \begin{minipage}[b]{0.48\textwidth}
      \includegraphics[width=\linewidth]{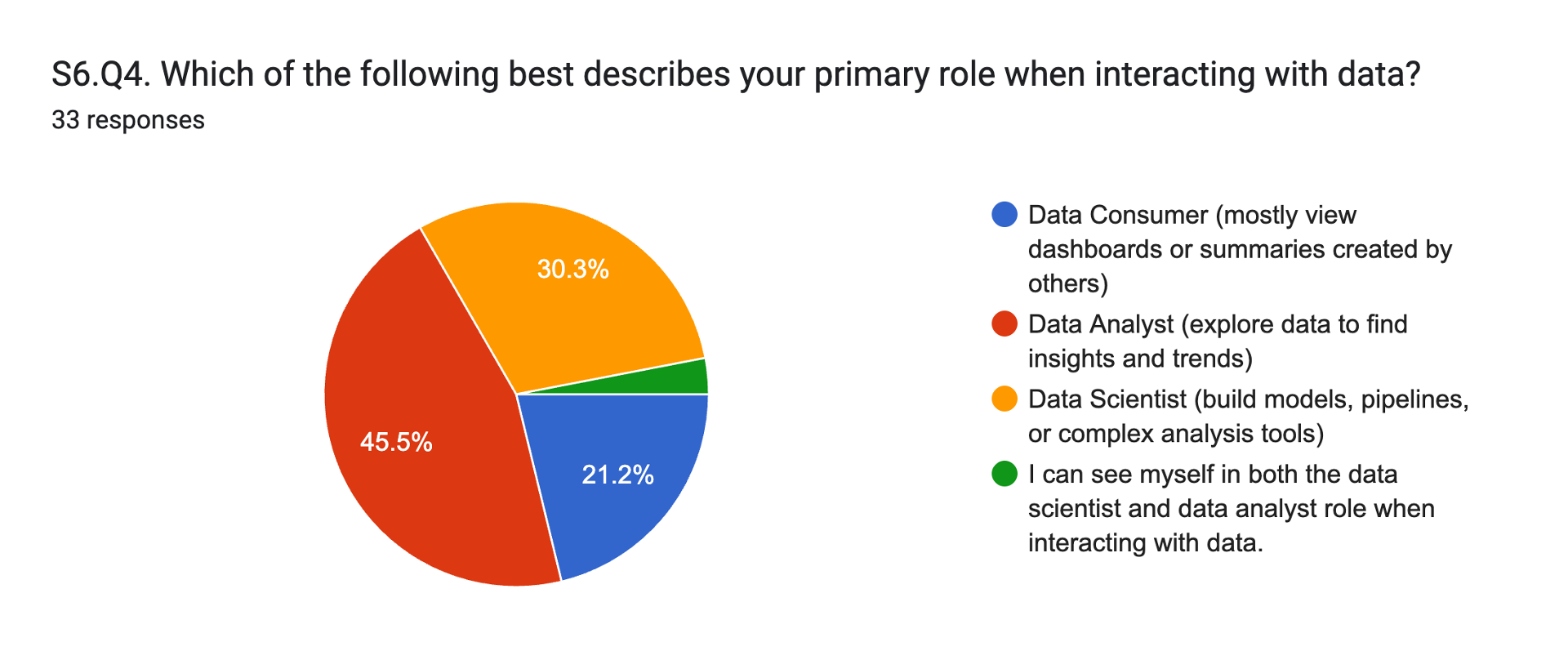}
    \end{minipage}

  \end{minipage}
  }

  \caption{User study response summaries for participants' background and experience.}
  \label{fig:user_study_section6}
\end{figure*}

Additionally, we provide participants' background and experience regarding data
summarization in Figure~\ref{fig:user_study_section6}. Most participants use
spreadsheets several times a week, yet the majority do not regularly apply data
summarization techniques in their work. Furthermore, many participants reported
low confidence in interpreting meaningful insights from summarized data. These
results highlight the need for effective data summarization tools. We also asked
participants to share their experience with automated summarization tools. Some
participants noted \textit{``I don't like when AI assistants often show
irrelevant summaries and visualizations, or hallucinates queries.'', ``AI
assistants can be poor at some of these things, but they are proficient at
interpreting data and knowing what kind of tables are useful based on data. I
love Tableau, but it can be frustrating because it isn't always the most
user-friendly.''} and \textit{I don't like when the tools show irrelevant
statistics, I also don't like when the heatmap overlay ends up providing the
same color or very similar for all values, I feel like this doesn't provide much
additional context. I did like how in certain cases it provided very relevant
information and important groupings straight to you without having to think
about it.} With these notes, we could know that participants had a clear
preference for tools that provide relevant, interpretable summaries and
highlight meaningful groupings, and they were frustrated when systems produced
irrelevant unhelpful outputs.

Moreover, we also asked participants to describe their ideal tool for
summarizing spreadsheet datasets and what features they would like to see.
Participants noted, \textit{``I would like to be able to specify which columns
to compare/summarize (i.e. a workflow that simplifies tasks that I can perform
with Python, SQL, or excel formulas), as well as receive a few potentially
insightful suggestions that have not yet been considered by my own analysis.'',
``Would like it to automatically and efficiently identify hidden trends and
correlations and present them in a readable way.''} and \textit {``Sometimes I
explore data with a specific goal. For example, I might like to investigate the
potential reason of a drop on a metric. It would be great if there is an
auto-summarizing tool providing some insights based on the given data and my
input questions in natural language.''} In summary, participants want tools that
let them specify what to analyze while also suggesting new insights they might
not have considered. They prefer systems that can automatically reveal hidden
trends or correlations and present them in a clear and readable way.

\subsection{Scalability} Figure~\ref{fig:scalability_plots_appendix} compares
the runtime performance of \sysName and \sysNameP as we vary \#tuples (left)
and \#attributes (right) across four datasets. As expected, runtime grows with
dataset size in both dimensions. When increasing the number of tuples, \sysName
shows a steady increase in runtime across all datasets, whereas \sysNameP
consistently achieves substantially lower runtime, with improvements
particularly pronounced on larger datasets such as Video and CoverType. When
increasing the number of attributes, the performance gap becomes even more
evident that \sysName's runtime rises sharply due to the rapidly expanding
search space, while \sysNameP scales more smoothly because it approximates
candidate combinations. Notably, on the full CoverType dataset, \sysNameP
completes execution in under 50 seconds, compared to over 8 minutes for
\sysName.

\subsection{Tuning the Diversity Threshold $\theta$} When the user is unable to
specify a desired diversity threshold $\theta$, either due to unfamiliarity of
the dataset or lack of expertise, the desired number of pivot tables $k$ can be
used as a guideline to derive the value of $\theta$. In that scenario, a
trivial extension to our approach would be to cluster the candidate pivot
tables into $k$ groups, and then apply a greedy strategy to maximize a linear
combination of \emph{Utility} (\S\ref{sec:three}) and \emph{Diversity}
(\S\ref{sec:four}) as was done in prior work~\cite{DBLP:conf/sigmod/HeMF25}.

\subsection{Brute Force}
For the brute-force algorithm, we first enumerate all possible pivot table
queries, then materialize them and compute their embeddings. Next, we generate
all possible sets of size $k$, resulting in $|\Universe|^k$ candidate sets. For
each candidate set, we verify whether it satisfies the pairwise distance
threshold and compute the sum of utility scores. The set with the highest
utility among those meeting the distance constraint is selected as the final
result. Algorithm~\ref{alg:brute_force} describes this brute-force procedure.
Its time complexity is $O(|P| \cdot (n^{2}m + nm^{2}) +  2 \cdot |P| \cdot
\texttt{embedding\_size} + |\mathbf{PT}| \cdot k)$, where the $n^{2}m + nm^{2}$
term accounts for outlier computation, $embedding\_size$ is the size of the
embeddings, and the $|\mathbf{PT}| \cdot k$ term accounts for evaluating all
size-$k$ sets over the pivot tables. As the number of distinct pivot tables
$|\mathbf{PT}|$ grows combinatorially with respect to the number of attributes
and aggregation choices, where $|\mathbf{PT}| = |\Universe|^k$. Therefore, the
final time complexity is $O(|\Universe|^k)$.

\subsection{Time Complexity of the Greedy Approach}

The overall time complexity of the greedy algorithm in Algorithm~\ref{alg:llm_proxy_diverse_selection} is
\(
O\big(|P_{\text{prun}}| \log |P_{\text{prun}}| \allowbreak
+ |P_{\text{prun}}| \cdot (n^{2} m + n m^{2} + \texttt{tree\_depth}) 
\) \\
\(+ 2 \cdot |P_{\text{prun}}| \cdot \texttt{embedding\_size}\big),
\)
where $|P_{\text{prun}}|$ is the number of tables after pruning,
\texttt{tree\_depth} is the depth of the LLM-proxy cache, and
\texttt{embedding\_size} is the dimensionality of the embedding vectors. The
term $|P_{\text{prun}}| |\log P_{\text{prun}}|$ corresponds to sorting the utility
scores. The term $n^{2}m + nm^{2}$ accounts for outlier computation, as in the
brute-force approach.  The overall complexity is dominated by the utility score
computation—specifically, the pairwise distance calculations—since both
\texttt{tree\_depth} and \texttt{embedding\_size} are constants. Therefore, the
dominant term is:
\(
O\left(|P_{\text{prun}}| \cdot (n^{2}m + nm^{2})\right).
\)

\begin{algorithm}[t]
\LinesNumbered
\small{
\Input{
    Database $D$\\
    Diversity threshold \( \theta \), \\
    Number of pivot tables to select \( k \)
}
\Output{
    A diverse, high-utility pivot table subset $T$ under the size budget $k$ and diversity constraint $d$
}

Generate all possible (unmaterialized) pivot-table queries \( P \)

\ForEach{\( p \in P \)}{
    Materialize \( p \) over $D$ \\
    Compute embedding \( e(p) \) (query + content) \\
    Compute utility score \( u(p) \)
}

\( \mathbf{T}^* \gets \emptyset \)
\( best\_score \gets -\infty \)

Generate all possible sets of $k$ pivot-table queries \( \mathbf{PT} \)

\ForEach{\( \mathbf{T} \in \mathbf{PT} \)}{
    \ForEach{\( T \in \mathbf{T} \)}{
        \If{\( \operatorname{Distance}(T, \mathbf{T} \setminus \{T\}) < \theta \)}{
            \textbf{continue}
        }
        Calculate utility score \( u \) for  \( \mathbf{T} \)
        \If{\( u > best\_score \)}{
            \( best\_score \gets u \) \\
            \( \mathbf{T}^* \gets \mathbf{T} \)
        }	
    }
}
\Return \( \mathbf{T}^* \) \tcc*[f]{Return final diverse, high-utility pivot table set}
}
\caption{Brute-Force Recommendation}
\label{alg:brute_force}
\end{algorithm}


\begin{table*}[t]
\centering
\resizebox{\textwidth}{!}{%
\begin{tabular}{p{3.2cm} p{9.5cm} p{4.5cm}}
\toprule
\textbf{Criterion} & \textbf{Participant Rationale} & \textbf{Implication} \\
\midrule

\multirow{2}{3.2cm}{\#1 Validating Insightfulness} 
& ``Specific values in summary 2 have significantly unexpected patterns, whereas in summary 1 they are less significant.'' 
& \multirow{2}{4.5cm}{Participants preferred insights that reveal unexpected patterns.} \\
& ``There are some interesting outliers that I would find interesting.'' & \\

\midrule

\multirow{2}{3.2cm}{\#2 Validating Interpretability}
& ``No Nulls in summary 1 increases readability and thus easiest to interpret and make sense of.'' 
& \multirow{2}{4.5cm}{Participants preferred clear tables and it improves user comprehension.} \\
& ``With so many null values, it's harder to understand the data well enough to derive insights.'' & \\

\midrule

\multirow{2}{3.2cm}{\#3 Validating Utility via Ablation}  
& ``Summary 2 is not only too large to understand, but the choice of grouping attributes is not really helpful (especially when we are supposed to be summarizing the data) since there are too many combinations of values which are already not very meaningful when used together in a GROUP BY in the first place.'' 
& \multirow{2}{4.5cm}{Participants preferred compact, insightful tables over larger but harder to interpret ones.} \\
& ``Although there is a more clear difference in some values of table two, table one has a more reasonable number of rows and columns.'' & \\

\midrule

\multirow{2}{3.2cm}{\#4 Validating Diversity}
& ``Summary 2 is more diverse. Summary 1 is just separating out the first summary table into it's separate parts while summary 2 provides new information that could be interesting.'' 
& \multirow{2}{4.5cm}{Participants preferred tables that provide non-redundant insights rather than variations of the same pattern.} \\
& ``Set 2 offers more variety in what it covers, as opposed to summary 1, which only covers 3 attributes.'' & \\

\midrule

\multirow{2}{3.2cm}{\#5 Contrasting with Other Baselines}
& ``Tool A is better than Tool B since a sum of birth year is not useful or insightful. Tool C is bad because of the hallucination mentioned.'' 
& \multirow{2}{4.5cm}{Participants preferred \sysName for providing insightful, interpretable, and diverse summaries.} \\
& ``A is the best option, as B is not very useful because of the sum of birth year. C could be useful if there were not hallucinated attributes.'' & \\
\bottomrule
\end{tabular}
}
\caption{Representative participant rationales and derived implications.}
\label{tab:participant_rationales}
\end{table*}


\begin{figure*}[h]
\centering

\begin{subfigure}[t]{0.5\textwidth}
\centering
\resizebox{\textwidth}{!}{%
\begin{tikzpicture}
\begin{axis}[
  ybar,
  bar width=12pt,
  width=20cm,
  height=10cm,
  ymin=0,
  ymax=500,
every axis/.append style={font=\fontsize{24}{28}\selectfont},
xlabel style={font=\fontsize{24}{28}\selectfont, yshift=-20pt},
ylabel style={font=\fontsize{24}{28}\selectfont, yshift= 20pt},
x tick label style={font=\fontsize{24}{28}\selectfont},
y tick label style={font=\fontsize{24}{28}\selectfont},
  xlabel={Tuples (\%)}, ylabel={Runtime (s)},
  xtick={0.2,0.4,0.6,0.8,1.0},
  xticklabels={20,40,60,80,100},
  ytick={0,100,...,500},
  grid=major,
  enlarge x limits={abs=0.7},
  legend style={
      at={(-0.05,1.02)}, 
      anchor=south west,
      legend columns=4,
      font=\large,
      fill=none,
      draw=none
  },
  grid=major,
  enlarge x limits=0.15,
  xtick pos=left,
  ytick pos=left,
]

=== Grouped bars per dataset ===
\addplot+[fill=purple!60!black, draw=black] coordinates {
  (0.2,2.5212)
  (0.4,2.4302)
  (0.6,2.3088)
  (0.8,2.5612)
  (1.0,2.4262)
};
\addplot+[fill=purple!20!white, draw=black, postaction={pattern=north east lines}] coordinates {
  (0.2,0.7345)
  (0.4,0.8113)
  (0.6,1.1677)
  (0.8,2.3631)
  (1.0,2.2551)
};
\addplot+[fill=blue!60, draw=black] coordinates {
  (0.2,32.9626)
  (0.4,45.8786)
  (0.6,57.9441)
  (0.8,65.0957)
  (1.0,69.0586)
};
\addplot+[fill=blue!20, draw=black, postaction={pattern=north east lines}] coordinates {
  (0.2,14.0848)
  (0.4,16.4765)
  (0.6,17.0372)
  (0.8,16.7144)
  (1.0,17.2702)
};
\addplot+[fill=green!60!black, draw=black] coordinates {
  (0.2,153.2561)
  (0.4,164.6201)
  (0.6,163.5402)
  (0.8,168.2446)
  (1.0,162.3688)
};
\addplot+[fill=green!20!white, draw=black, postaction={pattern=north east lines}] coordinates {
  (0.2,18.3847)
  (0.4,19.0369)
  (0.6,18.3386)
  (0.8,18.6775)
  (1.0,18.8035)
};
\addplot+[fill=orange!80!black, draw=black] coordinates {
  (0.2,113.4003)
  (0.4,204.8112)
  (0.6,307.0491)
  (0.8,395.9639)
  (1.0,491.3975)
};
\addplot+[fill=orange!30!white, draw=black, postaction={pattern=north east lines}] coordinates {
  (0.2,17.9051)
  (0.4,20.6787)
  (0.6,29.0860)
  (0.8,29.5663)
  (1.0,39.1373)
};

\end{axis}
\end{tikzpicture}
}
\end{subfigure}%
\hfill
\begin{subfigure}[t]{0.5\textwidth}
\centering
\resizebox{\textwidth}{!}{%
\begin{tikzpicture}
\begin{axis}[
  ybar,
  bar width=12pt,
  width=20cm,
  height=10cm,
  ymin=0,
  ymax=500,
every axis/.append style={font=\fontsize{24}{28}\selectfont},
xlabel style={font=\fontsize{24}{28}\selectfont, yshift=-20pt},
ylabel style={font=\fontsize{24}{28}\selectfont, yshift= 20pt},
x tick label style={font=\fontsize{24}{28}\selectfont},
y tick label style={font=\fontsize{24}{28}\selectfont},
  xlabel={Attributes (\%)},
  ylabel={\phantom{Runtime (s)}},
  xtick={0.2,0.4,0.6,0.8,1.0},
  xticklabels={20,40,60,80,100},
  ytick={0,100,...,500},
  yticklabels={},
  grid=major,
  enlarge x limits={abs=0.7},
  legend style={
      at={(-0.05,1.02)}, 
      anchor=south west,
      legend columns=4,
      font=\large,
      fill=none,
      draw=none
  },
    grid=major,
    enlarge x limits=0.15,
    xtick pos=left,
    ytick pos=left,
]

\addplot+[fill=purple!60!black, draw=black] coordinates {
  (0.2,0.0000)
  (0.4,0.0000)
  (0.6,0.0000)
  (0.8,2.4742)
  (1.0,2.5883)
};

\addplot+[fill=purple!20!white, draw=black, postaction={pattern=north east lines}] coordinates {
  (0.2,0.0000)
  (0.4,0.0000)
  (0.6,0.0000)
  (0.8,0.0000)
  (1.0,2.8619)
};
\addplot+[fill=blue!60, draw=black] coordinates {
  (0.2,0.0000)
  (0.4,0.0000)
  (0.6,3.3540)
  (0.8,37.2288)
  (1.0,68.0683)
};
\addplot+[fill=blue!20, draw=black, postaction={pattern=north east lines}] coordinates {
  (0.2,7.6348)
  (0.4,6.8682)
  (0.6,9.7580)
  (0.8,15.0182)
  (1.0,15.9796)
};
\addplot+[fill=green!60!black, draw=black] coordinates {
  (0.2,8.2704)
  (0.4,8.1352)
  (0.6,14.6462)
  (0.8,35.4537)
  (1.0,161.0900)
};
\addplot+[fill=green!20!white, draw=black, postaction={pattern=north east lines}] coordinates {
  (0.2,12.3594)
  (0.4,12.5614)
  (0.6,12.3169)
  (0.8,13.3029)
  (1.0,18.7541)
};
\addplot+[fill=orange!80!black, draw=black] coordinates {
  (0.2,30.0249)
  (0.4,247.6962)
  (0.6,483.3563)
  (0.8,460.0636)
  (1.0,488.6474)
};
\addplot+[fill=orange!30!white, draw=black, postaction={pattern=north east lines}] coordinates {
  (0.2,12.1158)
  (0.4,23.5718)
  (0.6,35.0790)
  (0.8,38.7217)
  (1.0,40.0975)
};

\end{axis}
\end{tikzpicture}
}
\end{subfigure}%

\vspace{0mm}
\begin{tikzpicture}
\begin{axis}[
  hide axis,
  xmin=0, xmax=1,
  ymin=0, ymax=1,
  area legend,
  legend columns=4,
  legend style={
      /tikz/every even column/.append style={column sep=0.3cm},
      draw=none,
      at={(0.5,1.2)},
      anchor=south,
      font=\large
  }
]

\pgfplotsset{
  legend cell align={left},
  legend image post style={xscale=0.8},
  legend image code/.code={
    \draw[#1, yshift=-0.2em] (0cm,0cm) rectangle (0.25cm,0.15cm);
  },
  legend style={
    /tikz/every even column/.append style={column sep=2pt},
    row sep=-2pt,
  },
}

\addlegendimage{ybar,fill=purple!60!black,draw=black}
\addlegendentry{\sysName (House)}
\addlegendimage{ybar,fill=purple!20!white,draw=black,postaction={pattern=north east lines}}
\addlegendentry{\sysNameP (House)}
\addlegendimage{ybar,fill=blue!60,draw=black}
\addlegendentry{\sysName (Video)}
\addlegendimage{ybar,fill=blue!20,draw=black,postaction={pattern=north east lines}}
\addlegendentry{\sysNameP (Video)}
\addlegendimage{ybar,fill=green!60!black,draw=black}
\addlegendentry{\sysName (Marketing)}
\addlegendimage{ybar,fill=green!20!white,draw=black,postaction={pattern=north east lines}}
\addlegendentry{\sysNameP (Marketing)}
\addlegendimage{ybar,fill=orange!80!black,draw=black}
\addlegendentry{\sysName (CoverType)}
\addlegendimage{ybar,fill=orange!30!white,draw=black,postaction={pattern=north east lines}}
\addlegendentry{\sysNameP (CoverType)}
\end{axis}
\end{tikzpicture}

\vspace{-7cm}
\caption{\sysName and \sysNameP runtime (s) w.r.t (Left) \#tuples, (Right)
\#attributes. We used $k=5$ and $\theta=0.1$ for these experiments.}
\label{fig:scalability_plots_appendix}
\end{figure*}

\subsection{Offline Preprocessing}

\begin{table}[t]
\centering
\caption{Accuracy comparison across datasets and metrics.}
\begin{tabular}{lrrr}
\toprule
\textbf{Dataset} & \textbf{Surprise} & \textbf{Correlation} & \textbf{Ratio} \\
\midrule
    Marketing  & 0.65 & 0.89 & 0.88 \\
    Video      & 0.61 & 0.92 & 0.91 \\
    House      & 0.63 & 0.88 & 0.87 \\
    Covertype  & 0.66 & 0.90 & 0.90 \\
\bottomrule
\end{tabular}
\label{tab:offline_accuracy}
\end{table}

The offline time for generating prompts and training a LLM-proxy-cache
classifier is as follows. We generated 10,000 prompts for correlation, ratio,
and surprise detection, and trained separate decision tree classifiers that
incorporate prompt variables as described in the likelihood prompt. The total
time taken was 2742.70 seconds for surprise and 3171.92 seconds for trend in
the marketing dataset, 1404.61 seconds for surprise and 3129.57 seconds for
trend in the video dataset, 3599.58 seconds for surprise and 3087.07 seconds
for trend in the house dataset, and 1473.68 seconds for surprise and 3569.50
seconds for trend in the cover type dataset. Table~\ref{tab:offline_accuracy}
reports the accuracy of our method across the three evaluation metrics for each
dataset. The results are consistent across datasets, with the surprise metric
showing moderate accuracy, while both the correlation and ratio metrics achieve
relatively high accuracy.

The offline pruning step takes 1.57 seconds for the house dataset (780{,}840
combinations), 0.04 seconds for the video dataset (2{,}145 combinations), 2.73
seconds for the marketing dataset (43{,}848 combinations), and 16.51 seconds for
the cover type dataset (2{,}785{,}120 combinations). Although the house dataset
contains far more combinations than the marketing dataset, its pruning time is
smaller. This is because the pruning process first computes attribute
significance, and only if the attribute is considered interesting does it
proceed to compute interpretability. For the house dataset, only a small number
of attributes are considered significant, resulting in far fewer
interpretability computations and therefore a shorter pruning time compared to
the marketing dataset.

\subsection{Experiments on a Large Dataset}

\begin{table}[t]
\centering
\begin{tabular}{l|l|rrrrrrr}
\toprule
 & & \#PT & T(s) & Ins & Int & Util & $\underset{\text{m-dist}}{\text{Div}}$ & $\underset{\text{heatmap}}{\text{Div}}$\\
\midrule
\multirow{2}{*}{\rotatebox{90}{$k{=}3$}} 
& \sysName  & 3  & 489 & 3.00 & 2.23 & 2.62 & 0.30 & \hmGreedyKThreeCover\\[2mm] 
& \sysNameP & 3  & 40  & 2.83 & 2.23 & 2.53 & 0.32 & \hmPGreedyKThreeCover\\

\midrule
\multirow{2}{*}{\rotatebox{90}{$k{=}5$}} 
& \sysName  & 5  & 489  & 5.00 & 3.72 & 4.36 & 0.20 & \hmGreedyKFiveCover\\[2mm] 
& \sysNameP & 5  & 40   & 4.98 & 3.72 & 4.35 & 0.21 & \hmPGreedyKFiveCover\\

\midrule
\multirow{2}{*}{\rotatebox{90}{$k{=}10$}} 
& \sysName  & 10  & 489  & 9.81 & 7.44 & 8.63 & 0.20 & \hmGreedyKTenCover\\[2mm] 
& \sysNameP & 10  & 40   & 6.77 & 7.44 & 7.11 & 0.20 & \hmPGreedyKTenCover\\

\bottomrule
\end{tabular}
\caption{Utility scores of \sysName and \sysNameP on the Cover Type dataset for different values of $k$. We set $\theta = 0.3$ for $k = 3$, $\theta = 0.2$ for $k = 5$, and $\theta = 0.1$ for $k = 10$.}
\label{tab:cover_type}
\end{table}

Table~\ref{tab:cover_type} shows the results on the cover type dataset. Although
the utility score for \sysNameP is slightly lower than that of \sysName (upto
18\% decrease), the runtime is reduced significantly, decreasing from 489
seconds to only 40 seconds. This demonstrates that \sysNameP provides a
substantial efficiency improvement while still maintaining competitive utility
performance.

\subsection{LLM Prompts}
Here we present the LLM prompts used for Attribute Significance, Semantic
Validity, and Likelihood. The details are shown in Figures~\ref{fig:prompt1},
\ref{fig:prompt2}, and \ref{fig:prompt3}. The placeholders \texttt{\{\}} in each
prompt represent variables that should be filled in when generating actual
prompts. For example, in Figure~\ref{fig:prompt1}, \texttt{\{table\}} represents
the pivot table. Figure~\ref{fig:prompt3} shows the ranking of aggregate
functions for the Semantic Validity score. The results of the marketing dataset
include the following attributes: [`Education', `Income', `NumWebPurchases',
`NumCatalogPurchases', `NumStorePurchases', `NumWebVisitsMonth', `AcceptedCmp1',
`AcceptedCmp2', `AcceptedCmp3', `AcceptedCmp4', `AcceptedCmp5', `Country',
`Response', `Year\_Birth', `Marital\_Status', `NumDealsPurchases', `Complain']
The results of the video dataset include: [`Genre', `Year', `NA\_Sales',
`EU\_Sales', `JP\_Sales', `Global\_Sales', `Publisher'] The results of the house
dataset include: [`MSSubClass', `OverallQual', `GrLivArea', `YearBuilt',
`YearRemodAdd', `FullBath', `SalePrice', `HalfBath', `BedroomAbvGr',
`KitchenAbvGr', `Id', `OverallCond']

\begin{figure*}[t]
\centering
\begin{tcolorbox}[
  title=Task: Attribute Significance,
  colback=gray!5,
  colframe=black!40,
  width=\textwidth,
  boxrule=0.8pt,
  fonttitle=\bfseries,
  breakable,
  sharp corners,
  enhanced jigsaw,
  before upper={\ttfamily}, 
  left=2mm,
  right=2mm,
  top=1mm,
  bottom=1mm
]

Task: Identify the attributes that are likely to be useful for grouping or aggregating the data in meaningful ways. Please identify at least one relevant attribute. Only return the result in the following JSON format: \{"chosen\_columns": "<a list of names for interesting column to be analyzed>"\}. \par

\# Input:\par

**Table:**  \par
   \{table\} \par

 Only return the result in the following JSON format: \{"chosen\_columns": "<a list of names for interesting column to be analyzed>"\}. \par

\# Output:\par
\end{tcolorbox}
\caption{Prompt for Attribute Significance}
\label{fig:prompt1}
\end{figure*}

\begin{figure*}[t]
\centering
\begin{tcolorbox}[
  title=Task: Likelihood,
  colback=gray!5,
  colframe=black!40,
  width=\textwidth,
  boxrule=0.8pt,
  fonttitle=\bfseries,
  breakable,
  sharp corners,
  enhanced jigsaw,
  before upper={\ttfamily}, 
  left=2mm,
  right=2mm,
  top=1mm,
  bottom=1mm
]
\textbf{Task Description:} \{groupA\}and \{groupB\}on \{aggregate function\}  \{value\_attribute\} have a correlation of \{magnitude\}. Based on this, how likely is it that this correlation is desirable? Choose the appropriate likelihood from following scale. Return the result as JSON in the following format:\{"likelihood": "<one likelihood from the scale>"\}. Please return only the JSON output. Do not include explanations, code, or the full table.\par

    \# Input:\par
    **Likelihood Scale:**\par
    Very Likely\par
    Likely\par
    Neutral\par
    Unlikely\par
    Very Unlikely\par
            
    Return the result as JSON in the following format:\{"likelihood": "<one likelihood from the scale>"\}.\par
            
    \# Output:\par
\end{tcolorbox}
\caption{Prompt for Likelihood}
\label{fig:prompt2}
\end{figure*}

\begin{figure*}[t]
\centering
\begin{tcolorbox}[
  title=Task: Semantic validity,
  colback=gray!5,
  colframe=black!40,
  width=\textwidth,
  boxrule=0.8pt,
  fonttitle=\bfseries,
  breakable,
  sharp corners,
  enhanced jigsaw,
  before upper={\ttfamily}, 
  left=2mm,
  right=2mm,
  top=1mm,
  bottom=1mm
]
\textbf{Task Description:} Given an input column of data and a list of candidate aggregation functions, rank the aggregation functions based on their suitability for summarizing the given column. Return only the ranked list of aggregation functions (from most to least suitable), using only functions from the candidate list. Do not return the entire table or any explanation only the ranked list. Return the result as JSON in the following format:\{"ranked\_aggregation\_functions": ["<aggregation function 1>", "<aggregation function 2>", ...]\}. Please return only the JSON output. \par

    \# Input: \par
    **Column:** \par
   \{column\} \par

   **Candidate aggregation function:** \par
    MEAN\par
    SUM\par
    COUNT\par
    MIN\par
    MAX\par
            
    Return the result as JSON in the following format:\{"ranked\_aggregation\_functions": ["<aggregation function 1>", "<aggregation function 2>", ...]\}.\par
            
    \# Output:\par
\end{tcolorbox}
\caption{Prompt for Semantic validity}
\label{fig:prompt3}
\end{figure*}

\subsection{Datasets}

\smallskip\noindent\textbf{Marketing.} The Marketing
dataset~\cite{marketing-campaigns-dataset} comprises demographic and behavioral
information about customers, including marital status and purchase history. It
consists of 2{,}240 tuples and 28 attributes, encompassing a diverse set of
features: 9 numerical attributes (e.g., \texttt{Year\_Birth}, \texttt{Income})
and 19 categorical attributes (e.g., \texttt{Education},
\texttt{Marital\_Status}, \texttt{Complaint}).

\smallskip\noindent\textbf{Video.} The Video
dataset~\cite{videogamesales-dataset} comprises video game sales from various
countries, platforms, or release years. It contains 16{,}600 tuples across 11
attributes. Out of these, 7 attributes are categorical (e.g.,
\texttt{Platform}, \texttt{Genre}, \texttt{Publisher}) and 4 are numerical
(e.g., \texttt{Rank}, \texttt{Year}, \texttt{NA\_Sales},
\texttt{Global\_Sales}).

\begin{sloppypar} \smallskip\noindent\textbf{House.} The House
dataset~\cite{house-sale-dataset} contains information about residential
property sales, comprising 1{,}460 tuples and 81 attributes. Among the
attributes, 23 are numerical (e.g., \texttt{SalePrice}, \texttt{LotArea},
\texttt{OverallQual}, \texttt{GrLivArea}) and 58 are categorical (e.g.,
\texttt{Street}, \texttt{HouseStyle}, \texttt{RoofStyle}, \texttt{GarageType}).
\end{sloppypar}

\smallskip\noindent\textbf{Cover Type.} The Covertype
dataset~\cite{covertype-dataset} is a large dataset with $581{,}000$ tuples and
$110$ attributes, containing tree observations across four areas of the
National Forest in Colorado for. We used this large dataset to stress test the
scalability of \sysNameP.

\subsection{Baselines}
Below, we list details regarding the baselines:

\smallskip\noindent{\textbf{Brute-Force.}} We employ an exhaustive Brute-Force
search, which considers all possible k-sized pivot table sets and follows a
naive exhaustive approach to solve Problem~\ref{eq:objective_function}. Note
that in the absence of an absolute ground-truth, the results obtained from this
technique can be treated as the optimal solution.

\smallskip\noindent{\textbf{Top-k.}} This baseline ranks candidate pivot tables
in descending order of their utility scores and selects the top-$k$ as the
recommended set, without accounting for the diversity constraint. To ensure a
fair comparison, we apply our optimization techniques (pruning and LLM-proxy)
prior to the selection phase. This allows it to generate recommendations within
a runtime comparable to that of \sysName.

\smallskip\noindent{\textbf{LLM.}} We used Llama-3-8B-Instruct
(Meta)~\cite{llama-3-1-8b-instruct-model} as a transformer-based large language
model (LLM) as another baseline. While LLMs are increasingly used for tabular
data tasks, few tools address our specific problem. We prompted the model with a
small data sample and asked for interesting and diversified pivot tables in
natural language, according to our problem setup.

\smallskip\noindent{\textbf{DAISY.}} DAISY~\cite{DAISYVLDB24Junjie} is a query
recommendation algorithm proposed in prior work. It first generates a set of
random queries and clusters them based on similarity. From each cluster,
representative queries are sampled. Human annotators are then employed to
evaluate pairs of queries: if the left query is more interesting, the pair is
labeled as 1; if the two are similar, as 0; and if they are too different, as
-1. However, the original training data and model used by DAISY are not
publicly available. To address this limitation, we developed our own version of
the DAISY model. Specifically, we constructed training data using the
Auto-Suggest Pivot Table framework~\cite{AutoSuggestSIGMODE2020Cong}. Since
DAISY's training approach resembles contrastive learning—by labeling pairs of
queries as more or less interesting—we mimicked this by treating pivot tables
crawled from the web as insightful. We then randomly selected columns to
replace parts of the original table, generating less-interesting alternatives.
Following DAISY's classification framework, we trained a model to distinguish
between insightful and less-insightful queries. While our implementation is not
identical to the original DAISY model, it captures its core idea. Our trained
model achieved 80\% accuracy on a dataset containing 541{,}599 instances, with
433{,}279 used for training and 108{,}320 for testing.

\smallskip\noindent{\textbf{Microsoft Excel.}} Microsoft
Excel~\cite{microsoft_excel} is a widely used commercial spreadsheet software
that offers built-in pivot table recommendations based on the underlying data.
We used the Windows version 2501, evaluated during February 2025.

\smallskip\noindent{\textbf{PowerBI.}} Microsoft
PowerBI~\cite{PowerBI,QuickInsightsDing19} is a business intelligence software
that offers ``quick insights''~\cite{QuickInsightsDing19} in various forms
including visualizations. For comparison, we focus only on the recommendations
where the underlying query excludes the \texttt{WHERE} clause, which aligns with
our settings.

\smallskip\noindent{\textbf{Google Sheets.}} Google Sheets~\cite{google_sheets}
is an online spreadsheet software known for easy collaboration. Google Sheets
provide automatic recommendation of pivot tables for a dataset. We used the
browser version during the month of February, 2025.

For Excel, PowerBI, and Google Sheets, $k$ cannot be controlled. Thus, we
consider the first-k items when more than k were returned. Also, Google Sheets
fail to produce more than 3 recommendations, this we omit it when $k > 3$.

\subsection{Case Studies} Here, we present case study results
comparing \sysName to commercial software and LLMs on marketing
data~\ref{fig:case_study}. As shown in our main paper case study, Excel returns
``SUM(Income) GROUP BY (Country)'' which produces predictable results since
higher GDP countries typically have higher incomes, while LLM suggests
``MEAN(Income) GROUP BY (Year\_Birth)'' which expectedly shows higher income
for mid-age groups. In contrast, \sysName recommends "SUM(Income) GROUP BY
(Complain, Education)" which reveals unexpected relationships between
complaints and education levels relative to income. Both LLM and PowerBI return
meaningless ``GROUP BY(ID)'' aggregations.

\begin{figure*}[t]
    \resizebox{\textwidth}{!}{%
    \centering
    \scriptsize
    \begin{tabular}{ll@{ }p{9cm}} 
		\toprule 
		\textbf{Tool} & \multicolumn{2}{l}{\textbf{Recommended Pivot Tables}} \\ 
		\midrule
		
		\multirow{2}{*}{Google Sheets} 
		& (1) & SUM(NumDealsPurchases, NumCatalogPurchases, NumStorePurchases) GROUP BY (Country) \\
        & (2) & COUNT(ID) GROUP BY (Country) \\
		
		\midrule
		
        \multirow{7}{*}{Microsoft Excel} 
        & (1) & SUM(Income) GROUP BY (Country) \\
        & (2) & SUM(MntWines) GROUP BY (Country) \\
        & (3) & SUM(MntFishProducts, MntSweetProducts, MntGoldProds) GROUP BY (Country) \\
        & (4) & SUM(MntFruits, MntMeatProducts, MntFishProducts) GROUP BY (Education) \\
        & (5) & SUM(MntMeatProducts, MntFishProducts, MntSweetProducts) GROUP BY (Marital\_Status) \\
        & (6) & SUM(MntWines, Income, Year\_Birth) GROUP BY (Marital\_Status) \\
        & (7) & SUM(Year\_Birth) GROUP BY (Country, Teenhome) \\
		
		\midrule 

        \multirow{9}{*}{PowerBI}
        & (1) & COUNT(Dt\_Customer) GROUP BY (ID, NumDealsPurchases) \\
        & (2) & SUM(Teenhome) GROUP BY (Country) \\
        & (3) & SUM(MntFruits) GROUP BY (Education, ID) \\
        & (4) & COUNT(Dt\_Customer) GROUP BY (ID, Income) \\
        & (5) & SUM(NumDealsPurchases) GROUP BY (Education) \\
        & (6) & SUM(Kidhome) GROUP BY (Education) \\
        & (7) & COUNT(Year\_Birth) GROUP BY (ID, Teenhome) \\
        & (8) & SUM(Teenhome) GROUP BY (Education) \\
        & (9) & COUNT(ID) GROUP BY (Education) \\
		
		\midrule
		
        \multirow{10}{*}{LLM} 
        & (1) & MIN(MntMeatProducts) GROUP BY (Recency) \\
        & (2) & MEAN(NumWebPurchases) GROUP BY (Teenhome) \\
        & (3) & SUM(NumStorePurchases) GROUP BY (Complain) \\
        & (4) & MAX(NumDealsPurchases) GROUP BY (Education) \\
        & (5) & MEAN(Income) GROUP BY (Year\_Birth) \\
        & (6) & COUNT(MntWines) GROUP BY (Country) \\
        & (7) & MEDIAN(Income) GROUP BY (NumWebVisitsMonth) \\
        & (8) & STD(Income) GROUP BY (AcceptedCmp1) \\
        & (9) & SUM(Income) GROUP BY (Marital\_Status) \\
        & (10) & COUNT(Recency) GROUP BY (MntFruits) \\

        \midrule
		
        \multirow{10}{*}{\sysName} 
        & (1) & COUNT(NumStorePurchases) GROUP BY (Complain) \\
        & (2) & COUNT(AcceptedCmp5) GROUP BY (Complain) \\
        & (3) & COUNT(Year\_Birth) GROUP BY (Complain) \\
        & (4) & SUM(AcceptedCmp5) GROUP BY (Complain) \\
        & (5) & MEAN(AcceptedCmp2) GROUP BY (AcceptedCmp4) \\
        & (6) & SUM(Income) GROUP BY (Complain, Education) \\
        & (7) & SUM(Complain) GROUP BY (AcceptedCmp4, AcceptedCmp5) \\
        & (8) & COUNT(Education) GROUP BY (Marital\_Status) \\
        & (9) & SUM(NumWebVisitsMonth) GROUP BY (Marital\_Status) \\
        & (10) & COUNT(NumStorePurchases) GROUP BY (AcceptedCmp3, Education) \\
        \bottomrule
		
    \end{tabular}}
	 \vspace{-3mm}
	 \caption{\small Results of \sysName, commercial software, and LLM on marketing data.}
	 \vspace{-4mm}	 
    \label{fig:case_study}
\end{figure*}


\subsection{Related Work}
We detailed the standards from Figure~\ref{fig:related_work} in
Table~\ref{tab:comparison_pivot_tables}. Current commercial software has
significant limitations in delivering smart recommendations. Microsoft Excel
and Google Sheets provide redundant \texttt{GROUP BY} attributes and prefer
numerical over categorical attributes, creating meaningless aggregations like
\texttt{SUM(YEAR)}. PowerBI identifies statistical trends but lacks diverse
selections and suffers from complexity issues. Tableau requires prior user
query logs and manual value selection. Large Language Models (LLMs) like
ChatGPT present additional challenges: they hallucinate attributes and result
tables, lack inherent scoring mechanisms, and fail to exhaustively evaluate
candidates without explicit user-defined logic in prompts. These systems
exhibit critical limitations including poor semantic understanding that leads
to meaningless aggregations and low-quality recommendations that focus on
schema rather than content.

\onecolumn
\begin{sidewaystable}
    \centering
    \vspace*{20cm}
    \resizebox{\textheight}{!}{
    \begin{tabular}{p{4cm}p{6cm}cccccccccc}
        \toprule
        \textbf{Property} & \textbf{Standards} & \textbf{Excel (Mac OS)} & \textbf{Excel (MS OS)} & \textbf{GSheet} & \textbf{Power BI} & \textbf{Tableau} & \textbf{ChatGPT} & \textbf{Llama3-instruct} & \textbf{DAISY} & \textbf{AutoSuggest} & \textbf{TableGPT} \\
        \midrule
        Recommends pivot tables & If it always recommends pivot tables as their RESULTS, then mark them as \always. If it only recommends queries, then it is marked as \sometimes. If it only recommends other visualizations, then it is marked as \rarely. If it never includes pivot tables, then mark it as \never. & \always & \always & \always & \rarely & \always & \always & \always & \sometimes & \always & \always \\
        Multiple recommendations & If it includes the functionality of recommending multiple tables, then mark it as \always, if not \never. & \never & \always & \always & \always & \always & \always & \always & \always & \never & \always \\
        Budgeted recommendations & If it has a functionality where users can choose the number of pivot tables, then mark it as \always, if not \never. & \never & \never & \never & \never & \never & \always & \always & \always & \never & \always \\
        Valid and Useful recommendations & Is there any recommendation including invalid or useless results such as hallucinations or summing IDs? If they explicitly care about this, we say \always. If they trained on user logs or create hallucinations, we say \sometimes. Otherwise, \never. & \sometimes & \sometimes & \sometimes & \sometimes & \unknown & \sometimes & \sometimes & \unknown & \unknown & \sometimes \\
        Interpretable Recommendations & Does the software explicitly consider interpretability (e.g., avoiding sparse or large tables)? If not, mark \never. If it requires explicit user input, mark \sometimes. Otherwise, mark \always. & \never & \always & \never & \never & \never & \always & \sometimes & \unknown & \always & \sometimes \\
        Adaptive recommendations & Does the software allow adaptive recommendations based on user input? If yes, mark \always; otherwise, \never. & \never & \never & \never & \never & \never & \always & \always & \never & \always & \always \\
        Allows user attribute specifications & Can the user specify attributes of interest? If yes, mark \always; otherwise, \never. & \never & \never & \never & \never & \always & \always & \always & \never & \never & \always \\
        Diversity Aware & Does the software explicitly support diverse recommendations? If yes, mark \always. If the system uses such functionality but we cannot verify it, mark \unknown. & \unknown & \unknown & \unknown & \never & \never & \always & \always & \never & \never & \always \\
        Attribute-name semantics aware & Does the software consider attribute names when generating pivot tables? If different names lead to different recommendations, mark \always; otherwise, \never. & \never & \always & \always & \always & \never & \always & \always & \never & \always & \always \\
        Attribute-order insensitive & Is the system insensitive to attribute order? If order changes affect results, mark \never. If the system is robust to ordering, mark \always. & \never & \never & \never & \always & \always & \never & \never & \never & \never & \never \\
        Data syntactic aware & Does the system recommend tables based on syntactic data patterns? If yes, mark \always; otherwise, \never. & \always & \always & \always & \always & \always & \always & \always & \always & \always & \always \\
        Data semantic aware & Does the system recommend tables based on data semantics? If it does so correctly, mark \always. If it attempts but often fails, mark \sometimes. Otherwise, \never. & \sometimes & \always & \never & \always & \never & \sometimes & \sometimes & \always & \never & \sometimes \\
        No additional requirements & Can users obtain pivot tables with a single action? If yes, mark \always; otherwise, \never. & \always & \always & \always & \always & \never & \never & \never & \never & \never & \never \\
        Free (no cost) & Does the system require additional resources such as GPU or quotas? If only hardware is needed, mark \sometimes. If limited by quotas or data collection, mark \never. & \never & \never & \never & \never & \never & \never & \sometimes & \never & \never & \sometimes \\
        Open-source (transparent) & Is the system's code openly available (e.g., GitHub)? If yes, mark \always; otherwise, \never. & \never & \never & \never & \never & \never & \never & \always & \never & \never & \always \\
        \bottomrule
    \end{tabular}
    }
    \caption{Comparison of Pivot Table Recommendation Systems with Standards Included}
    \label{tab:comparison_pivot_tables}
\end{sidewaystable}

\end{document}